\documentclass[12pt,a4paper]{article}

\usepackage{amsmath}
\usepackage{amssymb,xfrac,mathrsfs}
\usepackage{graphicx}
\usepackage{caption}
\usepackage{subcaption}
\usepackage{bm}
\usepackage[bbgreekl]{mathbbol}
\usepackage{cite,color,float}
\usepackage{enumitem}
\usepackage{color}
\usepackage{comment}
\usepackage[toc,page]{appendix}
\usepackage{tikz}
\usepackage{slashed}
\usetikzlibrary{decorations.markings}
\usepackage{manfnt}

\newcommand\shrink{\hspace{.4em}}

\def\qmb{\bm{q}\!\cdot\!\bm{M}}
\def\qmpb{\bm{q'}\!\!\cdot\!\bm{M}}
\def\sqmb{\slashed{\bm{q}}}
\def\sqmpb{\slashed{\bm{q}}\bm{'}}

\def\vred{\displaystyle{\widehat{\mathop{V}_-}}}
\def\rd{\mathrm{d}}
\def\ri{\mathrm{i}}
\def\I{\mathrm{I}}

\def\np{N_\mathrm{P}}
\def\nl{N_\mathrm{L}}

\tikzset{    arrow/.style={decoration={markings, mark=at position 1 with
    {\fill(-0.09*#1,-0.03*#1) -- (0,0) -- (-0.09*#1,0.03*#1) -- cycle;}}, postaction={decorate}},
    arrow/.default=1}

\def\linkcur{\tikz[baseline=-.6ex]{\draw[arrow=1.8] (-.5,0) -- (0.1,0);\draw[-] (.1,0) -- (.5,0);
}}

\def\linkcurinv{\tikz[baseline=-.6ex]{\draw[-] (-.5,0) -- (0.1,0);\draw[arrow=1.8] (.5,0) -- (-.1,0);
}}

\def\linkdash{\tikz[baseline=-.6ex]{\draw[arrow=1.8,dashed] (-.5,0) -- (0.1,0);\draw[dashed] (.1,0) -- (.5,0);
}}

\def\linkdashinv{\tikz[baseline=-.6ex]{\draw[dashed] (-0.1,0) -- (-.5,0);\draw[arrow=1.8,dashed] (.5,0) -- (-.1,0);
}}

\def\circ{\tikz[baseline=-.5ex]{
		\draw [dashed] circle (.412); \draw[arrow=1.8] (.32,.25) -- (.3,.28); \draw[arrow=1.8] (-.32,-.25) -- (-.3,-.28);}}
	
\def\circinv{\tikz[baseline=-.5ex]{
		\draw [dashed] circle (.412); \draw[arrow=1.8](.3,.28)  -- (.33,.25); \draw[arrow=1.8] (-.3,-.28) -- (-.33,-.25);}}

\def\rec{\tikz[baseline=-.5ex]{\draw (-.55,-.55) rectangle (.55,.55);
		\draw[arrow=1.8] (.1,.55) -- (-.1,.55); \draw[arrow=1.8] (-.1,-.55) -- (.1,-.55);
		\draw[arrow=1.8] (.55,-.1) -- (.55,.1); \draw[arrow=1.8] (-.55,.1) -- (-.55,-.1);}}
	
\def\recinv{\tikz[baseline=-.5ex]{\draw (-.55,-.55) rectangle (.55,.55);
		\draw[arrow=1.8] (-.1,.55) -- (.1,.55) ; \draw[arrow=1.8] (.1,-.55) -- (-.1,-.55);
		\draw[arrow=1.8] (.55,.1) -- (.55,-.1); \draw[arrow=1.8] (-.55,-.1) -- (-.55,.1);}}

\def\recdashcirc{\tikz[baseline=-.5ex]{\draw (-.55,-.55) rectangle (.55,.55);
\draw[arrow=1.8] (.1,.55) -- (-.1,.55); \draw[arrow=1.8] (-.1,-.55) -- (.1,-.55);
\draw[arrow=1.8] (.55,-.1) -- (.55,.1); \draw[arrow=1.8] (-.55,.1) -- (-.55,-.1);
\draw [dashed] circle (.412); \draw[arrow=1.8] (.3,.28) -- (.33,.25); \draw[arrow=1.8] (-.3,-.28) -- (-.33,-.25);}}

\def\recdashcircinv{\tikz[baseline=-.5ex]{\draw (-.55,-.55) rectangle (.55,.55);
\draw[arrow=1.8] (-.1,.55) -- (.1,.55); \draw[arrow=1.8] (.1,-.55) -- (-.1,-.55);
\draw[arrow=1.8] (.55,.1) -- (.55,-.1); \draw[arrow=1.8] (-.55,-.1) -- (-.55,.1);
\draw [dashed] circle (.412); \draw[arrow=1.8] (.33,.25) -- (.3,.28); \draw[arrow=1.8] (-.33,-.25) -- (-.3,-.28);}}

\def\dashreccirc{\tikz[baseline=-.5ex,scale=0.8]{\draw (-.55,-.55) [dashed] rectangle (.55,.55);
\draw[arrow=1.8] (-.1,.55) -- (.1,.55); \draw[arrow=1.8] (.1,-.55) -- (-.1,-.55);
\draw[arrow=1.8] (.55,.1) -- (.55,-.1); \draw[arrow=1.8] (-.55,-.1) -- (-.55,.1);
\draw [dashed] circle (.412); \draw[arrow=1.8] (.32,.25) -- (.3,.28); \draw[arrow=1.8] (-.32,-.25) -- (-.3,-.28);}}

\def\recdashrec{\tikz[baseline=-.5ex]{\draw (-.55,-.55) rectangle (.55,.55);
\draw[arrow=1.8] (.1,.55) -- (-.1,.55); \draw[arrow=1.8] (-.1,-.55) -- (.1,-.55);
\draw[arrow=1.8] (.55,-.1) -- (.55,.1); \draw[arrow=1.8] (-.55,.1) -- (-.55,-.1);
\draw (-.4,-.4) [dashed] rectangle (.4,.4);
\draw[arrow=1.8] (-.1,.4) -- (.1,.4); \draw[arrow=1.8] (.1,-.4) -- (-.1,-.4);
\draw[arrow=1.8] (.4,.1) -- (.4,-.1); \draw[arrow=1.8] (-.4,-.1) -- (-.4,.1);}}

\def\recdashreccirc{\tikz[baseline=-.5ex]{\draw (-.55,-.55) rectangle (.55,.55);
\draw[arrow=1.8] (.1,.55) -- (-.1,.55); \draw[arrow=1.8] (-.1,-.55) -- (.1,-.55);
\draw[arrow=1.8] (.55,-.1) -- (.55,.1); \draw[arrow=1.8] (-.55,.1) -- (-.55,-.1);
\draw (-.4,-.4) [dashed] rectangle (.4,.4);
\draw[arrow=1.8] (-.1,.4) -- (.1,.4); \draw[arrow=1.8] (.1,-.4) -- (-.1,-.4);
\draw[arrow=1.8] (.4,.1) -- (.4,-.1); \draw[arrow=1.8] (-.4,-.1) -- (-.4,.1);
\draw [dashed] circle (.288); \draw[arrow=1.8] (.21,.2) -- (.29,.05); \draw[arrow=1.8] (-.21,-.2) -- (-.29,-.05);}}

\def\rectwocirc{\tikz[baseline=-.5ex]{\draw (-.55,-.55) rectangle (.55,.55);
\draw[arrow=1.8] (.1,.55) -- (-.1,.55); \draw[arrow=1.8] (-.1,-.55) -- (.1,-.55);
\draw[arrow=1.8] (.55,-.1) -- (.55,.1); \draw[arrow=1.8] (-.55,.1) -- (-.55,-.1);
\draw [dashed] circle (.412); \draw[arrow=1.8] (.3,.28) -- (.33,.25); \draw[arrow=1.8] (-.3,-.28) -- (-.33,-.25);
\draw [dashed] circle (.288); \draw[arrow=1.8]  (.21,.2)-- (.29,.05); \draw[arrow=1.8] (-.21,-.2) -- (-.29,-.05);}}

\def\rectwocircinv{\tikz[baseline=-.5ex]{\draw (-.55,-.55) rectangle (.55,.55);
\draw[arrow=1.8] (-.1,.55) -- (.1,.55); \draw[arrow=1.8] (.1,-.55) -- (-.1,-.55);
\draw[arrow=1.8] (.55,.1) -- (.55,-.1); \draw[arrow=1.8] (-.55,-.1) -- (-.55,.1);
\draw [dashed] circle (.412); \draw[arrow=1.8] (.32,.25) -- (.3,.28); \draw[arrow=1.8] (-.32,-.25) -- (-.3,-.28);
\draw [dashed] circle (.288); \draw[arrow=1.8](.29,.05)  -- (.21,.2); \draw[arrow=1.8] (-.29,-.05) -- (-.21,-.2);}}

\def\recthreecirc{\tikz[baseline=-.5ex]{\draw (-.55,-.55) rectangle (.55,.55);
\draw[arrow=1.8] (.1,.55) -- (-.1,.55); \draw[arrow=1.8] (-.1,-.55) -- (.1,-.55);
\draw[arrow=1.8] (.55,-.1) -- (.55,.1); \draw[arrow=1.8] (-.55,.1) -- (-.55,-.1);
\draw [dashed] circle (.412); \draw[arrow=1.8] (.3,.28) -- (.33,.25); \draw[arrow=1.8] (-.3,-.28) -- (-.33,-.25);
\draw [dashed] circle (.288); \draw[arrow=1.8] (.21,.2) --  (.29,.05); \draw[arrow=1.8] (-.21,-.2) -- (-.29,-.05) ;
\draw [dashed] circle (.19);}}

\def\twocirc{\tikz[baseline=-.5ex]{
\draw [dashed] circle (.412); \draw[arrow=1.8] (.3,.28) -- (.33,.25); \draw[arrow=1.8] (-.3,-.28) -- (-.33,-.25);
\draw [dashed] circle (.288); \draw[arrow=1.8] (.29,.05) -- (.21,.2); \draw[arrow=1.8] (-.29,-.05) -- (-.21,-.2);}}

\def\backforth{\tikz[baseline=-.6ex]{\draw[arrow=1.8,dashed] (0,-.5) -- (0,0);\draw [-,dashed] (0,.1) -- (0,.5);
\draw[arrow=1.8,dashed] (.1,.42) -- (.1,-.1);\draw [-,dashed] (.1,-.2) -- (.1,-.5);}}

\def\backforthhor{\tikz[baseline=-.6ex]{\draw[arrow=1.8] (-.5,0) -- (0,0);\draw[-] (-.1,0) -- (.5,0);
 \draw[arrow=1.8,dashed] (.45,.12) -- (-.08,.12);\draw[-,dashed]  (-.0,.12) -- (-.5,.12);}}

\begin{document}

~
\vskip 1cm 

\begin{center}
\noindent {\large\textbf{Diagrammatic Strong Coupling Expansion of  U(1) \\
		Lattice Model in Fourier Basis 
}}
\\[1.5\baselineskip]
Afsaneh Kianfar
~~~~~and~~~~~ 
Amir H. Fatollahi~\footnote{Corresponding Author: fath@alzahra.ac.ir}
\\[1.5\baselineskip]
\textit{Department of Physics, Faculty of Physics and Chemistry, \\
Alzahra University, Tehran 1993891167, Iran}
\\[2\baselineskip]
\begin{abstract}

\noindent
The transfer-matrix of U(1) lattice gauge theory is investigated in the field Fourier 
space, the basis of which consists of the quantized currents on lattice links.
Based on a lattice version of the current conservation, 
the transfer-matrix elements are shown to be non-zero only between current-states that
differ in circulating currents inside plaquettes. 
In the strong coupling limit, a series expansion is developed for the elements of the transfer-matrix, 
to which a diagrammatic representation based on the occurrence of virtual link 
and loop currents can be associated. With $g$ as the coupling, the weight of each virtual 
current in the expansion is $1/g^2$, by which at any given order the relevant diagrams
are determined. 
Either by interpretation or through their role in fixing the relevant terms,
the diagrams are reminiscent of the Feynman ones of the perturbative small coupling expansions.
\end{abstract}
\end{center}

\vskip 2cm


 \newpage  

\section{Introduction}

\begin{figure}[H]
	\begin{center}
		\includegraphics[scale=.85]{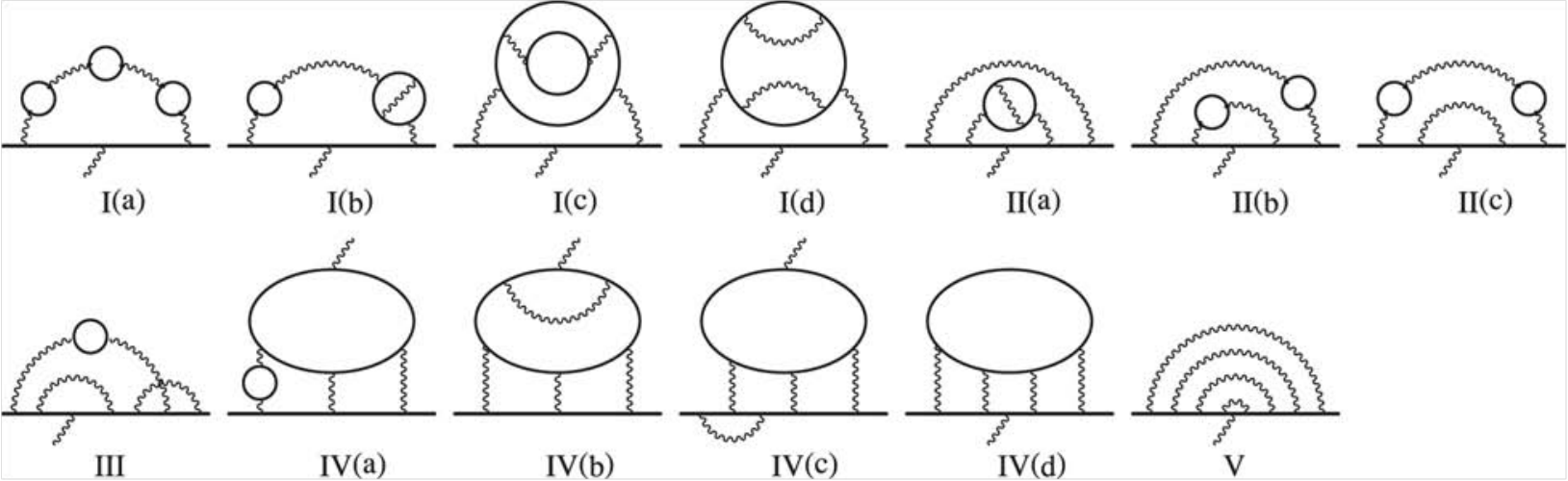}
\vskip 0cm  \caption{\small Thirteen representatives of 891 Feynman diagrams 
contributing the lepton's $g-2$ at the order $\alpha^4$. Reprinted from \cite{kinoshita}.}
		\label{fig1}
	\end{center}
\end{figure}

According to the perturbative formulation of quantum field theories,
the transition amplitudes of physical processes are expressed by series expansions 
in the small coupling constant. Whenever applicable, the perturbative series
makes it possible to calculate the transition rates up to the desired accuracy. 
Based on the interpretation that the expansion terms are representing the space-time 
\textit{virtual events} between the initial and final states \cite{fey}, the perturbative 
expansions are commonly represented by a set of graphs, the Feynman diagrams. 
Apart from pure theoretical interests, diagrammatic 
representations have found a crucial role in determining and managing the relevant terms 
at any given order of the coupling constant. 
As an example of the role of Feynman diagrams in managing the perturbative 
expansions, representatives of 891 diagrams contributing 
to the anomalous magnetic moment of leptons at the order $e^8$ are presented in Fig.~\ref{fig1} \cite{kinoshita}.

\vskip -.5cm

The main purpose of the present work is to introduce a diagrammatic expansion for 
lattice gauge theories \cite{wilson,kogut} in the strong coupling regime. In 
particular, we consider the transfer-matrix of the U(1) 
lattice gauge theory in the field Fourier space. In the common approach, 
the elements of the transfer-matrix are defined in the field space through the Euclidean action
between two adjacent times of the discrete space-time. Due to the angle-variable nature
of the gauge fields in the lattice formulation, the Fourier conjugates of fields turn out to be integer-valued, 
being identified as quantized currents on lattice links. It is for the matrix-elements 
between these currents that the series expansion in the strong coupling
and associated diagrammatic representation are derived. With $g$ as the gauge coupling, 
the expansion parameter is $1/g^2$, which is small in the strong coupling limit. 
As we will see in detail, the expansion of matrix-elements between two current-states is 
interpreted as the occurrence of all possible 
\textit{virtual link and loop currents} that transform the current-states to vacuum
(the state with no current). In this sense, the diagrammatic expansion may be 
considered as a `current expansion'. The weight of each virtual current 
is $1/g^2$, which serves as the expansion parameter.
Either by interpretation or through managing the relevant terms in 
a given order of the strong coupling expansion, the diagrams 
play the role of Feynman diagrams at the small coupling regime.

The use of a Fourier basis in lattice gauge models dates back almost 
to the time of appearance of these models. 
The transform of \textit{plaquette degrees} to Fourier ones known
as dual variables \cite{savit}, is used to qualitatively describe the underlying mechanism governing 
the phases of the U(1) model in three and four dimensions \cite{banks}. 
In particular, performing the dual variable transform in the partition function
and using a certain small coupling model known as Villain action \cite{villain}, 
the U(1) model is reduced to that of a gas of monopoles or monopole-loops \cite{banks}.
The dual formulation of U(1) lattice model  
shows also clear advantages for various numerical purposes; see \cite{gatt} for a recent review. 
The advantages include the replacement of the continuous variables by the more tractable
integer ones, and also the more efficient and accurate calculation 
of expectation values in the presence of multiple or separated source charges 
\cite{zach1,zach2,panero,boris,casel}. 

The present strong coupling expansion is different in two respects to 
that of the perturbative approach, in which the small coupling expansion is 
associated with the elements of the $S$-matrix between initial and final states,  
usually with an infinite time separation. First, the present strong 
coupling expansion is between current-states at two adjacent times, 
as the transfer-matrix elements are defined based on the Euclidean action between the adjacent times.
The second difference is related to the fact that, the $S$-matrix element between equal states is 
not zero but supposedly infinite. In the present expansion, however,
the matrix-element between equal current-states other than vacuum 
is vanishing in the extreme strong coupling limit $1/g^2\to 0$. As will be discussed 
in detail, this simply is related to the fact that, 
for a non-zero result, the Fourier integrals related to initial and final states 
are to be non-zero independently. This promotes the vacuum state
as a seemingly passed intermediate state in all transitions. 

A manageable strong coupling expansion for the elements of the transfer-matrix 
provides a basis to find the energy spectrum of the model 
in the strong coupling limit, especially when it is combined with numerical calculations.

The present diagrammatic expansion in the strong coupling limit
is based on \cite{vadfat}, in which the elements of the transfer-matrix were 
obtained in the Fourier basis through the plaquette-link matrix. Accordingly, 
it is observed that by a lattice version of current conservation, 
emerged through the construction, the transfer-matrix elements are non-zero 
only between current-states that differ in loop-currents circulating in one or more plaquettes. 
By the mentioned version of current conservation, it is found that
the transfer-matrix is block-diagonal in the Fourier basis \cite{vadfat}.
The current-states differing only in loop-currents belong to the same block. 

The paper is organized as follows. In Sec.~2, a short review of the formulation 
of the transfer-matrix in the field Fourier basis is presented, in line with \cite{vadfat}. 
In Sec.~3, a detailed description of the emerged notions 
in the Fourier basis is discussed. In particular, the exact connection between current-states 
belonging to the same block is presented. In sec.~4, first some examples of the strong coupling expansions and then 
their elementary graphical representations are presented, and then rules are set up for diagrammatic expansions. 
Sec.~5 presents the detailed application of the given rules at higher orders in several examples. In Sec.~6 
the present expansion is used to calculate the ground-state and some excited energies 
of the model in the strong coupling regime. 
In Sec.~7, based on the observation that the lattice size dependence of expansions 
can be factored out from the matrix-elements and eigenvalues, 
the physical interpretation of the results is discussed. 
Sec.~8 is devoted to concluding remarks. 
Some derivations and extended expressions, as well as more examples of the application of expansion 
rules are presented in Appendices A, B, and C. 

\section{Review: Transfer-Matrix in Fourier Basis}
In \cite{vadfat}, a formulation of the pure U(1) lattice gauge theory in the field Fourier 
basis is presented. In particular, based on the plaquette-link matrix the elements of the
transfer-matrix $\widehat{V}$ in the Fourier basis are obtained explicitly, for which some 
mathematical statements are expressed \cite{vadfat}. 
In this section, a short review of the mathematical derivation of matrix elements
is presented. 

Following \cite{luscher,seiler}, the formulation is presented in the temporal gauge $A^0\equiv 0$, 
in which the transfer-matrix takes a simple form.
The link at site $\bm{r}$ in spatial direction `$\, i\,$' is represented by $(\bm{r},i)$. 
Conveniently, the gauge variables on the spatial link $(\bm{r},i)$ at adjacent times 
$n_t$ and $n_t+1$ are replaced by the angle variables:
\begin{align}\label{1}
\begin{aligned}\theta^{(\bm{r},i)}&=a\,g\,A_{n_t}^{(\bm{r},i)}\cr
\theta'^{(\bm{r},i)}&=a\,g\,A_{n_t+1}^{(\bm{r},i)}
\end{aligned}
\end{align}
taking values in $[-\pi,\pi]$ \cite{wilson}. 
Above, `$g$' and `$a$' are gauge coupling and lattice spacing parameters, respectively.
The Euclidean action symmetrized between $\theta$ and $\theta'$ variables 
for pure U(1) theory in temporal gauge 
on a lattice with $d$ spatial dimensions is given explicitly by \cite{luscher,seiler}
\begin{align}\label{2}
S_E(n_t,n_t+1)=- \frac{1}{2\, g^2} \sum_{\bm{r}}\sum_{i\neq j =1}^d 
& \left[2-\cos\big(\theta^{(\bm{r},i)}+\theta^{(\bm{r}+\widehat{i},j)} 
-\theta^{(\bm{r}+\widehat{j},i)}-\theta^{(\bm{r},j)}\big) \right. 
\cr
& \left. -\cos\big(\theta'^{(\bm{r}, i)}+\theta'^{(\bm{r}+\widehat{i},j)} 
-\theta'^{(\bm{r}+\widehat{j},i)}-\theta'^{(\bm{r},j)}\big)\right]\cr
-\frac{1}{g^2} &\sum_{\bm{r}}\sum_{i=1}^d 
\left[1-\cos\big(\theta^{(\bm{r},i)}-\theta'^{(\bm{r},i)}\big)\right]
\end{align}
with $\widehat{i}$ as the unit-vector along the spatial direction $i$.
For a spatial lattice with $\nl$ number of links and
$\np$ number of plaquettes, it is convenient to define the  
plaquette-link matrix $\bm{M}$ of dimension $\np\times \nl$,
as following explicitly by its elements
\begin{align}\label{3}
M^p_{~l}=\begin{cases}
                     +1,& \mbox{link $l=(\bm{r},i)$ belongs to oriented plaquette $p$ } \\
                    -1,& \mbox{link $l=(\bm{r},-i)$ belongs to oriented plaquette $p$} \\
                     ~~0,& \mbox{otherwise.}
        \end{cases}
\end{align}
\begin{figure}[t]
		\includegraphics[scale=2.2]{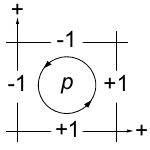}
\vskip -2cm  \caption{\small Graphical representation of definition (\ref{3}). }
\label{fig2}
\vskip 1.4cm
\end{figure}
\noindent In Fig.~\ref{fig2}, the above definition is presented graphically. 
The explicit representations for $d=2$ case will be given later.
The elements of the transfer-matrix $\widehat{V}$ is defined in terms of 
the Euclidean action between two adjacent times
\begin{align}\label{4}
\langle n_t+1 | \widehat{V} | n_t \rangle~ \propto~
e^{ S_E(n_t,n_t+1)}
\end{align}
Setting
\begin{align}\label{5}
\gamma=\frac{1}{g^2}
\end{align}
and labeling links as $l=(\bm{r},i)$ and plaquettes as `$p$', 
in terms of matrix $\bm{M}$ the elements of the transfer-matrix $\widehat{V}$
are given in the field basis by
\begin{align}\label{6}
\langle \bm{\theta'} |\widehat{V} | \bm{\theta}\rangle =~ \mathcal{A}~ &
\prod_p \exp\!\left\{-\frac{\gamma}{2}\left[2-\cos\big(M^p_{~l}\,\theta^l\big)
-\cos\big(M^p_{~l}\,\theta'^l\big)\right]\right\}\cr
&\times\prod_{l}\exp\!\left\{-\gamma\Big[1-\cos\big(\theta^l-\theta'^l\big)\Big]\right\}
\end{align}
in which the summations over repeated indices are understood. 
Above, $\mathcal{A}$ is inserted to fix the normalization \cite{normfixing}.
The aim is to formulate the theory in the
field Fourier basis $| k_l \rangle$, which is related to the compact $\theta$-basis by:
\begin{align}\label{7}
\langle\theta^{l'}|k_{l} \rangle&=\frac{\delta^{l'}_{~l}}{\sqrt{2\,\pi}}
\,\exp(\ri\,k_l\,\theta^l)
,~~~~~ 
k_l=0,\pm1,\pm2,\cdots
\end{align}
Using an identity that involves the modified Bessel function $\I_n(x)$ 
\begin{align}\label{8}
\exp(x\,\cos\phi)=\sum_n\I_n(x)\,\exp(\ri\,n\,\phi)
\end{align}
and the relation 
\begin{align}\label{9}
\int_{-\pi}^\pi\rd\theta\,\exp(\ri\,n\,\theta)=2\pi\,\delta(n)
\end{align}
one directly finds the matrix elements of $\widehat{V}$ in the field Fourier basis \cite{vadfat}
\begin{align}\label{10}
\langle\bm{k}'|\widehat{V}|\bm{k}\rangle=\mathcal{A}
\, e^{-\gamma(\np+\nl)}
\,(2\pi)^{\nl}&
\sum_{\{n_p\}}\sum_{\{n'_p\}}\prod_{p}
\I_{n_p}\!\left(\frac{\gamma}{2}\right)\,\I_{n'_p}\!\left(\frac{\gamma}{2}\right)
\nonumber\\
&\quad\;\times\prod_{l} \I_{m_l}\!(\gamma)\,
\delta\big[(n_p+n'_p)M^p_{~l}+k_l-k'_l\big],
\end{align}
in which $m_l=k_l+\sum_p n_p\,M^p_{~l}=\,k'_l-\sum_p n'_p\,M^p_{~l}$. 
Above $n_p$, $n'_p$, and $m_l$ are all integer-valued.
The general solutions of the 
$\delta$'s in the summations are given by means of the vectors satisfying \cite{vadfat}
\begin{align}\label{11}
\bm{n}^{\bm{0}}\cdot \bm{M}=\bm{0}.
\end{align}
These vectors include $\bm{n^0}=\bm{0}$. The general solution of the delta is then given as 
\begin{align}\label{12}
n_p+n'_p=Q_p+{n}^0_p,
\end{align}
with $Q_p$'s integer-valued. The above solution leads to the following condition to have a non-zero 
matrix element (\ref{10}) in the Fourier basis 
\begin{align}\label{13}
\bm{k'}=\bm{k}+\bm{Q}\cdot \bm{M}
\end{align}
Later the physical meaning of the above condition, by means of the current conservation 
between initial and final states at adjacent times, will be given. 
Accordingly, it is shown that $\widehat{V}$ in the Fourier basis is block-diagonal
\cite{vadfat}, and all elements of a block can be represented by a Fourier 
vector $\bm{k}_\ast$, whose co-blocks are all constructed as 
\begin{align}\label{14}
\bm{k}_{\ast\bm{q}}=\bm{k}_\ast + \qmb 
\end{align}
in which $\bm{q}$ is a vector with $\np$ integers as components. 
We will see that the condition that two co-blocks differ as above
is a manifestation of the current conservation, allowing us to have a non-zero matrix-element.
By setting $\bm{Q}=\bm{q'}-\bm{q}$, we have two co-block vectors 
\begin{align}\label{15}
\bm{k}_{\ast\bm{q}}=\bm{k}_\ast + \qmb,~~~~
\bm{k'}_{\ast \bm{q'}}=\bm{k}_\ast + \qmpb
\end{align}
for which the matrix-element, using $\I_m=\I_{-m}$ and $n_p+q_p\to n_p$, is given by \cite{vadfat}
\begin{align}\label{16}
\langle\bm{k'}_{\ast\bm{q'}}|\widehat{V}|\bm{k}_{\ast\bm{q}}\rangle_{\bm{k}_\ast}=\mathcal{A}\,
 e^{-\gamma(\np+\nl) }  (2\pi)^{\nl}
\sum_{\{n^0_p\}}   \sum_{\{n_p\}}
& \prod_{p}\I_{q_p-n_p}\!\left(\frac{\gamma}{2}\right)\I_{q'_p-n_p+n^0_p}
\!\left(\frac{\gamma}{2}\right) 
\cr &
\prod_{l} \I_{k_\ast+\sum_p \!\! n_p M^p_{~l}}\!(\gamma)
\end{align}
in which no constraint on the summations is present, except that $n^0_p$'s satisfy (\ref{11}).
The important fact is that the allowed ${n}^0_p$'s are not dependent on $\bm{k}_\ast$, 
$\bm{q}$, and $\bm{q'}$, but only on the matrix $\bm{M}$. 

As the final point in this part, since there are infinite 
possible choices for $\bm{q}$'s, each block is in fact infinite-dimensional. 
One of the most special blocks is the one represented by the 
vacuum state $\bm{k}_\ast=\bm{k_0}=\bm{0}$. 
In \cite{vadfat}, it is shown that, provided that the ground-state is unique,
it belongs to the vacuum block.
The reason is simply that, in the extreme large coupling 
limit $g \to \infty$ ($\gamma \to 0$), as all elements except $V_{00}$ are
approaching zero, using the fact the energy and $\widehat{V}$ eigenvalues are related as 
\begin{align}\label{17}
\varepsilon_i=-\frac{1}{a}\ln v_i 
\end{align}
the ground-state belongs to the vacuum's block. By uniqueness of the 
ground-state, upon lowering the coupling, no crossing between ground-state 
by other states occurs, leaving the ground-states in the vacuum block at any coupling \cite{vadfat}.

\section{Currents, States and Blocks}
In this section, the basic elements and notions that emerged in the derivation of the previous chapter are discussed. 
In particular, the role of $k_l$ and $q_p$ numbers, the meaning of conditions (\ref{11}) and (\ref{13}), 
as well as some graphical representations are presented. To cover the basic idea, the 
detailed presentation is for the case of 2d spatial lattice, although the general expectations
and differences of 3d case are discussed as well. 

The first step is to understand the physical interpretation of the Fourier 
vectors $\bm{k}$. That is most directly understood by the way that these vectors appear, 
namely by (\ref{1}) and (\ref{7}) and their similar expressions in the continuum theory, 
for the coupling of the current $J$ to the gauge field $A$
\begin{align}\label{18}
e^{\mathrm{i} \sum_l k_l \theta^l} = e^{\mathrm{i}\, a\,g\sum_l k_l A^l} \to
e^{\mathrm{i}\,g \int J\cdot A\, dx}
\end{align}
In above $k_l$ is interpreted as the number of current-quanta at link $l$, 
coupled to the gauge field $A^l$ associated with this link. 
So the current-vector $\bm{k}$ consists of link-currents $k_l$'s. 
The integer nature of $k_l$ reflects the fact that, due to the compact nature of gauge fields in the 
lattice formulation, the quantization of charge is satisfied automatically. 

By the above interpretation of $k_l$'s, the Fourier basis $|\bm{k}\rangle$ is representing 
the state with a set of current quanta $k_l$ on link $l$, with $l\in\mathrm{links}$.
Now by the definition of the transfer-matrix $\widehat{V}=\exp(-a\widehat{H})$, with the Hamiltonian $\widehat{H}$, the matrix element $\langle\bm{k}'|\widehat{V}|\bm{k}\rangle$
is the transition amplitude between states with $\bm{k}$ and $\bm{k'}$ currents 
during the imaginary time interval `$a$'. 

\begin{figure}[t]
	\begin{center}
		\includegraphics[scale=.25]{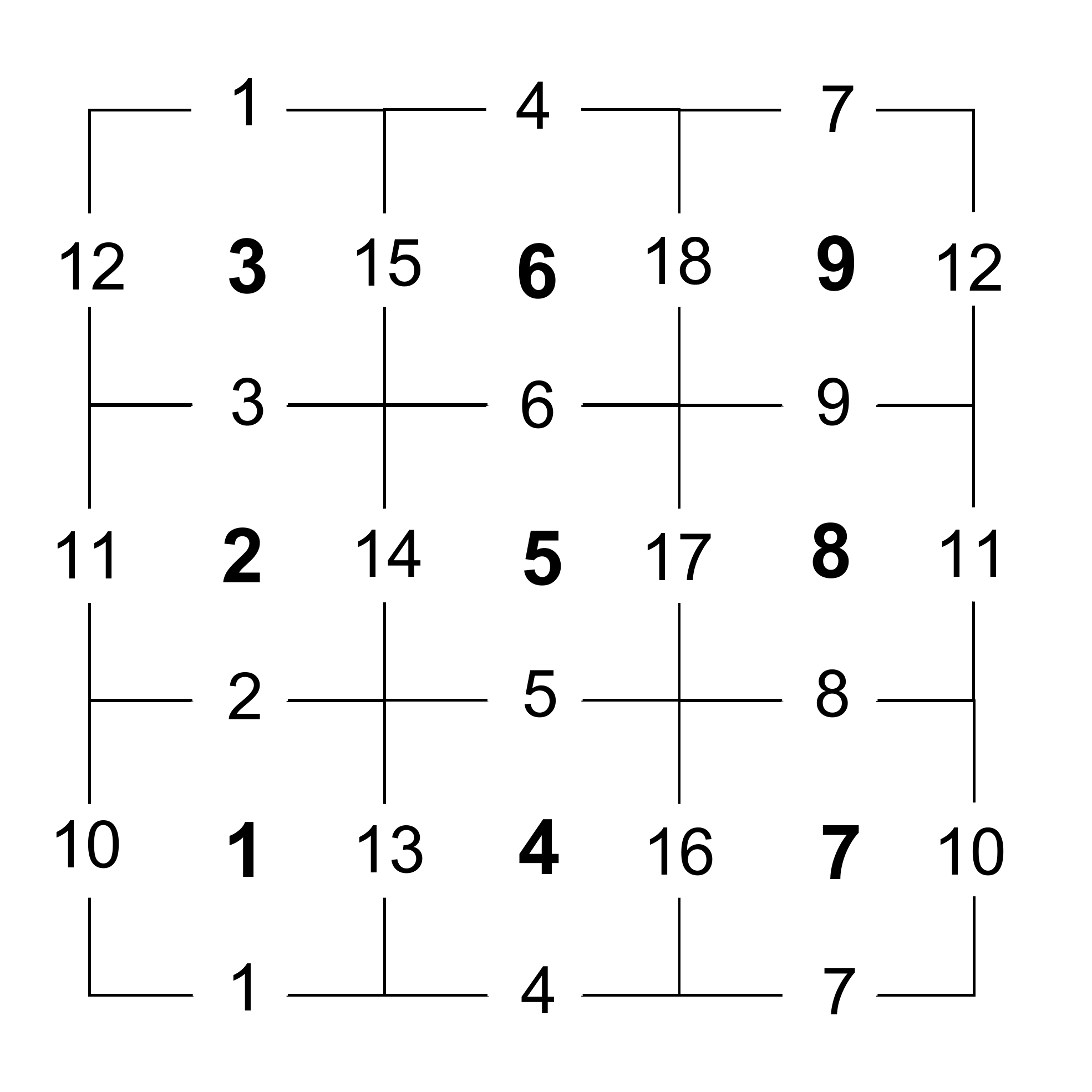}
\vskip -.6cm   \caption{\small The numbering of links and plaquettes for $3\times 3$ 2d
\textit{periodic} lattice used in (\ref{19}) as the representation of matrix $\bm{M}$ \cite{vadfat}. }
\label{fig3}
	\end{center}
\end{figure}

Before proceeding, it is useful to find an explicit representation for the plaquette-link matrix
$\bm{M}$. For the 2d periodic lattice with $N_s$ sites in 
each direction, there are $\np=N_s^2$ plaquettes and $\nl=2N_s^2$ links.
For the plaquette and link numbering of the $3\times 3$ periodic lattice given 
in Fig.~\ref{fig3}, using the definition (\ref{3}), one finds the following form for the $9\times 18$ dimensional 
matrix $\bm{M}$ \cite{vadfat}
\begin{align}\label{19}
\bm{M}=\left(\begin{array}{c@{\shrink}c@{\shrink}c@{\shrink}c@{\shrink}
c@{\shrink}c@{\shrink}c@{\shrink}c@{\shrink}c@{\shrink\shrink}
c@{\shrink}c@{\shrink}c@{\shrink}c@{\shrink}c@{\shrink}c@{\shrink}
c@{\shrink}c@{\shrink}c}
 + & - & 0 & 0 & 0 & 0 & 0 & 0 & 0 & - & 0 & 0 & + & 0 & 0 & 0 & 0 & 0 \\
 0 & + & - & 0 & 0 & 0 & 0 & 0 & 0 & 0 & - & 0 & 0 & + & 0 & 0 & 0 & 0 \\
 - & 0 & + & 0 & 0 & 0 & 0 & 0 & 0 & 0 & 0 & - & 0 & 0 & + & 0 & 0 & 0 \\
 0 & 0 & 0 & + & - & 0 & 0 & 0 & 0 & 0 & 0 & 0 & - & 0 & 0 & + & 0 & 0 \\
 0 & 0 & 0 & 0 & + & - & 0 & 0 & 0 & 0 & 0 & 0 & 0 & - & 0 & 0 & + & 0 \\
 0 & 0 & 0 & - & 0 & + & 0 & 0 & 0 & 0 & 0 & 0 & 0 & 0 & - & 0 & 0 & + \\
 0 & 0 & 0 & 0 & 0 & 0 & + & - & 0 & + & 0 & 0 & 0 & 0 & 0 & - & 0 & 0 \\
 0 & 0 & 0 & 0 & 0 & 0 & 0 & + & - & 0 & + & 0 & 0 & 0 & 0 & 0 & - & 0 \\
0 & 0 & 0 & 0 & 0 & 0 & - &
0 &  + & 
0 & 0 & + & 0 & 0 & 0 & 0  & 
0 & -  \\
\end{array}
\right)
\end{align}

\noindent In the 2d lattice, each link is being shared by two plaquettes, resulting in only 
one `$+$' and one `$-$' in each column. 
The general form of the matrix $\bm{M}$ can be given by means of the $N_s\times N_s$ 
translation-matrix $\bm{T}$ as follows \cite{vadfat}
\begin{align}\label{20}
\bm{M}=\left(\begin{array}{c|c}
  & \cr
 \mathbb{1}_{N_s}\otimes \bm{T} ~ &   - \bm{T}\otimes \mathbb{1}_{N_s} \cr 
  & 
\end{array}\right)
\end{align}
By construction, the matrix $\bm{M}$ is $N_s^2\times 2N_s^2$ dimensional, as it should.
The elements of $\bm{T}$ are given by \cite{vadfat}
\begin{align}\label{21}
T_{ab}= \delta_{ab}-\delta_{a+1,b}-\delta_{a,N_s}\,\delta_{b1},~~~~~
a,b=1,\cdots,N_s
\end{align}

\begin{figure}[H]
	\begin{center}
		\includegraphics[scale=.6]{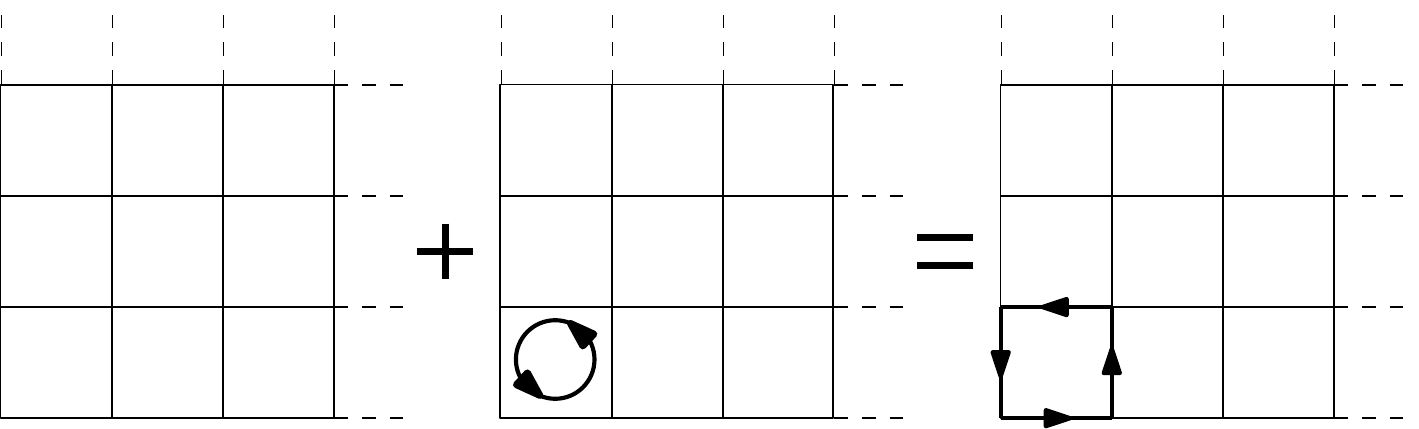}
\caption{\small The graphical representation of (\ref{23})  to construct $\bm{k_{0;1}}$ as 
co-block of $\bm{k_0}=\bm{0}$.}
\label{fig4}
\end{center}
\end{figure}

By the explicit form of the matrix $\bm{M}$, the meaning of the condition (\ref{13}) to have a 
non-zero matrix element can be understood by means of the current conservation, as follows. 
Let us begin with the vacuum state $\bm{k_0}=\bm{0}$, which according to (\ref{18}), 
represents the state with no current on any links. Now, by (\ref{13})  and using 
\begin{align}\label{22}
\bm{q_1}=\underbrace{(1,0,\cdots,0)}_{\np} 
\end{align}
a co-block of the vacuum state is found as the following current-vector
\begin{align}\label{23}
\bm{k_{0;1}}=\bm{k_0}+\bm{q_1}\cdot \bm{M}
\end{align}
in which only four links of the first plaquette have non-zero unit currents, 
namely two $+1$'s and two $-1$'s, making a circulating unit current in the first plaquette, as
represented in Fig.~\ref{fig4}. 
Then the non-zero matrix-element $\langle\bm{k_0}|\widehat{V}|\bm{k_{0;1}}\rangle$ 
is made possible since two states differ in a unit current circulating in a plaquette,
as required by current conservation. 
 \vskip -.6cm

\begin{figure}[H]
	\begin{center}
		\includegraphics[scale=.6]{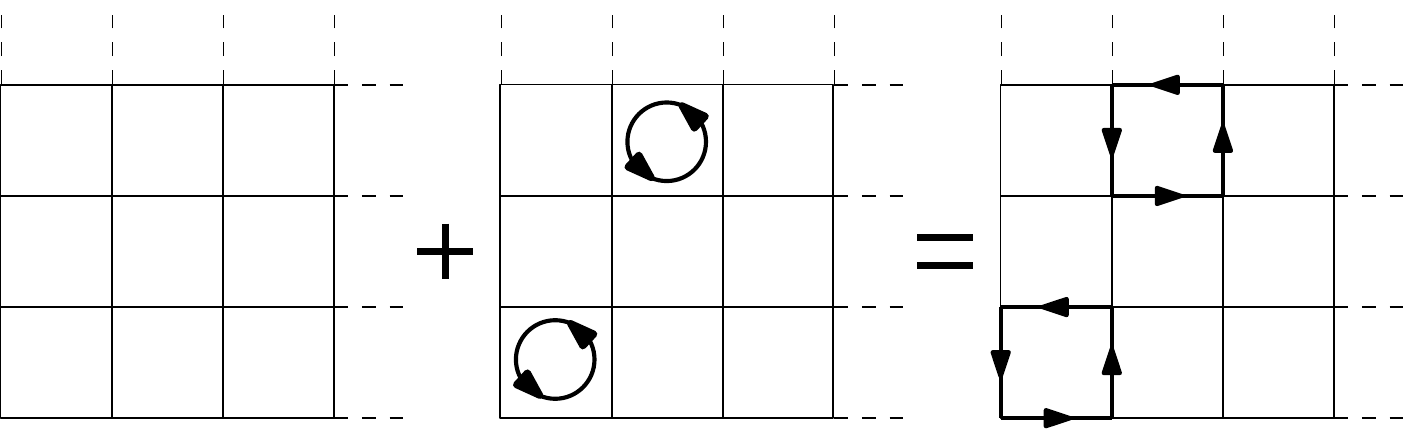}
\caption{\small Construction of a co-block of $\bm{k_0}$ with two non-adjacent plaquette-currents.}
\label{fig5}
	\end{center}
\end{figure}

As seen in the above example, $q_p$ is determining the number of current quanta
circulating in the plaquette $p$. It is befitted to call the $q_p$ numbers as plaquette-currents or loop-currents.
As another example from the vacuum block, consider the state constructed by all
$q_p$'s zero, except two of the non-adjacent plaquettes, as depicted in Fig.~\ref{fig5}. 

 \vskip -.6cm

\begin{figure}[H]
	\begin{center}
		\includegraphics[scale=.6]{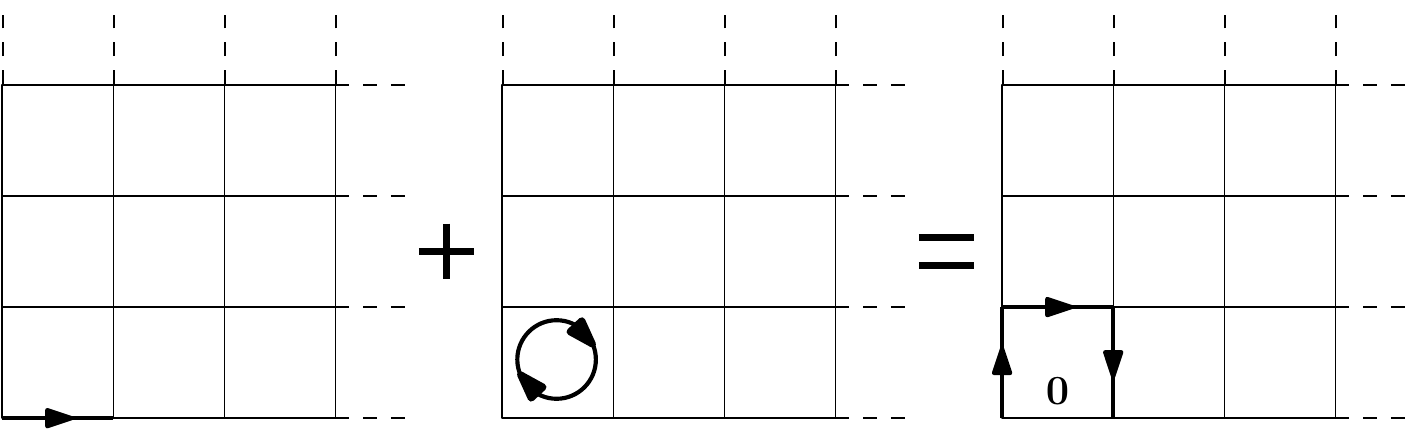}
\caption{\small $\bm{k_{1;-1}}=\bm{k_1}-\bm{q_1}\cdot \bm{M}$
as a co-block of $\bm{k_1}$ with 3 links having unit current.}
\label{fig6}
	\end{center}
\end{figure}

\begin{figure}[H]
	\begin{center}
		\includegraphics[scale=.6]{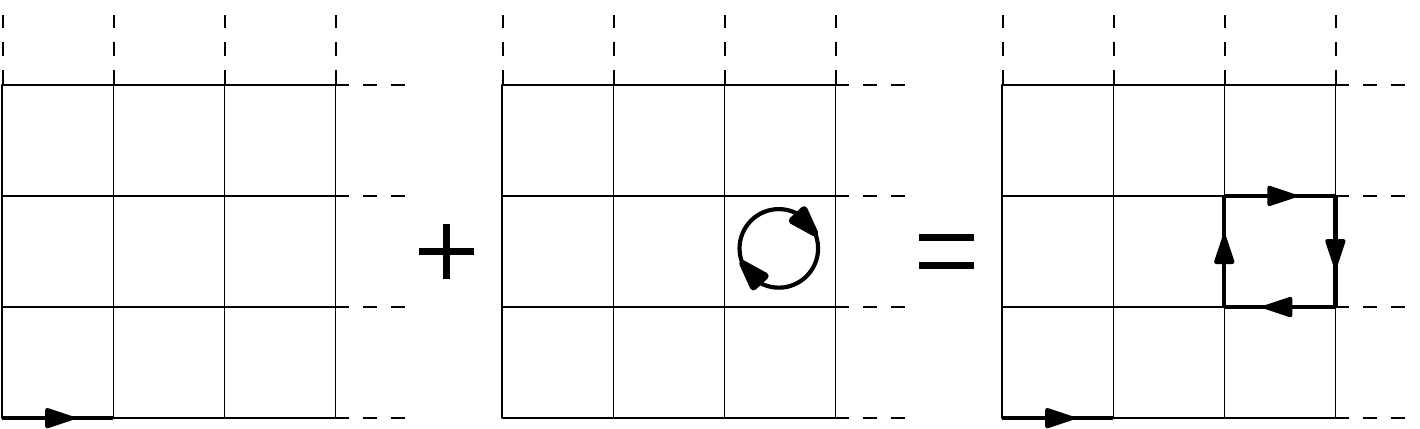}
\caption{\small A co-block of $\bm{k_1}$ with 5 links having unit current.}
\label{fig7}
	\end{center}
\end{figure}

Let us go beyond the vacuum block by considering the state represented by the vector-current 
\begin{align}\label{24}
\bm{k_1}=\underbrace{(1,0,\cdots,0)}_{\nl} 
\end{align}
which has one unit of current on the first link of the lattice, with all other link-currents zero. 
There is no set of plaquette-currents that could yield this vector from the vacuum. Equivalently,
one can say that there is no $\bm{q}\cdot\bm{M}$ equal to (\ref{24}). So the state (\ref{24}) does not belong
to the vacuum block. Two co-blocks of this state are presented in Figs. \ref{fig6} and \ref{fig7}, one with the 
loop-current in the first plaquette and one non-adjacent to the first plaquette, respectively. It is easy to check
that there is a plaquette-current that relates the resulting states in Figs. \ref{fig6} and \ref{fig7} as well. 
Again as the differences of these three current-states are just loop-currents circulating in plaquettes, 
the transition between the two of them is possible, leading to the non-zero $\widehat{V}$ elements between them. 

 \vskip -.5cm

 \vskip -.5cm

\begin{figure}[H]
	\begin{center}
		\includegraphics[scale=.6]{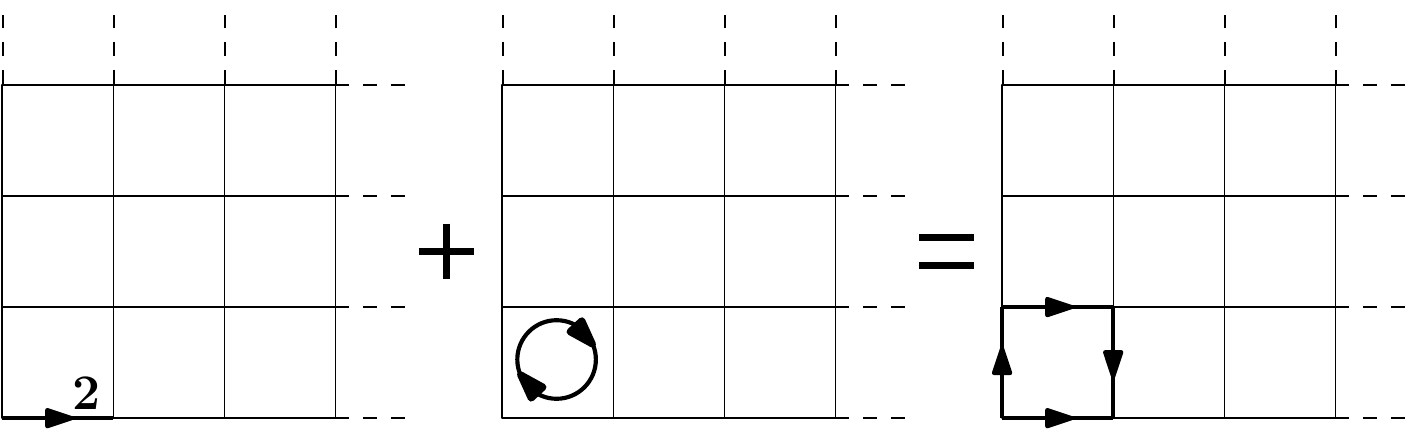}
\caption{\small$ \bm{k_{2;-1}}=\bm{k_2}-\bm{q_1}\cdot \bm{M}$ as a 
co-block of $\bm{k_2}$ with 4 links at the first plaquette having unit current.}
\label{fig8}
	\end{center}
\end{figure}

As another example, consider the state represented by the vector-current below
\begin{align}\label{25}
\bm{k_2}=\underbrace{(2,0,\cdots,0)}_{\nl} 
\end{align}
which has two units of current on the first link, with all others zero. Again it is easy to check that this 
state does not belong to the vacuum's and $\bm{k_1}$'s blocks. A co-block of this state is
presented in Fig.~\ref{fig8}. The difference between the direction of currents in 
the resulting state here with the last one in Fig.~\ref{fig4} is to be noted. 

 \vskip -.6cm

\begin{figure}[H]
	\begin{center}
		\includegraphics[scale=.6]{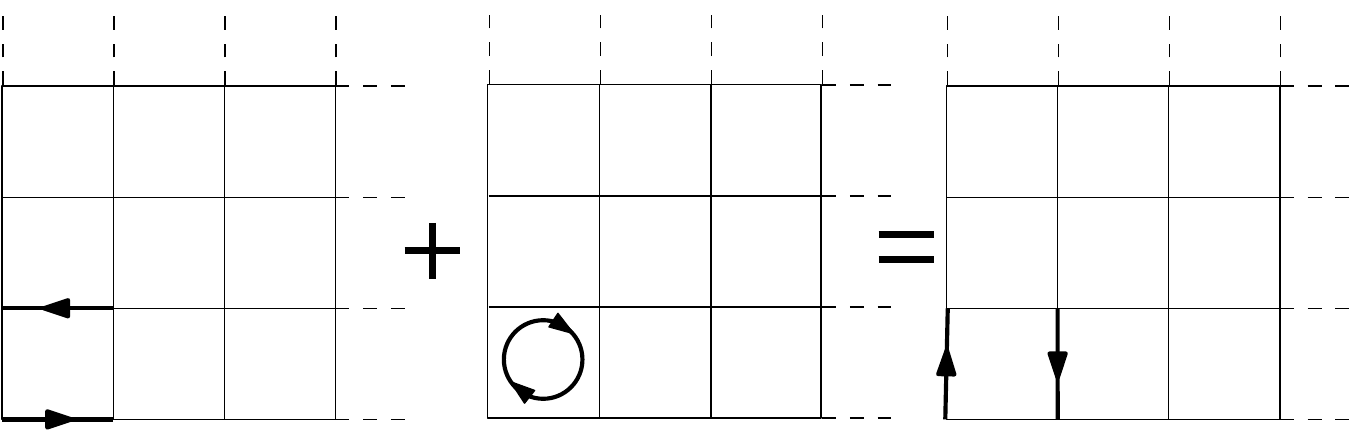}
\caption{\small$ \bm{k_{1,-1;-1}}=\bm{k_{1,-1}}-\bm{q_1}\cdot \bm{M}$ as a 
		co-block of $\bm{k_{1,-1}}$, both with 2 links having a unit current.}
	\label{fig9}
		\end{center}
\end{figure}

The last example, which will be used later, is the first vector in Fig.~\ref{fig9}, 
\begin{align}\label{26}
\bm{k_{1,-1}}=\underbrace{(1,-1,0,\cdots,0)}_{\nl} 
\end{align}
and its co-block on the right-hand side. The two vectors both have
one unit of current on only two links, and in this respect are equivalent to 
be selected as representative of the block. 

 \vspace{-.6cm}

By the representation of matrix $\bm{M}$ it is easy to check that 
for the periodic 2d lattices, the sub-space by vectors $\bm{n^0}$ satisfying 
(\ref{11}) is one dimensional with the general form
\begin{align}\label{27}
\bm{n^0}=n^0 \underbrace{(1,1,\cdots,1)}_{\np} 
\end{align}
leaving all link-currents zero by equal $n^0$ assigned to all plaquettes \cite{vadfat}.
The condition (\ref{11}) and its solution (\ref{27}) may be represented by the graphical representation we used
in different examples for $\bm{q}\cdot\bm{M}$ as depicted in Fig.~\ref{10}. 
\begin{figure}[!ht]
	\begin{center}
		\includegraphics[scale=.8]{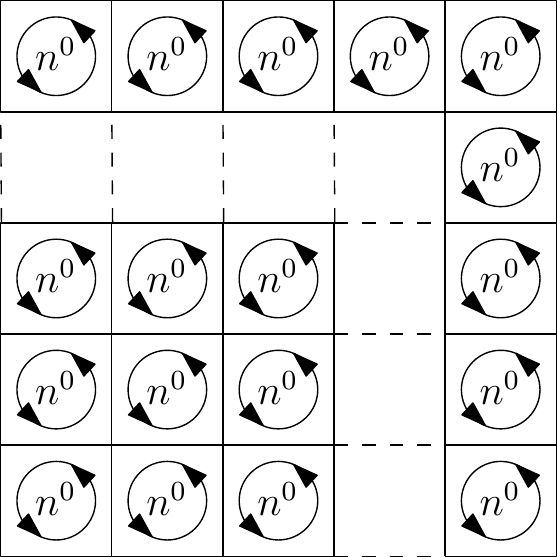}
\caption{\small Equally $n^0$ current-units in all plaquettes make the net current
at all links zero, provided that opposite edges being identified in the periodic 2d lattice. }
\label{fig10}
	\end{center}
\end{figure} 
The loop-current vectors $\bm{n^0}$ are specially important to manage 
the orders of Bessel's in (\ref{16}), and more related to this work, 
when the strong coupling expansion in $\gamma$ is being developed. 
We postpone the second for later when the expansion in $\gamma$ will be discussed. 
About the Bessel's, by above for the 2d lattices, ranks of all Bessel's $\I_{q'_p-n_p-n^0}$ in (\ref{16}) 
are affected by ${n^0\neq  0}$ equally. 
For definiteness, by setting $\bm{k}_\ast=\bm{q}=\bm{q'}=\bm{0}$ in (\ref{16}), 
let us consider the vacuum-to-vacuum element 
$\langle\bm{0}|\widehat{V}|\bm{0}\rangle$ for the 2d lattice case: 
\begin{align}\label{28}
\langle\bm{0}|\widehat{V}_\mathrm{2d}|\bm{0}\rangle_{\bm{0}}=\mathcal{A}\,
 e^{-\gamma(\np+\nl) }  (2\pi)^{\nl}
\sum_{n^0}   \sum_{\{n_p\}}
\prod_{p}\I_{-n_p}\!\left(\frac{\gamma}{2}\right)\I_{-n_p-n^0}
\!\left(\frac{\gamma}{2}\right) 
\prod_{l} \I_{\sum_p \!\! n_p M^p_{~l}}\!(\gamma)
\end{align}
in which by (\ref{27})  the index `$p$' is dropped from $n^0_p$.
The formal expansion on Bessel functions in the strong coupling 
$\gamma=1/g^2 \ll 1$ is mainly based on the ordering relation 
between the Bessel's as $\I_n(x)>\I_m(x)$ for $|n|<|m|$, coming in the strong form when
$x \ll 1$
\begin{align}\label{29}
\I_0(x) \gg \I_{\pm 1}(x)\gg \I_{\pm 2}(x)\gg \I_{\pm 3}(x) \gg \cdots
\end{align}
For example, let us compare two terms in (\ref{28}) with all $n_p$'s equal to zero, 
but in one $n^0=0$ and in the other $n^0=1$, 
\begin{align}\label{30}
\langle\bm{0}|\widehat{V}_\mathrm{2d}|\bm{0}\rangle_{\bm{0}}&=\mathcal{A}\,
 e^{-\gamma(\np+\nl) }  (2\pi)^{\nl}
\left[
\I_0^{2\np}\!\!\left(\frac{\gamma}{2}\right)
\I_0^{\nl}\!\!\left(\gamma\right)
+  
\I_0^{\np}\!\!\left(\frac{\gamma}{2}\right)
\I_1^{\np}\!\!\left(\frac{\gamma}{2}\right)
\I_0^{\nl}\!\!\left(\gamma\right)
\right]+\cdots
\cr
&=\mathcal{A}\,
 e^{-\gamma(\np+\nl) }  (2\pi)^{\nl}
\I_0^{2\np}\!\!\left(\frac{\gamma}{2}\right)
\I_0^{\nl}\!\!\left(\gamma\right)
\left[1+  
\left(\frac{\I_1\!\left(\gamma/2\right)}{\I_0\!\left(\gamma/2\right)}\right)^{\!\!\np}
\right]+\cdots
\end{align}
Now, for large lattice size $\np\gg 1$ the term with the power of ${\np}$
can be dropped safely in comparison to one. In fact, a simple inspection shows that,
depending on $\bm{q}$ or $\bm{q'}$ defining an element, there are one or two choices for 
$n^0$ in which the Bessel's of lowest order appear, and the other values of $n^0$ 
for large lattices can be ignored in the summation. 
The 2d latices may be contrasted to the 3d lattices, in which a 
finite number of non-zero 
loop-currents may generate a vector $\bm{n^0}$ satisfying $\bm{n}^{\bm{0}}\cdot \bm{M}=\bm{0}$.
One case with the lowest number of non-zero loop-currents is presented in Fig.~\ref{fig11},
in which six plaquettes making a cube and their currents are presented.
By the right-hand rule, all currents in Fig.~\ref{fig11} point outward the cube.
The only other case with the lowest number is that with all loop-currents reversed,
pointing inward the cube. As all link-currents by the 
configuration are zero, the loop-currents presented in Fig.~\ref{fig11} satisfy 
$\bm{n}^{\bm{0}}\cdot \bm{M}=\bm{0}$. 
As a consequence, the term $\I_1/\I_0$ in (\ref{30})  for a 3d lattice would have the power of 6 instead of 
$\np$. Any number of such cubes when sitting together also make a $\bm{n^0}$ vector that satisfies the condition 
(\ref{11}).

\begin{figure}[H]
	\begin{center}
		\includegraphics[scale=1.5]{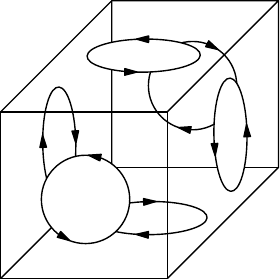}
\caption{\small The 6-plaquette cube configuration in a 3d lattice that satisfies
$\bm{n^0\cdot M}=\bm{0}$. All currents point outward by right-hand rule}
\label{fig11}
	\end{center}
\end{figure} 

\section{Current Expansion at Large Coupling}
The aim in this section is to provide the basic rules of the strong coupling expansion 
in $\gamma=1/g^2\ll 1$ for the elements of the transfer-matrix in the field Fourier basis.
In the next section and Appendix~C, several examples of the application 
of rules for the strong coupling expansion are given.
The samples of expansion in $\gamma$ presented in this work 
can be checked either by massive direct integration over field space or by 
the expression (\ref{16}), provided that the Bessel's with relevant ranks are identified. 

In the previous section, the role of integer numbers $k_l$, $q_p$, and $n^0_p$ 
together with their graphical representations are introduced.
Based on the notions developed before, it will be shown that the strong 
coupling expansion can be regarded as the summations on currents. 
In particular, a non-zero element of the transfer-matrix, as the transition amplitude between 
two current-states belonging to the same block, is interpreted as the summations on occurrences of
\textit{virtual} loop and link currents, each weighted by $\gamma$, that 
transform both states to the vacuum. 

The presentation is coming in two subsections. First, by working out some examples of 
the matrix-elements at lower orders, it is seen how the expansion based on loop and link 
currents would emerge. In the second subsection, the set of rules of current expansion in the strong 
coupling regime is presented. 

\subsection{Basic Observations on Current Expansion}
To begin with, it is convenient to define the reduced form of the transfer-matrix as:
\begin{align}\label{31}
\widehat{V}=\mathcal{A}\,  e^{-\gamma(\nl+\np) }  (2\pi)^{\nl}~\vred
\end{align}
for which the extra numerical factors are dropped, with the element in field-basis as:
\begin{align}\label{32}
\langle \bm{\theta'} |\vred | \bm{\theta}\rangle =  \frac{1}{(2\pi)^{\nl}}
&\prod_p \exp\!\left\{\frac{\gamma}{2}\left[\cos\big(M^p_{~l}\,\theta^l\big)
+\cos\big(M^p_{~l}\,\theta'^l\big)\right]\right\}\cr
&\times\prod_{l}\exp\!\left\{\gamma\cos\big(\theta^l-\theta'^l\big)\right\}
\end{align}
The element of the reduced form in the Fourier basis is given by
\begin{align}\label{33}
\langle\bm{k'}|\vred|\bm{k}\rangle_{\bm{k}_\ast} = \frac{1}{(2\pi)^{\nl}}
\int_{-\pi}^\pi \prod_l  \rd\theta'_l \rd\theta_l \, e^{-\ri\,\bm{k'}\cdot\bm{\theta'}}
e^{\ri\,\bm{k}\cdot\bm{\theta}}\langle \bm{\theta'} |\vred | \bm{\theta}\rangle
\end{align}
in which the representative vector of the block is $\bm{k}_\ast$, and the 
$\bm{k}$ and $\bm{k'}$ vectors are given by means of 
loop-currents $\bm{q}$ and $\bm{q'}$ as (\ref{15}).
Let us begin with the matrix-elements 
in the vacuum block by setting $\bm{k}_\ast=\bm{0}$, for which we have
\begin{align}\label{34}
\langle \qmpb|\vred|\qmb\rangle_{\bm{0}} = \frac{1}{(2\pi)^{\nl}}
\int_{-\pi}^\pi \prod_l  \rd\theta'_l \rd\theta_l \, e^{-\ri\,\bm{q'}\!\cdot\bm{M}\cdot\bm{\theta'}}\,
e^{\ri\,\bm{q}\cdot\bm{M}\cdot\bm{\theta}}\langle \bm{\theta'} |\vred | \bm{\theta}\rangle
\end{align}
At the first order in $\gamma$ it is easy to find the following terms:
\begin{align}\label{35}
\langle \qmpb|\vred|\qmb\rangle_{\bm{0}} &=\prod_l \delta((\qmpb)_l)\delta((\qmb)_l) 
\cr& +\frac{\gamma}{4} \prod_{l}\delta((\qmb)_{l})\sum_{p'} \bigg[\prod_{l'}\delta((\qmpb)_{l'}+M^{p'}_{\,l'})+\prod_{l'}\delta((\qmpb)_{l'}-M^{p'}_{\,l'})\bigg]
\cr& +\frac{\gamma}{4}\prod_{l'}\delta((\qmpb)_{l'}) \sum_{p}  \bigg[\prod_{l}\delta((\qmb)_l+M^{p}_{\,l})+\prod_{l}\delta((\qmb)_l -M^{p}_{\,l})\bigg]
\cr&+\frac{\gamma}{2}\sum_{l_1}\bigg[\prod_{l}\delta((\qmb)_{l}+\delta_{ll_1})\delta((\qmpb)_l+\delta_{ll_1})
\cr&~~~~~~~~~~+\prod_{l}\delta((\qmb)_{l}-\delta_{ll_1})\delta((\qmpb)_l-\delta_{ll_1}) \bigg]
+\mathrm{O}(\gamma^2)
\end{align}
in which $(\qmb)_l$ is the current on link $l$. 
It is noted that in the last term, the link index $l_1$ is common between two $\delta$'s by 
$\bm{q'}$ and $\bm{q}$, as originated from the term $\cos(\theta-\theta')$, in which
$\theta$ and $\theta'$ are appearing equally.
The other point about the last term is this: \textit{in the vacuum block} 
and for any plaquette vectors $\bm{q}$ and $\bm{q'}$, 
there is no chance that all multiplying $\delta$'s would be satisfied, leading to zero value for this term. 
This is simply because there are always non-zero arguments for some of the $\delta$'s, and so 
this term has no contribution. 
The vacuum-to-vacuum (v.t.v.) transition with $\bm{q}=\bm{q'}=\bm{0}$, by $\delta(m)=\delta(-m)$
takes the form
\begin{align}\label{36}
\langle \bm{0}|\vred|\bm{0}\rangle_{\bm{0}} =1+
\gamma \sum_{p} \prod_{l}\delta(M^{p}_{\,l})
+\mathrm{O}(\gamma^2)
\end{align}
Due to the non-zero elements of $M^{p}_{\,l}$
at each plaquette, the linear term in $\gamma$ vanishes, leading to
\begin{align}\label{37}
\langle \bm{0}|\vred|\bm{0}\rangle_{\bm{0}} =1+\mathrm{O}(\gamma^2)
\end{align}
At this order now, let us consider the element by vacuum and the current-vector (\ref{23}),
by setting $\bm{q'}=\bm{q_1}$ and $\bm{q}=\bm{0}$ 
with one unit of loop-current in the first plaquette as in Fig.~\ref{fig4}. By (\ref{35}) 
then the transition between vacuum and $|\bm{k_{0;1}}\rangle=|\bm{1}\rangle$ is given by 
\begin{align}\label{38}
\langle \bm{1}|\vred|\bm{0}\rangle_{\bm{0}} &=\prod_l \delta(M^{1}_{\,l}) 
+\frac{\gamma}{2} \sum_{p'} \prod_{l'}\delta(M^{p'}_{\,l'})\prod_{l}\delta(M^{1}_{\,l}) 
\cr& +\frac{\gamma}{4} \sum_{p'}  \bigg[\prod_{l'}\delta(M^{1}_{\,l'} +M^{p'}_{\,l'})
+\prod_{l'}\delta(M^{1}_{\,l'} -M^{p'}_{\,l'})\bigg]
+\mathrm{O}(\gamma^2)
\end{align}
in which the first and second terms do not contribute, due to the non-zero elements by 
$M^{1}_{\,l'}$ or $M^{p'}_{\,l'}$. The third term, however, contributes due to the term 
$\delta(M^{1}_{\,l'} -M^{p'}_{\,l'})$ for $p'=1$ in the summation, leading to 
\begin{align}\label{39}
\langle \bm{1}|\vred|\bm{0}\rangle_{\bm{0}}=\frac{\gamma}{4}
+\mathrm{O}(\gamma^2)
\end{align}
As a graphical representation for the cancellation between link-currents and loop-current in the same plaquette in the $\delta$, one may consider the diagram in Fig.~\ref{fig12} for the transition $\bm{1}\!\to\!\bm{0}$: 
\begin{figure}[H]
	\begin{center}
		\includegraphics[scale=.5]{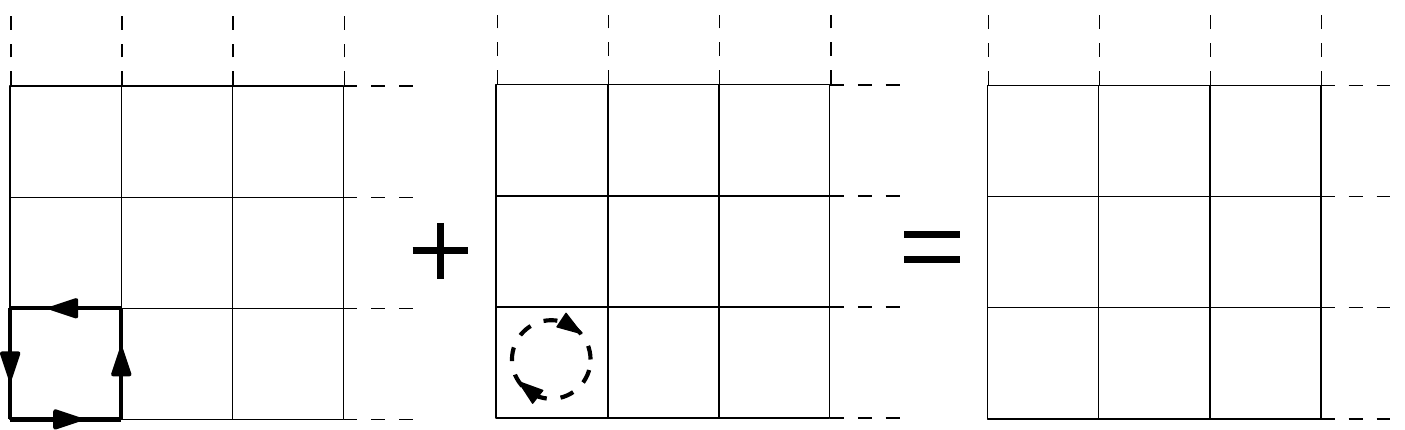}
\caption{\small The combination of $\bm{k_{0;1}}$ and virtual loop-current}
\label{fig12}
	\end{center}
\end{figure} \vskip -.6cm
\noindent In fact, Fig.~\ref{fig12} is similar 
to Fig.~\ref{fig4}, except that the loop-current is plotted in a reversed and dashed form 
in the $\bm{k_{0;1}}$'s side. The reason for being dashed is to emphasize its virtual nature. 
As a space-saving representation, one may suggest the following for Fig.~\ref{12}
\begin{align}\label{40}
\recdashcirc  \pmb{\longrightarrow} \bm{0}
\end{align}

It can be seen that at the first order of $\gamma$, the only non-zero elements are the third term of (\ref{38}),
like Fig.~\ref{12} or (\ref{40}), with one loop current to vacuum.
In the vacuum block, the order of $\gamma^2$ can be obtained by the expansion of the exponential 
as well. To present the long expressions in a compact form from now on, the alternative representations 
are used as:
\begin{align}\label{41}
\qmb &\pmb{\longrightarrow} \sqmb ,~~~~~~
\qmpb \pmb{\longrightarrow} \sqmpb  \cr
(\qmb)_l &\pmb{\longrightarrow} \slashed{q}_l,~~~~~~
(\qmpb)_l \pmb{\longrightarrow} \slashed{q}'_l
\end{align}
in which at order $\gamma^2$ in the vacuum block, we have
\begin{align}\label{42}
\Big[\langle \sqmpb |\vred|\sqmb \rangle_{\bm{0}}&\Big]_{\gamma^2}\!\!\! =
\frac{\gamma^2}{32}\prod_l\delta(\slashed{q}_l)
\sum_{{p'_1},{p'_2}}\bigg[\prod_{l'}\delta(\slashed{q}'_{l'}\!+\!M^{p'_1}_{\,l'}\!+\!M^{p'_2}_{\,l'})
\cr&~~~~~~~~~~\!+\!2\prod_{l'}\delta(\slashed{q}'_{l'}\!+\!M^{p'_1}_{\,l'}\!-\!M^{p'_2}_{\,l'}) 
\!+\!\prod_{l'}\delta(\slashed{q}'_{l'}\!-\!M^{p'_1}_{\,l'}\!-\!M^{p'_2}_{\,l'})\bigg]
\cr& ~~~~~~\!+\!\frac{\gamma^2}{32}\prod_{l'}\delta(\slashed{q}'_{l'})\sum_{{p_1},{p_2}}
\bigg[\prod_{l}\delta(\slashed{q}_l\!+\!M^{p_1}_{~l}\!+\!M^{p_2}_{~l})
\cr&~~~~~~~~~~\!+\!2\prod_{l}\delta(\slashed{q}_l\!+\!M^{p_1}_{~l}\!-\!M^{p_2}_{~l}) 
\!+\!\prod_{l}\delta(\slashed{q}_l\!-\!M^{p_1}_{~l}\!-\!M^{p_2}_{~l})\bigg]
\cr&~~~~~~\!+\!\frac{\gamma^2}{16}\sum_{p,p'}\Bigg[
\prod_{l}\delta(\slashed{q}_l\!+\!M^{p}_{\,l}) \delta(\slashed{q}'_l\!-\!M^{p'}_{\,l})
\!+\!\prod_{l}\delta(\slashed{q}_l\!+\!M^{p}_{\,l})\delta(\slashed{q}'_l\!+\!M^{p'}_{\,l})
\cr&~~~~~~~~~~\!+\! \prod_{l}\delta(\slashed{q}_l\!-\!M^{p}_{\,l})\delta(\slashed{q}'_l\!-\!M^{p'}_{\,l})
\!+\!\prod_{l}\delta(\slashed{q}_l\!-\!M^{p}_{\,l})\delta(\slashed{q}'_l\!+\!M^{p'}_{\,l})\Bigg]
\cr+\frac{\gamma^2}{8} & \sum_{{l_1},{l_2}}
\bigg[\prod_{l} \delta(\slashed{q}_l\!+\!\delta _{ll_1}\!+\!\delta _{ll_2}) \delta(\slashed{q}'_l\!+\!\delta _{ll_1}\!+\!\delta_{ll_2})
\!+\!\prod_{l} \delta(\slashed{q}_l\!+\!\delta _{ll_1}\!-\!\delta _{ll_2}) \delta(\slashed{q}'_l\!+\!\delta _{ll_1}\!-\!\delta_{ll_2})
\cr&~~~~
+\prod_{l} \delta(\slashed{q}_l\!-\!\delta _{ll_1}\!+\!\delta _{ll_2}) \delta(\slashed{q}'_l\!-\!\delta _{ll_1}\!+\!\delta_{ll_2})
\!+\!\prod_{l} \delta(\slashed{q}_l\!-\!\delta _{ll_1}\!-\!\delta _{ll_2}) \delta(\slashed{q}'_l\!-\!\delta _{ll_1}\!-\!\delta_{ll_2})\bigg]
\cr&\!+\!\frac{\gamma^2}{8}
\sum_{{p'},{l}}\bigg[\prod_{l'}\delta(\slashed{q}'_{l'}\!-\!M^{p'}_{\,l'}\!+\!\delta_{ll'})\delta(\slashed{q}_{l'}\!+\!\delta_{ll'})
\!+\!\prod_{l'}\delta(\slashed{q}'_{l'}\!+\!M^{p'}_{\,l'}\!+\!\delta_{ll'})\delta(\slashed{q}_{l'}\!+\!\delta_{ll'})\bigg]
\cr&\!+\!\frac{\gamma^2}{8}
\sum_{{p'},{l}}\bigg[\prod_{l'}\delta(\slashed{q}'_{l'}\!-\!M^{p'}_{\,l'}\!-\!\delta_{ll'})\delta(\slashed{q}_{l'}\!-\!\delta_{ll'})
\!+\!\prod_{l'}\delta(\slashed{q}'_{l'}\!+\!M^{p'}_{\,l'}\!-\!\delta_{ll'})\delta(\slashed{q}_{l'}\!-\!\delta_{ll'})\bigg]
\cr&\!+\!\frac{\gamma^2}{8}
\sum_{{p},{l'}}\bigg[\prod_{l}\delta(\slashed{q}_{l}\!-\!M^{p}_{\,l}\!+\!\delta_{ll'})\delta(\slashed{q}'_{l}\!+\!\delta_{ll'})
\!+\!\prod_{l}\delta(\slashed{q}_{l}\!+\!M^{p}_{\,l}\!+\!\delta_{ll'})\delta(\slashed{q}'_l\!+\!\delta_{ll'})\bigg]
\cr&\!+\!\frac{\gamma^2}{8}
\sum_{{p},{l'}}\bigg[\prod_{l}\delta(\slashed{q}_{l}\!-\!M^{p}_{\,l}\!-\!\delta_{ll'})\delta(\slashed{q}'_l\!-\!\delta_{ll'})
\!+\!\prod_{l}\delta(\slashed{q}'_{l}\!+\!M^{p'}_{\,l}\!-\!\delta_{ll'})\delta(\slashed{q}_l\!-\!\delta_{ll'})\bigg]
\end{align}
Again the last four sums, containing the combination of one $\cos(M^p_{\,l}\theta^l)$ or $\cos(M^p_{\,l}\theta'^l)$
with one $\cos(\theta-\theta')$, has $\delta$'s with common link indices $l$ or $l'$.
Also, in the vacuum block, there is no chance that these terms would survive, as 
it is not possible that all multiplying $\delta$'s would be satisfied simultaneously. 
Let us consider the $\gamma^2$ contribution for the v.t.v. transition with $\bm{q}=\bm{q'}=\bm{0}$
\begin{align}\label{43}
\Big[\langle \bm{0}|\vred|\bm{0}\rangle_{\bm{0}}&\Big]_{\gamma^2}\!\!\! =
\frac{\gamma^2}{8}\sum_{{p_1},{p_2}}
\bigg[\prod_{l}\delta(\!M^{p_1}_{~l}\!+\!M^{p_2}_{~l})
+\!\prod_{l}\delta(\!M^{p_1}_{~l}\!-\!M^{p_2}_{~l}) \bigg]
\cr&~~~~\!+\!\frac{\gamma^2}{4}\sum_{p,p'}
\prod_{l}\delta(\!M^{p}_{\,l}) \delta(\!M^{p'}_{\,l})
\cr&~~~+\frac{\gamma^2}{4} \sum_{{l_1},{l_2}}
\bigg[\prod_{l} \delta(\!\delta _{ll_1}\!+\!\delta _{ll_2}) \delta(\!\delta _{ll_1}\!+\!\delta_{ll_2})
\!+\!\prod_{l} \delta(\delta _{ll_1}\!-\!\delta _{ll_2}) \delta(\delta _{ll_1}\!-\!\delta_{ll_2})\bigg]
\end{align}
In the first term, only the second part may survive by cancellation between $M$'s, 
provided that $p_1$ and $p_2$ refer to the same plaquette,
developing a factor of $\np$ by the summation over all plaquettes. 
The second term vanishes, as for each plaquette there are four links for which $M^p_{\,l}$ is non-zero. 
In the last term, again, the second part may contribute by cancellation between $\delta_{ll_i}$ 
if $l_1$ and $l_2$ refer to the same link, giving a factor of $\nl$. With these, we have
\begin{align}\label{44}
\Big[\langle \bm{0}|\vred|\bm{0}\rangle_{\bm{0}}\Big]_{\gamma^2}\!\!\! =
\frac{\gamma^2}{8}\np+\frac{\gamma^2}{4}\nl
\end{align}
The above may be represented graphically as follows:
\begin{align}\label{45}
\Big[\langle \bm{0}|\vred|\bm{0}\rangle_{\bm{0}}&\Big]_{\gamma^2}:
\begin{cases}
\bm{0}+\twocirc ~ \pmb{\longrightarrow} \bm{0}&~ \frac{1}{2^4}\frac{1}{2!} 2\np\cr
\bm{0} ~ \pmb{\longrightarrow} \bm{0}+\twocirc&~ \frac{1}{2^4}\frac{1}{2!} 2\np\cr
\bm{0}+ \backforth~ \pmb{\longrightarrow} \bm{0}+\backforth &~ \frac{1}{2^2}\frac{1}{2!} 2\nl
\end{cases}
\end{align}
The rules for associated numerical factors and the powers of $\gamma$ will be 
discussed later. (\ref{37}) combined with (\ref{44}) leads to
\begin{align}\label{46}
\langle \bm{0}|\vred|\bm{0}\rangle_{\bm{0}} =1+
\left(\frac{\np}{8}+\frac{\nl}{4}\right)\gamma^2
+\mathrm{O}(\gamma^3)
\end{align}

For the $\bm{1}\!\to\!\bm{0}$ transition 
by setting $\bm{q'}=\bm{q_1}$ and $\bm{q}=\bm{0}$, it is easy to see that always unsatisfied 
$\delta$'s would remain, showing that this transition will not take contribution at order $\gamma^2$.
Now consider the $\gamma^2$ contribution for the $\bm{1}\to\bm{1}$, for which the only
surviving term is  
\begin{align}\label{47}
\Big[\langle \bm{1}|\vred|\bm{1}\rangle_{\bm{0}}&\Big]_{\gamma^2}\!\!\! =
\frac{\gamma^2}{16}\sum_{{p},{p'}}
\bigg[\prod_{l}\delta(\!M^{1}_{~l}\!-\!M^{p}_{~l})
\delta(\!M^{1}_{~l}\!-\!M^{p'}_{~l}) \bigg]
\end{align}
In the summation, the only contribution comes from $p=p'=1$, by which
$M^p_{~l}$ and $M^{p'}_{~l}$ are canceled by $M^{1}_{~l}$ in both $\delta$'s, leading to
\begin{align}\label{48}
\langle \bm{1}|\vred|\bm{1}\rangle_{\bm{0}}=\frac{\gamma^2}{16}+\mathrm{O}(\gamma^3)
\end{align}
The naive way of thinking about $\bm{1}\to\bm{1}$ may suggest that it would always be possible as both initial and final states are the same. However, as seen in the extreme large coupling 
limit $\gamma\to 0$, this transition vanishes as $\gamma^2$. The expression obtained after doing the Fourier integrals shows that the $\delta$'s related to 
initial and final states must be satisfied separately to obtain a non-zero result. In other words,
the final expression should be interpreted as the result of the transform of both initial and final states to the 
vacuum. Accordingly, as the transforms $\bm{1}\to\bm{0}$ and $\bm{0}\to\bm{1}$ at lowest order,
each with a loop-current transform to vacuum, are both proportional to $\gamma$, 
the $\bm{1}\to\bm{1}$ transition should be of order $\gamma^2$. 
This crucial role of vacuum that, as if it is being passed in $\bm{1}\to\bm{1}$ transition 
due to the nature of Fourier integrals, is true for all other transitions, including 
$\bm{0}\to\bm{0}$ by (\ref{46}) and even those in blocks other than the vacuum. 
To avoid misinterpretation, it is quite suitable to highlight 
this role by the vacuum by changing the notation slightly from $\bm{1}\to\bm{1}$ to 
$\bm{1}\to\bm{0}\to\bm{1}$, or simply $\bm{1}\displaystyle{\mathop{\to}^0}\bm{1}$.
By this, the graphical representation for the above transition may be given as below
\begin{align}\label{49}
\Big[\langle \bm{1}|\vred|\bm{1}\rangle_{\bm{0}~}\Big]_{\gamma^2}:
\recdashcirc  \mathop{\pmb{\longrightarrow}}^0 \recdashcirc  ~~~~~~~\frac{\gamma^2}{16}
\end{align}

The order of $\gamma^3$ in the vacuum block is given in Appendix~A with about 80 terms. 
It is obvious that the terms with one $\delta_{ll_i}$ or three $\delta_{ll_i}$ would not survive in 
the vacuum block,
as there is no chance that all multiplying $\delta$'s would be satisfied. So the surviving terms
at 3rd order in the vacuum block are those of the first three lines of (\ref{130}). 
For the v.t.v. transition with $\bm{q}=\bm{q'}=\bm{0}$, again we see that there is no way that all
multiplying $\delta$'s would be satisfied. By $\bm{q'}=\bm{q_1}$ and $\bm{q}=\bm{0}$ 
with one unit of loop-current in the first plaquette as in Fig.~\ref{fig4}, or when one or three $M$'s are 
in the $\bm{q_1}$'s side, there is the chance that all $\delta$'s would be satisfied. 
The only thing is to calculate the combinatorial factors for all possible states that would lead to satisfying all $\delta$'s.
For the first term, the possible ways would be three combinations of two $-M$'s and one $+M$
\begin{align}\label{50}
q_1,-M,-M,+M\pmb{\longrightarrow} 0
\end{align}
in which $q_1$ cancels with one of $-M$'s, and two others cancel out each other. Depending on whether all $M$'s are in the $q_1$ plaquette or not, we have the combinatorial factor 
\begin{align}\label{51}
3(2(\np-1)+1)=3(2\np -1)
\end{align}
The second term may contribute by
\begin{align}\label{52}
q_1,-M \pmb{\longrightarrow} 0,+M,-M
\end{align}
having two possibilities for $+/-$ signs on the second side, giving the combinatorial factor
\begin{align}\label{53}
2\np
\end{align}
The last term contribution comes from the combinations of:
\begin{align}\label{54}
q_1,-M,+\delta,-\delta \pmb{\longrightarrow} 0,+\delta,-\delta
\end{align}
Having in mind that the place of link $\delta$'s in the lattice in two sides are the same, together with 
two possibilities for $+/-$ signs for $\delta$'s, leads to the combinatorial factor
\begin{align}\label{55}
2\nl
\end{align}
It can be shown by the following graphical representation: 
\begin{align}\label{56}
\Big[\langle \bm{1}|\vred|\bm{0}\rangle_{\bm{0}}\Big]_{\gamma^3}:
\begin{cases}
\recdashcirc +\twocirc  ~\displaystyle{\mathop{\pmb{\longrightarrow}}^0}~ \bm{0} &~ \frac{1}{2^6}\frac{1}{3!} 3(2\np-1) \cr
\recdashcirc ~\displaystyle{\mathop{\pmb{\longrightarrow}}^0}~ \bm{0} + \twocirc &~ \frac{1}{2^6}\frac{1}{2!} 2\np \cr
\recdashcirc + \backforth~\displaystyle{\mathop{\pmb{\longrightarrow}}^0}~ \bm{0} + \backforth &~ \frac{1}{2^4}\frac{1}{2!} 2\nl
\end{cases}
\end{align}
All together, using (\ref{39}) and (\ref{56}), we have the result as follows:
\begin{align}\label{57}
\langle \bm{1}|\vred|\bm{0}\rangle_{\bm{0}}=\frac{\gamma}{4}
+\left(\frac{-1}{128}+\frac{\np}{32}+\frac{\nl}{16}\right)\gamma^3
+\cdots
\end{align}

Before proceeding to the rules in the next subsection, it is instructive to consider examples of the
matrix-elements in the non-vacuum block by setting the representative 
state $\bm{k}_\ast\neq\bm{0}$ in (\ref{15}) and (\ref{33}). 
In particular, we consider the block with a representative state of (\ref{24})
$\bm{k}_\ast=\bm{k}_{1}=(1,0,...,0)$ from the previous section, with just one unit of current on the first link. 
It is easy to see that the zeroth-order $\gamma^0$ belongs only to the v.t.v. of the vacuum block, and so 
up to the first order of $\gamma$ in $\bm{k_1}$'s block we find 
\begin{align}\label{58}
\langle\bm{k_1}\!\!+\!\sqmpb|\vred|\bm{k_1}\!\!+\!\sqmb\rangle_{\bm{1}} &\!=\!
\frac{\gamma}{4} \prod_{l}\delta(k_{1l}+\slashed{q}_l)\sum_{p'} \bigg[\prod_{l'}\delta({k}_{1l}+\slashed{q}'_{l'}
+M^{p'}_{\,l'})
+\prod_{l'}\delta({k}_{1 l}+\slashed{q}'_{l'}
-M^{p'}_{\,l'})\bigg]
\cr& +\frac{\gamma}{4}\prod_{l'}\delta({k}_{1l}+\slashed{q}'_{l'}) \sum_{p}  \bigg[\prod_{l}\delta({k}_{1l}
+\slashed{q}_l+M^{p}_{\,l})
+\prod_{l}\delta({k}_{1l}+\slashed{q}_l 
-M^{p}_{\,l})\bigg]
\cr&+\frac{\gamma}{2}\sum_{l_1}\bigg[\prod_{l}\delta({k}_{1l}+\slashed{q}_{l}+\delta_{ll_1})
\delta({k}_{1l}+\slashed{q}'_l+\delta_{ll_1})
\cr&~~~~~~~~~~+\prod_{l}\delta({k}_{1l}+\slashed{q}_{l}-\delta_{ll_1})\delta({k}_{1l}+\slashed{q}'_l-\delta_{ll_1}) \bigg]+\mathrm{O}(\gamma^2)
\end{align}
Using the fact that $\sqmb$ and $\sqmpb$ add loop-currents to the representative state $\bm{k_1}$, it is
shown that the first order in this block survives only for $\bm{q}=\bm{q'}=\bm{0}$, 
leading to the matrix-element 
\begin{align}\label{59}
\langle\bm{k}_{1}|\vred|\bm{k}_{1}\rangle_{\bm{1}} &=\frac{\gamma}{2}\sum_{l_1}\bigg[\prod_{l}\delta({k}_{1l}-\delta_{ll_1})\delta({k}_{1l}-\delta_{ll_1}) \bigg]
+\mathrm{O}(\gamma^2)
\end{align}
for which, setting $l_1=1$ by the $\delta$'s, we find
\begin{align}\label{60}
\langle \bm{k}_{1}|\vred|\bm{k}_{1}\rangle_{\bm{1}} =\frac{\gamma}{2}+\mathrm{O}(\gamma^2)
\end{align}
The expression (\ref{59}) shows explicitly how the separate multiplying 
$\delta$'s connect the initial and final states to the vacuum state, although
it does not belong to $\bm{k_1}$'s block. The crucial role of the vacuum state, as mentioned before,
is related to the fact that both $\delta$'s by initial and final states are 
to be satisfied, as if both passed through the vacuum state. Accordingly, 
the graphical representation of the above is expressed as follows:
\begin{align}\label{61}
\Big[\langle \bm{k_1}|\vred|\bm{k_1}\rangle_{\bm{1}}\Big]_{\gamma}:
 \backforthhor ~\mathop{\pmb{\longrightarrow}}^0~ \backforthhor~~~~~~ \frac{\gamma}{2}
\end{align}
At the next order now let us consider the element 
by setting $\bm{q'}=\bm{0}$ and $\bm{q}=\bm{q_1}=(1,0,\cdots,0)$ of (\ref{22}).
The current-vector $\bm{k_1}+\slashed{\bm{q}}_{\bm{1}}=\bm{k_{1;1}}$ 
with one unit of loop-current and link-current in the first plaquette,
may be represented as
\begin{align}\label{62}
\mathop{\rec}_{\linkcur}~~~~~\mathrm{or}~~~~~~\rec\hspace{-4mm}\raisebox{-6mm}{2}
\end{align}
In the following, for the sake of clarity, we use the first of the above.
Then the surviving matrix-element is found to be 
\begin{align}\label{63}
\langle\bm{k_1}|\vred|\bm{k_{1;1}}\rangle_{\bm{1}} =\frac{\gamma^2}{8} \sum_{p_1} \sum_{l_1}\bigg[\prod_l \delta(k_{1l}-\delta_{ll_1}) 
\delta(k_{1l}+M^{1}_{\,l}-M^{p_1}_{\,l}-\delta_{ll_1})\bigg]
+\mathrm{O}(\gamma^3)
\end{align}
Above, cancellation of link-currents in both $\delta$'s and loop-currents in the second $\delta$ by $p_1=1$, 
leads to
\begin{align}\label{64}
\langle \bm{k}_{1}|\vred|\bm{k_{1;1}}\rangle_{\bm{1}} =\frac{\gamma^2}{8}+\mathrm{O}(\gamma^3)
\end{align}
Once again, the $\delta$'s in expression (\ref{63}) by the initial and final states, 
show explicitly how the connections between two states and the vacuum, 
which do not even belong to this block, determine the strength of transition. Graphically the way of satisfying $\delta$'s 
may be represented as 
\begin{align}\label{65}
\Big[\langle \bm{k_1}|\vred|\bm{k_{1;1}}\rangle_{\bm{1}}\Big]_{\gamma^2}\!:~ \backforthhor ~\mathop{\pmb{\longrightarrow}}^0\, \mathop{\recdashcirc}_{\backforthhor}~~~~~~~~~~ \frac{\gamma^2}{8}
\end{align}
Let us consider the diagonal element by setting $\bm{q'}=\bm{q}=\bm{q_1}$,  
with one unit of loop-current and link-current in the first plaquette, 
given by 
\begin{align}\label{66}
\langle\bm{k_{1;1}}|\vred|\bm{k_{1;1}}\rangle_{\bm{1}}\!=\!\frac{\gamma^3}{32} \sum_{p,p',l_1}\prod_l \delta(k_{1l}+M^{1}_{\,l}-M^{p'}_{\,l}-\delta_{ll_1}) 
\delta(k_{1l}+M^{1}_{\,l}-M^{p}_{\,l}-\delta_{ll_1})
\!+\!\mathrm{O}(\gamma^4)
\end{align}
This is the only term for this order. Four link-currents are canceled by loop-currents if $p=p'=1$, however, $\bm{k_1}$ should be canceled by $\delta_{ll_1}$'s in both sides, leading to
\begin{align}\label{67}
\langle \bm{k_{1;1}}|\vred|\bm{k_{1;1}}\rangle_{\bm{1}} =\frac{\gamma^3}{32}+\mathrm{O}(\gamma^4)
\end{align}
with the graphical representation as follows:
\begin{align}\label{68}
\Big[\langle \bm{k_{1;1}}|\vred|\bm{k_{1;1}}\rangle_{\bm{1}}\Big]_{\gamma^3}:
~ \mathop{\recdashcirc}_{\backforthhor}~
\mathop{\pmb{\longrightarrow}}^0~\mathop{\recdashcirc}_{\backforthhor} ~~~~~~~~~~ \frac{\gamma^3}{32}
\end{align}

\subsection{Rules of Current Expansion in Strong Coupling}

As announced earlier, in the present subsection the set of rules is given in which at any order of 
$\gamma$ in principle, one can write the transfer-matrix element between two states. The rules are 
based on determining the ways that transform the initial and final states to the vacuum, accompanied by the 
associated numerical and combinatorial factors of each transition. As mentioned, the 
distinguished role of the vacuum state simply comes back to the fact that 
the $\delta$'s related to the initial and final states' Fourier integrals are to be satisfied
separately. Accordingly, and as seen in previous examples,
the transform can be represented by a set of graphs in which a proper combination of 
\textit{virtual} loop and link currents would make the required pass through the vacuum. 
Also, by the given examples, the transform to vacuum is to be considered even for 
states that do not belong to the vacuum block. This is because the co-blocks
of a state are determined by adding loop-currents, via $\bm{q}\cdot\bm{M}$ in (\ref{15}),
but the concerned transform to vacuum is due to both link and loop currents, the former via
`$\cos(\theta-\theta')$' term that is irrelevant for making co-blocks. It is due to these
link-currents that transform of any state into the vacuum is made possible, even for those 
states in blocks other than the vacuum. 

For any state in the Fourier basis, there are infinite ways to transform it into the
vacuum. As seen before, this is correct for the vacuum state itself.
For two given states of $|\bm{k}\rangle$ and $|\bm{k'}\rangle$ in the same block, consider
the case that they transform to vacuum by $m$ and $m'$ numbers of virtual loop-currents, 
respectively, accompanied by $\ell$ numbers of virtual link-currents for both states. 
By (\ref{32}) the mentioned numbers of 
currents appear through  the integration of $m$, $m'$ and $\ell$ numbers of 
$\cos\!\big(M^p_{~l}\,\theta^l\big)$'s, $\cos\!\big(M^p_{~l}\,\theta'^l\big)$'s and $\cos\!\big(\theta^l-\theta'^l\big)$'s,
respectively. Now, as it can be derived easily (see Appendix~B), the numerical factor 
associated with the matrix-element of transition through the considered transform is 
\begin{align}\label{69}
\left[\langle\bm{k'}|\vred|\bm{k}\rangle\right]_{m,m',\ell} =
 \mathcal{K}_{m,m',\ell}~  \frac{1}{2^{2m+2m'+\ell}}\frac{1}{m!\,m'!\,\ell!} \,\gamma^{m+m'+\ell}
\end{align}
in which $\mathcal{K}_{m,m',\ell}$ is the combinatorial factor representing the number of ways that
loop and link currents can be combined, regarding the initial and final transforms to the vacuum.

In the previous subsection, graphical representations are suggested to each term contributing to 
the transition at the lowest orders of $\gamma$.
In fact, and as we will see in several examples, 
these graphical representations can be used to determine and manage the channels that contribute 
to a transition at a given order. In this respect, these graphical representations 
can serve as the Feynman diagrams in perturbative quantum field theory. 
The elements being used in graphs are simply the currents, loop or link ones, being 
characterized by their real or virtual natures. Accordingly, the initial and final states, being 
determined only by link-currents, are interpreted as real and presented by a combination 
of solid lines as 
\begin{align} \label{70}
\linkcur~~~~~~\mathrm{or}~~~~~~~\linkcurinv
\end{align}
and their rotated versions. 
Instead, the loop and link-currents that occurred during the transforms
are interpreted as virtual. The virtual loop-currents, representing the $\pm M^p_{\,l}$'s 
inside multiplying $\delta(\cdot)$'s, are coming as 
\begin{align}\label{71}
\circ ~~~~~~~~\mathrm{or}~~~~~~~~ \circinv
\end{align}
The virtual link-currents, representing the $\delta_{ll'}$'s 
inside multiplying $\delta(\cdot)$'s, are drawn below and their rotated versions
\begin{align}\label{72}
\linkdash~~~~~~~\mathrm{or}~~~~~~~\linkdashinv
\end{align}
The important point about the virtual link-currents is, as mentioned earlier, that 
they come in both sides of initial and final transforms to vacuum equally,   
since they are originated from Fourier integration over $\cos\!\big(\theta-\theta'\big)$'s.

Some general statements about the expansion for an arbitrary element can be made and come in order.
First, it can be easily shown that two subsequent orders of $\gamma$ 
in the expansion of an element differ by two. By the previous examples as well as 
several ones given later, that is evident. 
This simply comes back to the fact that any order for an element differs 
from a higher-order one by adding an even number of virtual currents
to transform both initial and final states to vacuum. 
By this, the expansion for an element in $\bm{k}_\ast$-block looks like 
\begin{align}\label{73}
\langle\bm{k'}_{\ast\bm{q'}}|\vred|\bm{k}_{\ast\bm{q}}\rangle_{\bm{k}_\ast} = \gamma^h \,
	(c_0 + c_{2}\,\gamma^{2}+ c_{4}\,\gamma^{4} +c_{6}\,\gamma^{6}+\cdots)
\end{align}
in which $h$ is the lowest order at which the transforms of both initial and final states 
to vacuum, by adding virtual link and loop currents, are made possible. So by 
$\gamma=1/g^2$, the subsequent increase of order is in fact $1/g^4$, 
which makes the expansion fairly reliable for even not so large 
values of $g$. The value of $h$ can be determined as well, once $\bm{k}_\ast$, $\bm{q}$, and 
$\bm{q'}$ are given. By Sec.~3, we know that each vector-current inside a block can be selected as
the representative. To make things systematically, we use the convention that
the representative vector would have the minimum value of 
\begin{align}\label{74}
|\bm{k}|=\sum_{l=1}^{\nl} |k_l|
\end{align}
In some blocks, the above specifies just one vector-current. For example, in 
the vacuum block, it is only the vacuum and not, say, its co-block (\ref{23}) in Fig.~\ref{fig4} 
with larger (\ref{74}). The other block with $\bm{k_1}$ in (\ref{24}) and those in 
Figs. \ref{fig6} and \ref{fig7} as co-blocks, 
again $\bm{k_1}$ has a minimum (\ref{74}), is being taken as representative. 
The same is true for the block with $\bm{k_2}$ and its co-block in Fig.~\ref{fig8}. 
For some blocks, however, the condition would not identify just one candidate 
as the representative. An example is given in Fig.~\ref{fig9}, in which both current-vectors have 
equal (\ref{74}), and so both can be taken as representatives. The least order of $h$ 
can be obtained simply by realizing how many link and loop currents are necessary to 
transform the given states to the vacuum. It is here that the representative 
$\bm{k}_\ast$ with a minimum (\ref{74}) is needed. By $\bm{q}$ and $\bm{q'}$ of 
(\ref{15}) we then have 
\begin{align}\label{75}
h=|\bm{k}_\ast|+|\bm{q}|+|\bm{q'}|
\end{align}
in which $|\bm{k}_\ast|$ as (\ref{74}) and 
\begin{align}\label{76}
|\bm{q}|=\sum_{p=1}^{\np} |q_p|
\end{align}
and a similar one for $|\bm{q'}|$. The relation for `$h$' is correct for all blocks, including those with
more than one candidate as representative with a minimum (\ref{74}), provided that 
$\bm{q}$ and $\bm{q'}$ take those loop-vectors which satisfy (\ref{15}) 
with the selected representative. The above for `$h$' can be checked easily 
by the previous examples and several ones given later. 

As the last feature, it is befitting to discuss here the effect of lattice dimension on the expansion. 
The first footstep of lattice dimension is seen by the number of 
links and plaquettes, $\nl$ and $\np$, in the expansion. For example, 
for the 2d and 3d periodic spatial lattices with $N_s$ sites in each direction, we have the following 
\begin{align}\label{77}
\mathrm{2d}:~ \nl&=2N_s^2,~~~~~~~\np=N_s^2 \\
\label{78}
\mathrm{3d}:~ \nl&=3N_s^3,~~~~~~~\np=3N_s^3 
\end{align}
The other impact of lattice dimension is related to the difference of 
null condition (\ref{11}) for $\bm{n^0}$ in different dimensions. In Sec.~3 for the 2d and 3d cases,
the difference is mentioned and its effect on managing Bessel's orders in (\ref{16}) 
is discussed. Here the effect of this difference in $\gamma$-expansion is 
pointed. The vector $\bm{n^0}$ affects the expansion as it, via $\bm{n^0\cdot M}=\bm{0}$,
can contribute to the transformations of states to vacuum. As seen in the 2d case, $\bm{n^0}$
represents loop-currents in all $\np$ plaquettes. So its presence affects the order 
$\gamma^{\np}$, which is quite suppressed for supposedly large lattices $\np\to \infty$ 
in the strong coupling limit $\gamma \ll 1$. In the 3d case, however, vector $\bm{n^0}$ may
represent a 6-plaquette cubic configuration with net zero current, as shown in Fig.~\ref{fig11}.
Of course, there are many other finite-plaquette configurations with zero net currents, simply by 
combining the lower number ones. By these, we conclude that in form (\ref{73}) 
the orders up to $\gamma^{4+h}$ are formally the same in all dimensions, 
however, the number of links and plaquettes should be replaced 
according to dimensions as in (\ref{77}) and (\ref{78}).
The first difference due to the configurations that
do not fit in the 2d appears at order $6+h$.
As an example, in the next section, the orders $\gamma^4$ and $\gamma^6$
of transition $\bm{0}\,\displaystyle{\mathop{{\to}}^0}\,\bm{0}$ 
are considered, which consist of the ways that combinations of virtual currents can take place in 2d. 
The $\gamma^4$ order result is formally the same in both dimensions, 
but for the $\gamma^6$ order, one has to take into account also the cube configuration 
in Fig.~\ref{fig11}. Depending on whether the virtual currents on cube configuration 
are on the first or second side, we have for numbers in (\ref{69}) the following possibilities
\begin{align}\label{79}
\bm{k'}	\mathrm{~side:}~m=0,~~m'=6,~~\ell=0 \\ 
\label{80}
\bm{k}	\mathrm{~side:}~m=6,~~m'=0,~~\ell=0 
\end{align}
The evaluation of the combinatorial factor is given in the next section. 
In general, as the difference between 2d and 3d comes in the fourth term in the expansion of (\ref{73}), 
namely in `$\,c_6\,\gamma^{6+h}\,$', the difference between 3rd and 4th 
terms is of order $\gamma^2=1/g^4$, which for many practical purposes is not significant 
even for not so large coupling $g$. 

\section{Application of Rules}

The following examples are presented to see how the expression (\ref{69}) and 
the associated graphical representation work in practice. 
The matrix-elements expansions up to order $\gamma^3$ are already presented, 
together with the associated graphical representations.
Similar to order $\gamma^3$ in the vacuum block, the $\delta$-expansion 
of order $\gamma^4$ is given in Appendix~A.
Again in the vacuum block the terms with an odd number of $\delta_{ll_i}$'s have no chance to survive,
specifically those in the last three lines in (\ref{131}). 
The v.t.v. transition with $\bm{q}=\bm{q'}=\bm{0}$, also terms with an odd number of $M$'s 
at $q$ and $q'$ sides have no chance, those in lines four and five of (\ref{131}). 
So only the terms in lines one to three contribute to v.t.v. transition, 
for which the counting of combinatorial factors are given. 
For the first term, the possible ways would be six combinations 
of two $-M$'s and two $+M$, in which one $+M$ cancels out one of $-M$'s, and 
two others cancel out each other. Depending on whether all $M$'s are in the same plaquette or not
\begin{align}\label{81}
\left.\begin{matrix}
0\pmb{\longrightarrow} 0,-M,-M,+M,+M \cr
0,-M,-M,+M,+M\pmb{\longrightarrow} 0
\end{matrix}\right\}:~  C^2_4\left(2\np(\np-1)+\np\right) 
\end{align}
in which
\begin{align}\label{82}
C^m_n=\left(\begin{matrix}n \cr m \end{matrix}\right) = \frac{n!}{m!(n-m)!}
\end{align}
The possible ways for the second term would be four combinations of two $M$'s and two $\delta$'s, 
in which $+M$ cancels with one of $-M$'s, and $\delta$'s cancel out each other in two sides. 
So we have the combinatorial factor as:
\begin{align}\label{83}
\left.\begin{matrix}
0,-M,+M,+\delta,-\delta\pmb{\longrightarrow} 0,+\delta,-\delta\cr
0,+\delta,-\delta\pmb{\longrightarrow} 0,-M,+M,+\delta,-\delta
\end{matrix}\right\}:~
 C^1_2 C^1_2 \np \nl  
\end{align}
The third term may contribute as
\begin{align}\label{84}
0,+M,-M\pmb{\longrightarrow} 0,+M,-M:~ C^1_2 C^1_2 \np \np 
\end{align}
And the last combination at this order is:
\begin{align}\label{85}
0,+\delta,+\delta,-\delta,-\delta\pmb{\longrightarrow} 0,+\delta,+\delta,-\delta,-\delta:~ C^2_4\left(2\nl(\nl-1)+\nl\right)
\end{align}
which comes by considering whether all $\delta$'s are 
in the same link or not. All of these can be expressed in the graphical representation below:
\begin{align}\label{86}
\Big[\langle \bm{0}|\vred|\bm{0}\rangle_{\bm{0}}\Big]_{\gamma^4}:
\begin{cases}
\bm{0}+\twocirc~\twocirc ~\displaystyle{\mathop{\pmb{\longrightarrow}}^0}~ \bm{0} &~ \frac{1}{2^8}\frac{1}{4!}C^2_4 (2\np^2-\np) \cr
\bm{0} ~\displaystyle{\mathop{\pmb{\longrightarrow}}^0}~ \bm{0}+\twocirc~\twocirc &~ \frac{1}{2^8}\frac{1}{4!}C^2_4 (2\np^2-\np) \cr
\bm{0} +\twocirc~\displaystyle{\mathop{\pmb{\longrightarrow}}^0}~ \bm{0}+\twocirc &~ \frac{1}{2^8}\frac{1}{2!2!}C^1_2C^1_2 \np^2 \cr
\bm{0} +\twocirc~\backforth ~\displaystyle{\mathop{\pmb{\longrightarrow}}^0}~ \bm{0}+\backforth &~ \frac{1}{2^6}\frac{1}{2!2!}C^1_2C^1_2 \np\nl \cr
\bm{0} +\backforth ~\displaystyle{\mathop{\pmb{\longrightarrow}}^0}~ \bm{0}+\twocirc~\backforth &~  \frac{1}{2^6}\frac{1}{2!2!}C^1_2C^1_2 \np\nl \cr
\bm{0} +\backforth~\backforth ~\displaystyle{\mathop{\pmb{\longrightarrow}}^0}~ \bm{0}+\backforth~\backforth &~ \frac{1}{2^4}\frac{1}{4!}C^2_4 (2\nl^2-\nl) 
\end{cases}
\end{align}
Using (\ref{46}) results in:
\begin{align}\label{87}
\langle \bm{0}|\vred|\bm{0}\rangle_{\bm{0}}=& 1+ \left(\frac{\np}{8} + \frac{\nl}{4}\right) \gamma^2 \cr
&+\left(-\frac{\nl}{64}+\frac{\nl^2}{32}-\frac{\np}{512}+ \frac{\np \nl}{32} +\frac{\np^2}{128}
\right)\gamma^4+\cdots
\end{align}
It is easy to check that at order $\gamma^4$ the transition $\bm{1}\displaystyle{\mathop{\to}^0}\,\bm{0}$
does not find any contribution. Instead, the transition $\langle \bm{1}|\vred|\bm{1}\rangle_{\bm{0}}$ 
gets contributions at this order by the combination of loop and link currents as before. 
However, at this order, a new combination for cancellation between 
the initial real loop-current and four virtual link-currents takes place, graphically represented as 
below:
\begin{align}\label{88}
\recdashrec
\end{align}
Let us start with already known combinations. First, we have the possibility that one $-M$ 
on one side and $+M,-M,-M$ on the other side would result in full cancellation toward vacuum. 
The combinatorial factor is then
\begin{align}\label{89}
C^1_3  (2\np-1)
\end{align}
The other possibility is one $-M$ and two $\pm\delta_{ll'}$'s in each side, by the factor
\begin{align}\label{90}
C^1_2 \nl
\end{align}
The last is the new one (\ref{88}), for which counting the ways that four $\pm\delta_{ll'}$'s take place 
gives the factor 
\begin{align}\label{91}
C^2_4 \,4.
\end{align}
These all are summarized below 
\begin{align}\label{92}
\left[\langle \bm{1}|\vred|\bm{1}\rangle_{\bm{0}}\right]_{\gamma^4}:
\begin{cases}
 \recdashcirc  ~\displaystyle{\mathop{\pmb{\longrightarrow}}^0}~ \recdashcirc ~ \twocirc &~ \frac{1}{2^8}\frac{1}{3!} C^1_3 (2\np-1) \cr
\recdashcirc ~ \twocirc ~\displaystyle{\mathop{\pmb{\longrightarrow}}^0}~ \recdashcirc    &~ \frac{1}{2^8}\frac{1}{3!}C^1_3 (2\np-1) \cr
\recdashcirc~\backforth ~\displaystyle{\mathop{\pmb{\longrightarrow}}^0}~ \recdashcirc  ~ \backforth &~ \frac{1}{2^6}\frac{1}{2! } C^1_2 \nl \cr
\recdashrec ~\displaystyle{\mathop{\pmb{\longrightarrow}}^0}~ \recdashrec  &~ \frac{1}{2^4}\frac{1}{4! } C^2_4 \,4
\end{cases}
\end{align}
combined with the result of order $\gamma^2$ with this order giving
\begin{align}\label{93}
\langle \bm{1}|\vred|\bm{1}\rangle_{\bm{0}}= \frac{\gamma^2}{16}
+\left(\frac{15}{256}+\frac{\nl}{64}+\frac{\np}{128}\right)\gamma^4 + \cdots 
\end{align}

At 5th order, it is easy to see that v.t.v. does not get a contribution. 
Instead, the $\bm{1}\displaystyle{\mathop{\to}^0}\,\bm{0}$ transition takes the contribution from 
combinations of the initial loop-current $q_1$ and virtual loop and link currents as seen in previous examples.
However, apart from (\ref{40}), a new combination takes place; that is 
four link-currents may cancel out a virtual loop-current as below:
\begin{align}\label{94}
\dashreccirc
\end{align}
Again we start with more familiar cases. The first possibility is
\begin{align}\label{95}
q_1,-M,-M,-M,+M,+M \pmb{\longrightarrow} 0
\end{align}
in which $q_1$ cancels with one of $-M$'s, and four others cancel out each other. Depending on 
how many $M$'s sitting in the $q_1$ plaquette or are separated, we have the combinatorial factor 
\begin{align}\label{96}
C^3_5 &\Big(3(2(\np-1)(\np-2)+\np-1)\cr
&+3\times 2(\np-1)+1\Big)=C^3_5 (6\np^2-9\np+4)
\end{align}
The second may contribute as
\begin{align}\label{97}
q_1,-M,-M,+M \pmb{\longrightarrow} 0,+M,-M
\end{align}
having possibilities for $+/-$ signs in each of two sides, giving the combinatorial factor
\begin{align}\label{98}
C^1_2C^2_3\np(2(\np-1)+1)=C^1_2C^2_3\np(2\np-1)
\end{align}
The third combination is:
\begin{align}\label{99}
 q_1,-M \pmb{\longrightarrow} 0,+M,+M,-M,-M
\end{align}
as $q_1$ cancels $-M$, and on the other side, four virtual currents cancel each other, with the combinatorial factor:
\begin{align}\label{100}
C^2_4(2\np(\np-1)+\np)=C^2_4(2\np^2-\np)
\end{align}
By
\begin{align}\label{101}
q_1,-M,-M,+M,+\delta,-\delta \pmb{\longrightarrow} 0,+\delta,-\delta
\end{align}
again $q_1$ is canceled by $-M$, and for two other $M$'s depending on sitting in the $q_1$ plaquette or not, we have 
the combinatorial factor:
\begin{align}\label{102}
C^2_3C^1_2 \nl(2(\np-1)+1)=C^2_3C^1_2\nl(2\np-1)
\end{align}
The fifth term at this order is by the combination:
\begin{align}\label{103}
q_1,-M,+\delta,-\delta \pmb{\longrightarrow} 0,+M,-M,+\delta,-\delta
\end{align}
In the second side, two $M$'s cancel out each other, and the $\delta$'s each other too, with the factor: 
\begin{align}\label{104}
C^1_2C^1_2 \nl\np
\end{align}
By the combination
\begin{align}\label{105}
q_1,-M,+\delta,+\delta,-\delta,-\delta \pmb{\longrightarrow} 0,+\delta,+\delta,-\delta,-\delta
\end{align}
the ways for cancellation of the four $\delta$'s give the combinatorial factor 
\begin{align}\label{106}
C^2_4 (2\nl(\nl-1)+\nl)=C^2_4 (2\nl^2-\nl)
\end{align}
In the last one, the four currents by $q_1$ are canceled with the four $\delta$'s 
\begin{align}\label{107}
q_1,+\delta,+\delta,-\delta,-\delta \pmb{\longrightarrow} 0,+M,+\delta,+\delta,-\delta,-\delta
\end{align}
Having possibilities for $+/-$ with a fixed place for the $\delta$'s in the first plaquette the number of 
combinations is
\begin{align}\label{108}
C^2_4 \, 2\times 2. 
\end{align}
The combinations associated with the transition are summarized below:
\begin{align}\label{109}
\Big[\langle \bm{1}|\vred|\bm{0}\rangle_{\bm{0}}\Big]_{\gamma^5}:
\begin{cases}
\recdashcirc~\twocirc~\twocirc ~\displaystyle{\mathop{\pmb{\longrightarrow}}^0}~ \bm{0} &~ \frac{1}{2^{10}}\frac{1}{5!} C^3_5 (6\np^2-9\np+4) \cr
\recdashcirc~\twocirc ~\displaystyle{\mathop{\pmb{\longrightarrow}}^0}~ \bm{0} + \twocirc &~ \frac{1}{2^{10}}\frac{1}{3!2!} C^1_2C^2_3(2\np^2-\np) \cr
\recdashcirc~\displaystyle{\mathop{\pmb{\longrightarrow}}^0}~ \bm{0} +\twocirc ~\twocirc &~ \frac{1}{2^{10}}\frac{1}{4!} C^2_4 (2\np^2-\np)\cr
\recdashcirc~\twocirc~\backforth~\displaystyle{\mathop{\pmb{\longrightarrow}}^0}~ \bm{0} +\backforth  &~ \frac{1}{2^{8}}\frac{1}{3!2!} C^2_3C^1_2\nl(2\np-1) \cr
\recdashcirc~\backforth~\displaystyle{\mathop{\pmb{\longrightarrow}}^0}~ \bm{0} +\backforth ~\twocirc  &~ \frac{1}{2^{8}}\frac{1}{2!2!}  C^1_2C^1_2 \nl\np \cr
\recdashcirc~\backforth~\backforth~\displaystyle{\mathop{\pmb{\longrightarrow}}^0}~ \bm{0} +\backforth ~\backforth  &~ \frac{1}{2^{6}}\frac{1}{4!}  C^2_4  (2\nl^2-\nl)\cr
\recdashrec~\displaystyle{\mathop{\pmb{\longrightarrow}}^0}~ \bm{0} + \dashreccirc &~ \frac{1}{2^{6}}\frac{1}{4!} C^2_4 \, 4 
\end{cases}
\end{align}
all together up to 5th order 
\begin{align}\label{110}
\langle \bm{1}|\vred|\bm{0}\rangle_{\bm{0}}=&\frac{\gamma}{4}
+\left(\frac{-1}{128}+\frac{\np}{32}+\frac{\nl}{16}\right)\gamma^3\cr
&+\left( \frac{49}{3072}-\frac{3\nl}{512}+\frac{\nl^2}{128}-
\frac{3\np}{2048}+\frac{\nl\np}{128}+\frac{\np^2}{512} \right) \gamma^5+\cdots
\end{align}

Next is the 6th order of $\bm{0}\displaystyle{\mathop{\to}^0}\bm{0}$, for which, as mentioned in Sec.~4, 
the 6-plaquette cube configurations as in Fig.~\ref{fig11} are to be considered. 
Let us first consider the contribution in the 2d case. 
Here three $+M$'s and $-M$'s cancel each other, and, depending on whether $M$'s sitting in 
one plaquette or not, we have:
\begin{align}\label{111}
\left.\begin{matrix}
0\pmb{\longrightarrow} 0,-M,-M,-M,+M,+M,+M \cr
0,-M,-M,-M,+M,+M,+M\pmb{\longrightarrow} 0
\end{matrix}\right\}:&
\cr
 C^3_6(6\np(\np-1)(\np-2)&+9\np(\np-1)+\np) 
\end{align}
The other is with 12 combinations of two $M$'s on one side and four $M$'s on another side;
they cancel each other out:
\begin{align}\label{112}
\left.\begin{matrix}
0,-M,+M\pmb{\longrightarrow} 0,-M,-M,+M,+M \cr
0,-M,-M,+M,+M\pmb{\longrightarrow} 0,-M,+M
\end{matrix}\right\}:~  C^1_2 C^2_4\left(2\np(\np-1)+\np\right) \np
\end{align}
The third term may contribute by cancellation between the four $M$'s and two $\delta$'s in two sides:
\begin{align}\label{113}
\left.\begin{matrix}
0,-M,-M,+M,+M,+\delta,-\delta\pmb{\longrightarrow} 0,+\delta,-\delta\cr
0,+\delta,-\delta\pmb{\longrightarrow} 0,-M,-M,+M,+M,+\delta,-\delta
\end{matrix}\right\}:~
 C^1_2 C^2_4 \left(2\np(\np-1)+\np\right) \nl  
\end{align}
The fourth may contribute by canceling two $M$'s
in each side, and also $\delta$'s in two sides as:
\begin{align}\label{114}
0,+M,-M,+\delta,-\delta\pmb{\longrightarrow} 0,+M,-M,+\delta,-\delta:~ C^1_2 C^1_2 C^1_2\np \np\nl 
\end{align}
By this form, having possibilities for $+/-$ signs on the first side, we have the combinatorial factor  
\begin{align}\label{115}
\left.\begin{matrix}
0,-M,+M,+\delta,+\delta,-\delta,-\delta\pmb{\longrightarrow} 0,+\delta,+\delta,-\delta,-\delta\cr
0,+\delta,+\delta,-\delta,-\delta\pmb{\longrightarrow} 0,-M,+M,+\delta,+\delta,-\delta,-\delta,
\end{matrix}\right\}:~
 C^1_2 C^2_4 \left(2\nl(\nl-1)+\nl\right)\np  
\end{align}
The other is six $\delta$'s canceling each other out as follows:
\begin{align}\label{116}
0,+\delta,+\delta,+\delta,-\delta-\delta,-\delta\pmb{\longrightarrow}&\, 0,+\delta,+\delta,+\delta,-\delta-\delta,-\delta:~
\cr
 C^3_6&\left(6\nl(\nl-1)(\nl-2)+9\nl(\nl-1)+\nl\right) 
\end{align}
In the last one, cancellation is between four link-current with the virtual loop-current:
\begin{align}\label{117}
\left.\begin{matrix}
0,M,+\delta,+\delta,-\delta,-\delta\pmb{\longrightarrow} 0,M+\delta,+\delta,-\delta,-\delta\cr
0,-M,+\delta,+\delta,-\delta,-\delta\pmb{\longrightarrow}  0,-M+\delta,+\delta,-\delta,-\delta
\end{matrix}\right\}: ~ C^1_2 C^1_2 C^2_4 \np
\end{align}
The graphical representation with all the above at this order will be as follows:
\begin{align}\label{118}
\Big[\langle \bm{0}|\vred|\bm{0}\rangle_{\bm{0}}\Big]_{\gamma^6}:
\begin{cases}
\bm{0}+\twocirc ~\twocirc ~\twocirc ~\displaystyle{\mathop{\pmb{\longrightarrow}}^0}~ \bm{0} &~ \frac{1}{2^{12}}\frac{1}{6!}C^3_6 (6 \np^3-9\np^2+4\np) \cr
\bm{0} ~\displaystyle{\mathop{\pmb{\longrightarrow}}^0}~ \bm{0}+\twocirc~\twocirc~\twocirc &~ \frac{1}{2^{12}}\frac{1}{6!}C^3_6 (6 \np^3-9\np^2+4\np) \cr
\bm{0} +\twocirc~\displaystyle{\mathop{\pmb{\longrightarrow}}^0}~ \bm{0}+\twocirc~\twocirc &~ \frac{1}{2^{12}}\frac{1}{2!4!}C^1_2C^2_4 (2\np^2-\np)\np \cr
\bm{0} +\twocirc~\twocirc~\displaystyle{\mathop{\pmb{\longrightarrow}}^0}~ \bm{0}+\twocirc &~ \frac{1}{2^{12}}\frac{1}{2!4!}C^1_2C^2_4 (2\np^2-\np)\np \cr
\bm{0} +\backforth ~\displaystyle{\mathop{\pmb{\longrightarrow}}^0}~ \bm{0}+\twocirc~\twocirc~\backforth &~ \frac{1}{2^{10}}\frac{1}{2!4!}C^1_2C^2_4 (2\np^2-\np)\nl \cr
\bm{0} +\twocirc~\twocirc~\backforth~\displaystyle{\mathop{\pmb{\longrightarrow}}^0}~ \bm{0}+\backforth &~  \frac{1}{2^{10}}\frac{1}{2!4!}C^1_2C^2_4 (2\np^2-\np)\nl \cr
\bm{0} +\twocirc~\backforth~\displaystyle{\mathop{\pmb{\longrightarrow}}^0}~ \bm{0}+\twocirc~\backforth &~  \frac{1}{2^{10}}\frac{1}{2!2!2!}C^1_2C^1_2 C^1_2 \np\np\nl \cr
\bm{0} +\twocirc~\backforth~\backforth~\displaystyle{\mathop{\pmb{\longrightarrow}}^0}~ \bm{0}+\backforth~\backforth &~ \frac{1}{2^8}\frac{1}{2! 4!} C^1_2 C^2_4 (2\nl^2-\nl) \np \cr
\bm{0} +\backforth~\backforth~\displaystyle{\mathop{\pmb{\longrightarrow}}^0}~ \bm{0}+\twocirc~\backforth~\backforth &~ \frac{1}{2^8}\frac{1}{2! 4!} C^1_2 C^2_4 (2\nl^2-\nl) \np \cr
\bm{0} +\backforth~\backforth~\backforth~\displaystyle{\mathop{\pmb{\longrightarrow}}^0}~ \bm{0}+\backforth~\backforth~\backforth &~ \frac{1}{2^6}\frac{1}{6!}C^3_6 (6 \nl^3-9\nl^2+4\nl) \cr
\bm{0}+\dashreccirc~\displaystyle{\mathop{\pmb{\longrightarrow}}^0}~ \bm{0} + \dashreccirc &
\mathrm{and~reversed}~ \frac{1}{2^{8}}\frac{1}{4!} C^1_2 C^1_2 C^2_4 \np 
\end{cases}
\end{align}
all together leading to
\begin{align}\label{119}
\langle \bm{0}|\vred|\bm{0}\rangle_{\bm{0}}=& 1+ \left(\frac{\np}{8} + \frac{\nl}{4}\right) \gamma^2 \cr
&+\left(-\frac{\nl}{64}+\frac{\nl^2}{32}-\frac{\np}{512}+ \frac{\np \nl}{32} +\frac{\np^2}{128}
\right)\gamma^4\cr
&+\left(\frac{\nl}{576}-\frac{\nl^2}{256}+\frac{\nl^3}{384}+\frac{145\np}{18432}-\frac{5\nl\np}{2048}+\frac{\nl^2\np}{256}
\right.\cr
&\left.~~~-\frac{\np^2}{4096}+ \frac{\np^2 \nl}{512} +\frac{\np^3}{3072}
\right)\gamma^6 +\cdots
\end{align}
As mentioned earlier, for the 3d lattice, the above result should be added by the contribution 
of cube configuration like that in Fig.~\ref{fig11}. The numerical factor 
is known by cases (\ref{79}) or (\ref{80}), via (\ref{69}).
The last step is to find the relevant combinatorial factor that counts the possible ways
that the cubic setup may take place. 
A simple inspection shows that in either the configuration in Fig.~\ref{fig11} or 
its all currents reversed version, there are three $+M$'s and three $-M$'s,
by the convention introduced for $\bm{M}$ matrix. 
This rises a $3!\times 3!$ factor in which three $+M$'s and three $-M$'s
can sit on cube facets.
Also, the number of $\delta$'s containing six $M$'s by mentioned signs is $C^3_6$.
As for the position of the cube, for the 3d periodic lattice with $N_s$ sites in each direction, 
the cube may sit in $N_s^3$ places. These all together determine
the contribution of the cube configuration at 6th order by (\ref{69}) as
\begin{align}\label{120}
	\Big[\langle \bm{0}|\vred|\bm{0}\rangle_{\bm{0}}\Big]_{\mbox{\mancube}}^{\mathrm{3d}}&= 
	2\times 2 \times 3!\times 3!\,C^3_6\, N_s^3\, \frac{1}{2^{12}} \frac{1}{6!} \gamma^6 
	\cr
	& = \frac{1}{2^{10}} N_s^3 \gamma^6 
\end{align}

In the previous section, examples of the non-vacuum block at the lowest orders were given by explicit  
$\delta$-expressions by Fourier integrals that matched with given rules. 
It is useful to see how the given rules work for these blocks at higher orders.
At $\gamma^3$ order in $\bm{k_1}$ block, we consider terms by the virtual current 
in this block for $\bm{q'}=\bm{q}=\bm{0}$.
For the first term, the possible combination would be the cancellation of  $\bm{k_1}$ and `$-\delta$'
despite the presence of the virtual loop-current
\begin{align}\label{121}
\left.\begin{matrix}
{k_1}-\delta \pmb{\longrightarrow}{k_1}-\delta -M+M\cr
{k_1}-\delta-M+M \pmb{\longrightarrow}{k_1}-\delta 
\end{matrix}\right\}:~
 C^1_2 \np 
\end{align}
For the second term, the possible ways would be three combinations of two $-\delta$'s and one $+\delta$, in which 
$\bm{k_1}$ cancels one of $-\delta$'s, and two others cancel each other out. 
Depending on whether all $\delta$'s are in the same link or not, we have the combinatorial factor 
\begin{align}\label{122}
{k_1}-\delta -\delta+\delta\pmb{\longrightarrow}{k_1}-\delta -\delta+\delta:~  C^1_3\left(2(\nl-1)+1\right) 
\end{align}
These can be represented as
\begin{align}\label{123}
\Big[\langle \bm{k_1}|\vred|\bm{k_1}\rangle_{\bm{1}}\Big]_{\gamma^3}:
\begin{cases}
\backforthhor ~\displaystyle{\mathop{\pmb{\longrightarrow}}^0}~ \backforthhor+\twocirc &~ \frac{1}{2^5 2!}C^1_2 \np \cr
\backforthhor +\twocirc ~\displaystyle{\mathop{\pmb{\longrightarrow}}^0}~\backforthhor &~ \frac{1}{2^5 2!}C^1_2 \np \cr
  \backforthhor+\backforth ~\displaystyle{\mathop{\pmb{\longrightarrow}}^0}~\backforthhor+\backforth &~ \frac{1}{2^3 3!}C^1_3 (2\nl-1) 
\end{cases}
\end{align}
All together leading to
\begin{align}\label{124}
\langle \bm{k_1}|\vred|\bm{k_1}\rangle_{\bm{1}}= \frac{\gamma}{2}
+\left(-\frac{1}{16}+\frac{\nl}{8}+\frac{\np}{16}\right)\gamma^3+\cdots
\end{align}

At $\gamma^4$ order with  $\bm{q'}=\bm{0}$ and $\bm{q}=\bm{q_1}$ the graphical representations and combinatorial factors will be as follows:
\begin{align}\label{125}
\Big[\langle \bm{k_1}|\vred|\bm{k_{1;1}}\rangle_{\bm{1}}\Big]_{\gamma^4}:
\begin{cases}
\backforthhor ~\displaystyle{\mathop{\pmb{\longrightarrow}}^0}~
\mathop{\recdashcirc}_{\backforthhor}
+\twocirc &~ \frac{1}{2^7 3!}C^1_3 (2\np-1) \cr
\backforthhor +\twocirc ~\displaystyle{\mathop{\pmb{\longrightarrow}}^0}~ 
\mathop{\recdashcirc}_{\backforthhor} &~ \frac{1}{2^7 2!}C^1_2 \np \cr
 \backforthhor +\backforth ~\displaystyle{\mathop{\pmb{\longrightarrow}}^0}~\mathop{\recdashcirc}_{\backforthhor}+\backforth &~ \frac{1}{2^5 3!}C^1_3 (2\nl-1) 
\end{cases}
\end{align}
by which and (\ref{64}), it ends with
\begin{align}\label{126}
\langle \bm{k_1}|\vred|\bm{k_{1;1}}\rangle_{\bm{1}}=  \frac{\gamma^2}{8}
+\left(-\frac{5}{256}+\frac{\nl}{32}+\frac{\np}{64}\right)\gamma^4+\cdots
\end{align}

It is easy to check that at order $\gamma^5$ the element by $\bm{q'}=\bm{q}=\bm{q_1}$ gets contributions as
\begin{align}\label{127}
\Big[\langle \bm{k_{1;1}}|\vred|\bm{k_{1;1}}\rangle_{\bm{1}}\Big]_{\gamma^5}:
\begin{cases}
\displaystyle{\mathop{\recdashcirc}_{\backforthhor}}
 ~\displaystyle{\mathop{\pmb{\longrightarrow}}^0}~\mathop{\recdashcirc}_{\backforthhor}+\twocirc & \frac{1}{2^9 3!}C^1_3 (2\np-1) 
\vspace{2mm} \cr 
\displaystyle{\mathop{\recdashcirc}_{\backforthhor}} +\twocirc ~\displaystyle{\mathop{\pmb{\longrightarrow}}^0}~\mathop{\recdashcirc}_{\backforthhor} &  \frac{1}{2^9 3!}C^1_3 (2\np-1) 
\vspace{2mm} \cr 
\displaystyle{\mathop{\recdashcirc}_{\backforthhor}} +\backforth ~\displaystyle{\mathop{\pmb{\longrightarrow}}^0}~ \mathop{\recdashcirc}_{\backforthhor}+\backforth & \frac{1}{2^7 3!}C^1_3 (2\nl-1) 
\vspace{2mm} \cr 
\displaystyle{\mathop{\recdashrec}_{\backforthhor}} ~\displaystyle{\mathop{\pmb{\longrightarrow}}^0}~ \mathop{\recdashrec}_{\backforthhor} & \frac{1}{2^5 5!}C^3_5 \,6
\end{cases}
\end{align}
Using (\ref{67}) and (\ref{127}) results in:
\begin{align}\label{128}
\langle \bm{k_{1;1}}|\vred|\bm{k_{1;1}}\rangle_{\bm{1}}=  \frac{\gamma^3}{32}
+\left(\frac{5}{512}+\frac{\nl}{128}+\frac{\np}{256}\right)\gamma^5+\cdots
\end{align}

\section{Spectrum in Strong Coupling Limit}
In the present section, it is shown how the expansion of the elements of the transfer-matrix 
in the Fourier basis can be used directly to calculate the energy spectrum in the strong coupling regime. 
In particular, here we calculate the ground-state and the first excited energies using the expansion obtained 
for the transfer-matrix of the U(1) model. 
The calculation is analytical, using simple 
matrix and quantum perturbation methods, and the eigenvalues of the 
mentioned states are calculated up to the fourth order in $\gamma$. 
To calculate far beyond this order, the numerical methods are needed to work with large matrices. 

As it was seen in the previous sections, in the extreme strong coupling limit $\gamma=0$ all the 
matrix elements are zero, except $\langle\bm{0}|\vred|\bm{0}\rangle$ which is 1.
The $\gamma=0$ limit is considered as the unperturbed case, corresponding to the following eigenvalues 
for $\widehat{V}^{0}$:
\begin{align}\label{s6-1}
\gamma=0:~ v^{(0)}_0=1,~~~~ v^{(0)}_{q\neq 0}=0
\end{align}
Using (\ref{17}), the unperturbed energy values are obtained as $\varepsilon^{(0)}_0=0$ for the ground-state and 
$\varepsilon^{(0)}_{q\neq 0}\to +\infty$ for all other states. 
The full transfer-matrix, corresponding to a non-zero but small $\gamma$ is 
\begin{align}\label{s6-2}
\vred=\widehat{V}^{0}+\bar{V}
\end{align}
It is useful to have the explicit expressions of perturbative corrections to the eigenvalues up to the fourth 
order in the present case 
\begin{align}
\label{s6-3}   
&v_q^{(1)}=\bar{V}_{qq} \\
\label{s6-4}
&v_q^{(2)}=\frac{\bar{V}_{qq'}^2}{v_{qq'}} \\
\label{s6-5}
&v_q^{(3)}=\frac{\bar{V}_{qq'} \bar{V}_{q'q''} \bar{V}_{q''q} }{v_{qq'}v_{qq''}}-\bar{V}_{qq}
\frac{\bar{V}_{qq'}^2}{v_{qq'}^2}\\
\label{s6-6}
&v_q^{(4)}=\frac{\bar{V}_{qq'} \bar{V}_{q'q''}\bar{V}_{q''q'''} \bar{V}_{q'''q} }{v_{qq'}v_{qq''}v_{qq'''}} 
-\frac{\bar{V}_{qq'}^2}{v_{qq'}^2}\frac{\bar{V}_{qq''}^2}{v_{qq''}} -2 \bar{V}_{qq} 
\frac{\bar{V}_{qq'}\bar{V}_{q'q''}\bar{V}_{q''q}}{{v_{qq'}^2}v_{qq''}} + \bar{V}_{qq}^{2}\frac{\bar{V}_{qq'}^2}{v_{qq'}^3}
\end{align}
in which $v_{qq'}=v^{(0)}_{q}-v^{(0)}_{q'}$ and $\bar{V}_{qq'}=\langle \bm{q}|\bar{V}|\bm{q'}\rangle$. 
The summations in the above expressions are understood over those values of $q'$, $q''$, and $q'''$, for which 
the denominators do not vanish.
It was mentioned at the end of Sec.~2 that the ground-state of the U(1) model belongs 
to the vacuum block \cite{vadfat}. 
The lowest order of $\gamma$ in a matrix element was already obtained in (\ref{75}). Up to
the order $\gamma^2$, the only non-vanishing elements in the vacuum block are 
those with at most two units of loop-currents in one or two plaquettes, namely
\begin{align}
\label{s6-7}
&\bar{V}_{0,\pm1} = \frac{\gamma}{4}+\mathrm{O}(\gamma^3) \\
\label{s6-8}
&\bar{V}_{\pm 1,\pm 1' \,\mathrm{or}\,\pm1},\bar{V}_{\pm 1', \,\mathrm{or}\,\mp 1},
\bar{V}_{0,\pm 2}, \bar{V}_{0,\pm 1\pm 1},
\bar{V}_{0,\pm 1\mp 1}  \propto \gamma^2
\end{align} 
We also have 
\begin{align}\label{s6-9}
\bar{V}_{00}=\left(\frac{\np}{8} + \frac{\nl}{4}\right) \gamma^2 +\mathrm{O}(\gamma^4)
\end{align}
Using (\ref{s6-4}), it is easily understood that at the order of $\gamma^2$ there are $2\np$ elements  
$\bar{V}_{0,q}$ with $|q|=1$ that contribute to the ground-state.
Also by (\ref{s6-1}), in the perturbative corrections to the ground-state we have $v_{0q}=1-0=1$ in the denominator.
All together, it is seen that up to the order $\gamma^2$
\begin{align}\label{s6-10}
v_0&=1+ \bar{V}_{00}+2\np \bar{V}_{0,\pm1}^2   \\
\label{s6-11}
&= 1+ \frac{1}{4}(\np+\nl ) \gamma^2 +\mathrm{O}(\gamma^4)
\end{align}
Further, using the well-known expression for the correction to the eigenvectors, we find for the ground-state
\begin{align}\label{s6-12}
\vec{v}_0 =& (1,0,0,\cdots 0)+\sum_{|q|=1}\bar{V}_{0,q} ~ \vec{q} \\
\label{s6-13}
=&\Big(1,\underbrace{\frac{\gamma}{4},\frac{\gamma}{4},\cdots, \frac{\gamma}{4}}_{2\np} \Big)
+\mathrm{O}(\gamma^2)
\end{align}
Interestingly, we see that the expansion of the transfer-matrix elements in the Fourier basis can be used 
to develop an expansion for the energy and eigenvector of the ground-state.
The physical meaning of the above vector based on the current-states is simple. In the extreme strong 
limit $\gamma=0$, the ground-state is simply the vacuum. For non-zero but small $\gamma$,
the ground-state is represented by a linear combination of the vacuum state and $2\np$ states constructed 
by $\pm 1$ unit of loop-current in one plaquette, such as (\ref{23}) in Fig.~\ref{fig4}. 

It is instructive to find also the correction of order $\gamma^2$ to the ground-state by 
a truncated version of the vacuum block as following 
\begin{align}\label{s6-14}
\vred\!\!\!~_\mathrm{vac|\gamma^2}=\left(
\begin{array}{c | c }
\mathrm{A} & ~~~~~\mathrm{B}~~~~~  \\ \hline
~& ~ \\
\!\mathrm{B}^T\!  & ~\mathrm{C}   \\ 
~ & ~ 
\end{array}\right)
\end{align}
with $\mathrm{B}^T$ being the transpose of B. 
The sub-blocks A, B, and C are introduced in Table~1. 

\vskip .2cm
\begin{table}[H]
  \begin{center}
    \label{tab:table1}
    \begin{tabular}{c|c|c}
Sub-block & Dimension  & Elements \\
      \hline
A & $1\times 1$ & $a=1+\bar{V}_{00}$ \\ \hline
B & $1\times 2\np$ & $b=\bar{V}_{0,\pm 1}=\gamma/4$ \\ \hline
C & $2\np\times 2\np$ & $ c=\begin{cases}
                                 \bar{V}_{\pm 1,\pm 1' \,\mathrm{or}\,\pm1}=\gamma^2/16 \cr
                                  \bar{V}_{\pm 1,\mp 1' \,\mathrm{or}\,\mp1}=\gamma^2/16   \end{cases}  $
    \end{tabular}
\caption{The sub-blocks in the truncated vacuum block.}
  \end{center}
\end{table}
Fortunately, the eigenvalues and eigenvectors of (\ref{s6-14}) can be obtained analytically. 
For the eigenvalues we find:
\begin{align}
\label{s6-15}
&v_0=\frac{1}{2}\left(a+2\np\, c + \sqrt{\left(a-2\np\, c\right)^2 +8 \np\, b^2}\right) \\
\label{s6-16}
&v_1=\frac{1}{2}\left(a+2\np\, c - \sqrt{\left(a-2\np\, c\right)^2 +8 \np\, b^2}\right) \\
\label{s6-17}
&v_{i=2,\cdots,2\np+1}=0
\end{align}
and the following for the ground-state eigenvector
\begin{align}\label{s6-18}
\vec{v}_0 = \left( 
\frac{1}{2}\left(a-2\np\, c + \sqrt{\left(a-2\np\, c\right)^2 +8 \np\, b^2}\right)
,\underbrace{b,b,\cdots, b}_{2\np} \right)
\end{align}
Expanding (\ref{s6-15}) and (\ref{s6-18}) up 
to $\gamma^2$, the expressions (\ref{s6-11}) and (\ref{s6-13}) are obtained.
Also as a by-product, using (\ref{s6-16}), one also finds for the next eigenvalue in the vacuum block
\begin{align}\label{s6-19}
v_1=0\times \gamma^2+\mathrm{O}(\gamma^4)
\end{align}
This is of the order $\gamma^4$, so the first excited energy is not in the vacuum block. 
As we will see, there are blocks with eigenvalues of order $\gamma$, which are in fact the first excited energies
of the model. Before considering the excited-states, let us consider the ground-state energy at the order 
$\gamma^4$. First, we have the element
\begin{align}\label{s6-20}
\bar{V}_{00}=\left(\frac{\np}{8} + \frac{\nl}{4}\right) \gamma^2 +\left(-\frac{\nl}{64}+\frac{\nl^2}{32}-\frac{\np}{512}+ \frac{\np \nl}{32} +\frac{\np^2}{128}\right)\gamma^4+\mathrm{O}(\gamma^6)
\end{align}
and the elements
\begin{align}\label{s6-21}
\bar{V}_{0,q}&=\frac{\gamma}{4} +\left(\frac{-1}{128}+\frac{\np}{32}+\frac{\nl}{16}\right)\gamma^3
+\mathrm{O}(\gamma^5),~~~~|q|=1  \\
\label{s6-22}
\bar{V}_{0,q}&=
\frac{\gamma^2}{32}
	+\left(\frac{-1}{768}+\frac{\nl}{128}+\frac{\np}{256}\right)\gamma^4
	+\mathrm{O}(\gamma^6), ~~~|q|=2
\end{align}
each having $2\np$ cases, by considering $\pm$ signs. Second, there are the elements 
\begin{align}\label{s6-23}
\bar{V}_{q,q}&= \frac{\gamma^2}{16}
+\left(\frac{15}{256}+\frac{\nl}{64}+\frac{\np}{128}\right)\gamma^4 
	+\mathrm{O}(\gamma^6),~~~~|q|=1 \\
\label{s6-24}
\bar{V}_{q,-q}&=\frac{\gamma^2}{16}
	+\left(\frac{-1}{256}+\frac{\nl}{64}+\frac{\np}{128}\right)\gamma^4 
	+\mathrm{O}(\gamma^6),~~~~|q|=1
\end{align}
each with $2\np$ numbers. The other relevant elements are
\begin{align}\label{s6-25}
\bar{V}_{q,q'}&=\frac{\gamma^2}{16}
	+\left(-\frac{1}{256}+\frac{\nl}{64}+\frac{\np}{128}\right)\gamma^4 
	+\mathrm{O}(\gamma^6),~~~~q\neq q', ~|q|=|q'|=1\\
\label{s6-26}
\bar{V}_{0,q\,q'}&= \frac{\gamma^2}{16} 	+\left(-\frac{1}{256}+\frac{\nl}{64}+\frac{\np}{128}\right)\gamma^4 
	+\mathrm{O}(\gamma^6),~~~~q\neq q', ~|q|=|q'|=1
\end{align}
with $4\np(\np-1)$ and $2\np(\np-1)$ numbers, respectively. 
There are other elements at the order $\gamma^3$ such as:
\begin{align}\label{s6-27}
\bar{V}_{0,\pm 3},~\bar{V}_{0,\pm1\pm 2},~\bar{V}_{\pm1,\pm 2},~\bar{V}_{0,\pm1\pm 1\pm1},
~\bar{V}_{\pm1,\pm 1\pm1}
\end{align}
and also at the order $\gamma^4$:
\begin{align}\label{s6-28}
\bar{V}_{0,\pm 4},~\bar{V}_{\pm 1,\pm 3},~\bar{V}_{0,\pm 1\pm 3},~
\bar{V}_{\pm 2,\pm 2},~\bar{V}_{0,\pm 2\pm 2},~ \bar{V}_{1\pm,\pm 1\pm 2},~\cr
\bar{V}_{0,\pm 1\pm 1\pm 2}; \bar{V}_{0,\pm 1\pm 1\pm 1\pm 1},~
\bar{V}_{\pm 1,\pm 1\pm 1\pm 1},~\bar{V}_{\pm 1\pm 1,\pm 1\pm 1}
\end{align}
However, a simple inspection based on corrections (\ref{s6-3})-(\ref{s6-6}) shows that none of the terms of the order 
$\gamma^3$ in (\ref{s6-27}), and 
none of $\gamma^4$'s in (\ref{s6-22})-(\ref{s6-28}) contribute at the order $\gamma^4$ to the ground-state.
It is also easy to check that the correction at the order $\gamma^3$ vanishes. 
So the next correction is of the order $\gamma^4$, which finds contributions as following
\begin{align}\label{s6-29}
v_0^{(1)}&=\bar{V}_{00} -(\gamma^2~\mathrm{order})\! =\!
\left(-\frac{\nl}{64}+\frac{\nl^2}{32}-\frac{\np}{512}+ \frac{\np \nl}{32} +\frac{\np^2}{128}\right)\!\gamma^4 
\end{align}
\begin{align}\label{s6-30}
v_0^{(2)}&=2 \np \bar{V}_{0,1}^2 +2\np \bar{V}_{0,2}^2 
+2\np(\np-1) \bar{V}_{0,1\,1}^2 \cr
&= 2\np\! \times \!
2\frac{\gamma}{4}\left(\frac{-1}{128}+\frac{\np}{32}+\frac{\nl}{16}\right)\!\gamma^3\!
+2\np\! \left(\frac{\gamma^2}{32}\right)^2\!\!\!+2\np(\np-1) \left(\frac{\gamma^2}{16}\right)^2 \cr 
&=\left( \frac{-7\np}{512}+\frac{\nl\np}{16}+ \frac{5\np^2}{128}  \right)\gamma^4 
\end{align}
\begin{align}\label{s6-31}
v_0^{(3)}&= 2\times 2\np  \bar{V}_{0,1} \bar{V}_{1,1} \bar{V}_{0,1} 
 + 4\np(\np-1) \bar{V}_{0,1} \bar{V}_{1,1'} \bar{V}_{0,1'}
 -2\np \bar{V}_{00}\bar{V}_{0,1}^2  \cr
&= 4\np (\frac{\gamma}{4})^2(\frac{\gamma^2}{16})
+4\np(\np-1)(\frac{\gamma}{4})^2(\frac{\gamma^2}{16})
-2\np(\frac{\gamma}{4})^2(\frac{\nl}{4}+\frac{\np}{8})\gamma^2\cr 
& =-\frac{\nl\np}{32}\gamma^4
\end{align}
\begin{align}\label{s6-32}
v_0^{(4)}= - 2\np \times 2 \np \bar{V}_{0,1}^2 \bar{V}_{0,1}^2 
=-4\np^2 (\frac{\gamma}{4})^2 (\frac{\gamma}{4})^2=-\frac{\np^2}{64}\gamma^4
\end{align}
All together we have for the ground-state energy 
\begin{align}\label{s6-33}
v_0= 1+ \frac{1}{4}(\np+\nl ) \gamma^2 +\left(\frac{-\nl}{64}+\frac{\nl^2}{32}-\frac{\np}{64}+\frac{\nl \np}{16}+\frac{\np^2}{32}\right)\gamma^4+\mathrm{O}(\gamma^6)
\end{align}

Now let us consider the first excited state by the transfer-matrix of the model. 
Using the previously obtained expansions, the state 
with one unit of current on one link, such as $\bm{k_1}$ of (\ref{24}),
leads to a diagonal element linear in $\gamma$ at the lowest order
as (\ref{124}). The first element of the block up to order $\gamma^5$, 
expressed in the notation of present section, is:
\begin{align}\label{s6-34}
\bar{V}^{ \bm{k_1}}_{00}=& \frac{\gamma}{2}+\left( -\frac{1}{16}+\frac{\nl}{8}+\frac{\np}{16}\right)\gamma^3
\cr
&+\left(\frac{7}{384}-\frac{3\nl}{128}+\frac{\nl^2}{64}-\frac{9\np}{1024}+\frac{\nl\np}{64}+\frac{\np^2}{256}\right)
\gamma^5 +\mathrm{O}(\gamma^7)
\end{align}
So the first excited-states belongs to the blocks with 
current-states such as $\bm{k_1}$. Considering 
$\pm 1$ units of current on one link as the representative state,
the number of such blocks are $2\nl$. So the first excited energy has a degeneracy of $2\nl$.
In fact, finding the corrections to the first excited energy is simple and can be done in the same way
done for the ground-state. Factoring out $\gamma/2$ from the elements of the $\bm{k_1}$-block,
the very same calculation for the ground-state can be repeated. 
Similar to the method used for the vacuum block, we have the following elements in the $\bm{k_1}$-block
\begin{align}\label{s6-35}
\bar{V}^{ \bm{k_1}}_{00}&= \frac{\gamma}{2}\left( 1+( -\frac{1}{8}+\frac{\nl}{4}+\frac{\np}{8})\gamma^2+\mathrm{O}(\gamma^4)\right)\\
\label{s6-36}
\bar{V}^{\bm{k_1}}_{0,\pm1}&=\frac{\gamma}{2}\left( \frac{\gamma}{4}+\mathrm{O}(\gamma^3)\right)
\end{align}
in which we find for the first excited energy
\begin{align}\label{s6-37}
v^{\bm{k_1}}_0 &= \frac{\gamma}{2}\left[ 1+( -\frac{1}{8}+\frac{\nl}{4}+\frac{\np}{8})\gamma^2
+2\np( \frac{\gamma}{4}) ^2\right] + \mathrm{O}(\gamma^5) \\
\label{s6-38}
&= \frac{\gamma}{2}+ \left(-\frac{1}{16}+\frac{\nl}{8}+\frac{\np}{8}\right)\gamma^3+ \mathrm{O}(\gamma^5)
\end{align}
The gap of energy by the model can be calculated by the ground-state energy (\ref{s6-33}) and the 
above as the excited one.
The eigenvector of the energy (\ref{s6-38}) can be obtained using the perturbation method as well 
\begin{align}\label{s6-39}
\vec{v\,}^{\bm{k_1}}_0 =& (1,0,0,\cdots 0)+
\frac{1}{\gamma/2}\sum_{|q|=1}\bar{V}^{\bm{k_1}}_{0,q} ~ \vec{q} \\
=&\Big(1,\underbrace{\frac{\gamma}{4},\frac{\gamma}{4},\cdots, \frac{\gamma}{4}}_{2\np} \Big)
+\mathrm{O}(\gamma^2)
\end{align}
It is noticed that the vector is exactly in the form of (\ref{s6-13}) in
the vacuum block, however, regarding the current-states, with a different physical meaning. 
Here, the eigenvector is the linear combination of the $\bm{k_1}$-state and $2\np$ states 
constructed by $\bm{k_1}$-state added by $\pm1$ unit of loop-current in one plaquette, such as 
Fig.~\ref{fig7}. 

The next correction can be obtained along the lines of (\ref{s6-29})-(\ref{s6-32}) but in the 
$\bm{k_1}$-block, leading to 
\begin{align}
v^{\bm{k_1}}_0 =& \frac{\gamma}{2}\!+\!\left(\frac{-1}{16}+\frac{\nl}{8}+\frac{\np}{8}\right)\!\gamma^3\!\cr
&+\!\left(\frac{7}{384}-\frac{3\nl}{128}+\frac{\nl^2}{64}-\frac{3\np}{128}+\frac{\nl\np}{32}+\frac{\np^2}{64}\right)\!\gamma^5\!+\!\mathrm{O}(\gamma^7)
\end{align}

The above procedure to find the eigenvalues can be used for the other blocks as well. For later use
in the next section, we present the lowest eigenvalue in the $\bm{k_2}$-block, 
\begin{align}
v^{\bm{k_2}}_0 =& \frac{\gamma^2}{8}\!+\!\left(-\frac{1}{48}+\frac{\nl}{32}+\frac{\np}{32}\right)\!\gamma^4\!
\cr
&+\!\left(\frac{23}{3072}-\frac{11\nl}{1536}+\frac{\nl^2}{256}-\frac{11\np}{1536}+\frac{\nl\np}{128}+\frac{\np^2}{256}\right)\!\gamma^6\!+\!\mathrm{O}(\gamma^8)
\end{align}
the $\bm{k_{11'}}$-block,
\begin{align}
v^{\bm{k_{11'}}}_0 =& \frac{\gamma^2}{4}\!+\!\left(-\frac{1}{16}+\frac{\nl}{16}+\frac{\np}{16}\right)\!\gamma^4\!
\cr
&+\!\left(\frac{7}{384}-\frac{5\nl}{256}+\frac{\nl^2}{128}-\frac{5\np}{256}+\frac{\nl\np}{64}+\frac{\np^2}{128}\right)\!\gamma^6\!+\!\mathrm{O}(\gamma^8)
\end{align}
and the $\bm{k_{3}}$-block,
\begin{align}
v^{\bm{k_3}}_0 =& \frac{\gamma^3}{48}\!+\!\left(-\frac{1}{256}+\frac{\nl}{192}+\frac{\np}{192}\right)\!\gamma^5\!
\cr
&+\!\left(\frac{17}{10240}-\frac{\nl}{768}+\frac{\nl^2}{1536}-\frac{\np}{768}+\frac{\nl\np}{768}+\frac{\np^2}{1536}\right)\!\gamma^7\!+\!\mathrm{O}(\gamma^9)
\end{align}

\section{Lattice Size and Observable Values}

In the presented expansions of the transfer-matrix elements and eigenvalues,  
all of the terms except the first ones have positive powers of $\np$ and $\nl$.
So for an infinite lattice, in which $\np$ and $\nl$ go to infinity,
a proper interpretation of the expansions is necessary. 
The purpose of this section is twofold. First, to show how the presence 
of the mentioned numbers in the expansions is expected. Second, to 
present a proper interpretation of the resulted expansions,
which are formally divergent in the large lattice limit.  

Let us consider the matrix-element $\langle \bm{k'}|\widehat{V}|\bm{k}\rangle$
representing the transition $\bm{k'}\!\to\!\bm{k}$. 
It is shown that in the strong coupling regime, 
the $\gamma$-expansion of this matrix-element 
is represented by virtual occurrences of loop and link currents that transform both 
states to the vacuum. The weight of each virtual current in expansions is $\gamma$.
The first term represents the lowest number of virtual currents, 
in \textit{definite} links and plaquettes,
necessary to perform the mentioned transform to the vacuum. So, no $\nl$ and $\np$ is expected 
in the first term. The other terms, however, represent all virtual currents in \textit{any} 
link and plaquettes that may contribute to the transform. As a result, 
the number of links and plaquettes enter the expansions. As seen in the previous examples, 
this is exactly the way that powers of $\np$ and $\nl$ appear in the expansions. 

The way that $\nl$ and $\np$ appear in the expansions
provides a physical basis to treat the behavior of expansions in the infinite lattice limit. 
In particular, due to the mentioned role by the virtual currents in the transitions, 
a distinguished role is expected to be given to the element 
$\langle \bm{0}|\widehat{V}|\bm{0}\rangle$, as the amplitude of the 
vacuum-to-vacuum (v.t.v.) transition. 
The reason is that, the virtual events that derive the v.t.v. transition are expected to take place 
in the transitions between other current-states as well, leading to the appearance of  $\nl$ and $\np$ in the expansions. 
As a specific example, let us consider a 2d lattice, in which $\nl=2\np$. 
By replacing $\nl$ by $\np$, the v.t.v. transition (\ref{119}) in the 
2d lattice finds the following form
\begin{align}
\langle \bm{0}|\vred|\bm{0}\rangle_{\bm{0}}^{\mathrm{2d}}= 1+& \left(\frac{5\gamma^2}{8} - \frac{17 \gamma^4}{512}+\frac{209 \gamma^6}{18432}\right)\np +\left(\frac{25 \gamma^4}{128}-\frac{85 \gamma^6}{4096}\right)\np^2
\cr
&+\left(\frac{125 \gamma^6}{3072}\right)\np^3 +\cdots
\end{align}
Based on the above explanations, one expects that a footprint of the v.t.v. transition can be 
traced in other transitions. In particular, it is seen that the ratio of the other matrix elements
to the v.t.v. one is independent of $\np$. As examples in the vacuum block, 
we have the following for the 2d lattice:
\begin{align}
\langle \bm{0}|\vred|\bm{1}\rangle_{\bm{0}}=&\langle \bm{0}|\vred|\bm{0}\rangle\cdot
\frac{\gamma}{4}
\left(1-\frac{\gamma^2}{32}+\frac{49 \gamma^4}{768}-\frac{7691 \gamma^6}{196608}+\cdots\right)
\\
\langle \bm{1}|\vred|\bm{1}\rangle_{\bm{0}}=&\langle \bm{0}|\vred|\bm{0}\rangle\cdot
\frac{\gamma^2}{16}
\left(1+\frac{15\gamma^2}{16}-\frac{1525 \gamma^4}{3072}+\frac{22861 \gamma^6}{98304}+\cdots\right)
\\
\langle \bm{0}|\vred|\bm{2}\rangle_{\bm{0}}=&\langle \bm{0}|\vred|\bm{0}\rangle\cdot 
\frac{\gamma^2}{32}
\left(1-\frac{\gamma^2}{24}+\frac{1163 \gamma^4}{6144}-\frac{9073 \gamma^6}{81920}+\cdots\right)
\\
\langle \bm{0}|\vred|\bm{3}\rangle_{\bm{0}}=&\langle \bm{0}|\vred|\bm{0}\rangle\cdot
\frac{\gamma^3}{384}
\left(1-\frac{3\gamma^2}{64}+\frac{1941 \gamma^4}{10240}+\frac{168611 \gamma^6}{1474560}+\cdots\right)
\end{align}
in which we see that the $\np$ dependence is totally factored out as the v.t.v. transition element.
As more examples by composite current-vectors in the vacuum block we have:
\begin{align}
\langle \bm{0}|\vred|\bm{1,1}\rangle_{\bm{0}}=&\langle \bm{0}|\vred|\bm{0}\rangle\cdot
\frac{\gamma^2}{16}
\left(1-\frac{\gamma^2}{16}+\frac{203 \gamma^4}{3072}+\frac{1325 \gamma^6}{98304}+\cdots\right)
\\
\langle \bm{0}|\vred|\bm{1,2}\rangle_{\bm{0}}=&\langle \bm{0}|\vred|\bm{0}\rangle\cdot
\frac{\gamma^3}{128}
\left(1-\frac{7\gamma^2}{96}+\frac{137 \gamma^4}{2048}+\frac{73319 \gamma^6}{1474560}+\cdots\right)
\end{align}
The above observation is expected to be true in blocks other than the vacuum block as well. 
As examples, from the $\bm{k_1}$ and $\bm{k_2}$ blocks we have
\begin{align}
\langle \bm{k_1}|\vred|\bm{k_1}\rangle_{\bm{1}}=&\langle \bm{0}|\vred|\bm{0}\rangle\cdot
\frac{\gamma}{2}
\left(1-\frac{\gamma^2}{8}+\frac{7 \gamma^4}{192}-\frac{25 \gamma^6}{1536}+\cdots\right)
\\
\langle \bm{k_1}|\vred|\bm{k_{1;1}}\rangle_{\bm{1}}=&\langle \bm{0}|\vred|\bm{0}\rangle\cdot
\frac{\gamma^2}{8}
\left(1-\frac{5\gamma^2}{32}+\frac{25 \gamma^4}{384}-\frac{6619 \gamma^6}{196608}+\cdots\right)
\\
\langle \bm{k_{1;1}}|\vred|\bm{k_{1;1}}\rangle_{\bm{1}}=&\langle \bm{0}|\vred|\bm{0}\rangle\cdot
\frac{\gamma^3}{32}
\left(1+\frac{5\gamma^2}{16}-\frac{721 \gamma^4}{3072}+\frac{8369 \gamma^6}{98304}+\cdots\right)
\\
\langle \bm{k_2}|\vred|\bm{k_2}\rangle_{\bm{2}}=&\langle \bm{0}|\vred|\bm{0}\rangle\cdot
\frac{\gamma^2}{8}
\left(1-\frac{\gamma^2}{6}+\frac{11\gamma^4}{384}-\frac{11 \gamma^6}{960}+\cdots\right)
\\
\langle \bm{k_2}|\vred|\bm{k_{2;1}}\rangle_{\bm{2}}=&\langle \bm{0}|\vred|\bm{0}\rangle\cdot
\frac{\gamma^3}{32}
\left(1-\frac{19\gamma^2}{96}+\frac{55\gamma^4}{768}-\frac{103957 \gamma^6}{2949120}+\cdots\right)
\\
\langle \bm{k_{2;1}}|\vred|\bm{k_{2;1}}\rangle_{\bm{2}}=&\langle \bm{0}|\vred|\bm{0}\rangle\cdot
\frac{\gamma^4}{128}
\left(1+\frac{5\gamma^2}{48}-\frac{421\gamma^4}{3072}+\frac{75107 \gamma^6}{1474560}+\cdots\right)
\end{align}
It is interesting that the above expressions are valid for any value of $\np$. 
Despite the fact that the whole dependence on $\np$
can be extracted equally from all elements, it does not make them finite
in the large lattice limit. The basic idea to define physically finite values comes 
from the field theory approach to Feynman diagrams. As mentioned before, 
the present diagrammatic expansion
has features similar to the those of the Feynman diagrams of perturbative quantum
field theory. Here we see another example of these common features. 
The $n$-point functions in the field theory approach, supposedly contain
vacuum diagrams; those with no external legs originated from the theory with no source term. 
Having no external legs, the vacuum diagrams are the only ones that contribute
to the v.t.v. transition.
The absence of external legs has two important consequences. First, the vacuum diagrams 
are infinite, as they can take place in the infinite extent of the space \cite{rohrlich,peskin}. 
Second, the vacuum diagrams contribute equally as a multiplicative factor to 
all transitions \cite{rohrlich,ryder,peskin}. 
Hence, the total contribution of these diagrams can be factored out 
from the $S$-matrix elements, leaving only physically observable contributions to 
the transitions \cite{rohrlich,peskin,ryder}. In the field theory approach,
the contribution by the vacuum diagrams is extracted from the 
$n$-point functions by defining the \textit{normalized} generating functional,
through dividing the path-integral expression with source `$J$' by the sourceless one \cite{ryder}
\begin{align}
Z[J] = \frac{\displaystyle{\int\!\! \mathscr{D}\phi\, \exp\!\left[\mathrm{i}\!\int\!\! dx\,\big(\mathscr{L}+ J\, \phi \big)\right]}}
{\displaystyle{\int\!\! \mathscr{D}\phi\, \exp\!\left[\mathrm{i}\!\int\!\! dx\, \mathscr{L}\right]}}
\end{align}
It then can be shown that, the $n$-point functions derived from the normalized functional
are free from the vacuum diagrams \cite{ryder,peskin}. 
In a quite similar way, we see that the v.t.v. contribution 
can be extracted from the present strong coupling expansion of matrix-elements,
leaving a finite physical transition amplitude between the current-states. 
As mentioned in Sec.~3, the current-vector $\bm{k}$ plays the role of the source. 
The origin of the divergent behavior of $\langle \bm{0}|\vred|\bm{0}\rangle$ in
the infinite lattice limit is the same as its counterpart in the field theory approach. 

We see that by extracting $\langle \bm{0}|\vred|\bm{0}\rangle$ from the transfer-matrix,
all elements become independent of $\np$. However, that is not enough
to guarantee that the eigenvalues are independent of $\np$. 
This can be checked explicitly by the eigenvalues obtained in the previous section. 
The reason for this is simply that, the number of 
off-diagonal elements contributing to an eigenvalue may depend on $\np$. 
We already have seen examples of this in the previous section, for example 
(\ref{s6-10}) and (\ref{s6-30})-(\ref{s6-32}). 
Theoretically, for a system with finite energy density we expect 
infinite energy in the infinite size limit \cite{peskin}. So, there is no surprise about the divergent
behavior of the obtained eigenvalues. However, it is still expected that
the observable values related to energy would be finite. 
Fortunately the solution is quite known that, as far as measurements are concerned, the relevant quantity 
is the energy difference rather than the energy itself.  
In our case, using $v_i=\exp(-a\, \varepsilon_i)$, it is enough
to check the behavior of the ratio $v_i/v_0=\exp\big[-a(\varepsilon_i-\varepsilon_0)\big]$. 
If the mentioned ratio is independent of $\np$, then the difference between every two energies will be too. 
As explicit examples, for eigenvalues obtained in the $\bm{k_1}$, $\bm{k_2}$, $\bm{k_{11'}}$ 
and $\bm{k_3}$ blocks,
\begin{align}
\frac{v^{\bm{k_1}}_0 }{v_0 } &=\frac{\gamma}{2}\left( 1-\frac{\gamma^2}{8}+ \frac{7\gamma^4}{192}+\cdots\right)
\\
\frac{v^{\bm{k_2}}_0 }{v_0 } &= \frac{\gamma^2}{8} \left( 1-\frac{\gamma^2}{6}+ \frac{23\gamma^4}{384}+\cdots\right)
\\
\frac{v^{\bm{k_{11'}}}_0}{v_0 } &=\frac{\gamma^2}{4} \left( 1-\frac{\gamma^2}{4}+ \frac{7\gamma^4}{96}+\cdots\right)
\\
\frac{v^{\bm{k_3}}_0 }{v_0 } &= \frac{\gamma^3}{48} \left(1-\frac{3\gamma^2}{16}+ \frac{51\gamma^4}{640}+\cdots\right)
\end{align}
in which all ratios are independent of $\np$. By the above observation, to excite the system from the ground-state 
a finite amount of energy, independent of the lattice size, is needed. 

\section{Conclusion and Discussion}
The formulation of the transfer-matrix of the U(1) lattice model in the field Fourier basis 
was studied. It was discussed in detail how the states in the Fourier basis correspond to 
quantized currents on links. The constraint to have a non-vanishing element
by two states was shown to be in fact the condition that, as
a lattice version of current conservation, the two states differ in
loop-currents circulating inside plaquettes. 
These features provide a basis to develop the strong coupling 
expansion and its diagrammatic representation for the elements of the transfer-matrix 
in the field Fourier basis. Each term of the expansion represents 
the occurrence of virtual loop and link currents 
that transform the initial and final states to the vacuum state.
Accordingly, the diagrams correspond to combinations of the initial and 
final current-states and the occurred virtual currents that transform both states to the vacuum. 
The weight of each virtual current is $1/g^2$, which is small in the strong coupling regime. 
Either by interpretation or through managing the relevant terms at 
a given order of the strong coupling expansion, the diagrams 
play the role of Feynman diagrams at the small coupling regime.
The present expansion of the transfer-matrix elements are used to develop the
expansion of the ground-state and some excited energies at the lowest orders in the strong
coupling regime. Based on the observation that the lattice size dependence of expansions 
can be factored out from the matrix-elements and eigenvalues, 
the physical interpretation of the results is discussed. 

The existence of the manageable strong coupling expansion,
combined with the available perturbative small coupling expansion,
might be useful as it could provide some knowledge about the phase structure of the model. 
A classic example is the 2d Ising model, for which it can be 
shown that the small coupling (high temperature) expansion 
of the partition function is equivalent to the strong coupling 
(low temperature) expansion of the same model but on the dual lattice \cite{krawan,huang}. 
Accordingly, 
this leads to the fact that the 2d Ising model exhibits a phase transition with a known 
critical coupling (temperature) \cite{krawan}, even before the exact solution at any coupling 
is found \cite{onsager}. These kinds of extra benefits of a strong coupling expansion 
are especially important in the case of lattice gauge theories, as they are expected to 
capture the essential features of phase transitions of gauge models. In particular, any 
relation between  the small and the strong coupling expansions may be considered as a piece 
of evidence for a phase transition in an intermediate coupling value. 

\begin{appendices}

\section{Third and Fourth Orders of $\gamma$ in Vacuum Block}

Presenting the contribution at order $\gamma^3$ needs a more compact notation,
for which we define the new $\Delta$ as following with the summations on all $\pm$-sign
combinations 
\begin{align}\label{129}
\Delta\left(\frac{\slashed{q}_l ;\,M^{p_1}_{\,l},\cdots,M^{p_r}_{\,l}}
{\slashed{q}'_{l};\,M^{p'_1}_{\,l},\cdots,M^{p'_{s}}_{\,l}}; \,\delta_{ll_1},\cdots,\delta_{ll_{t}}\right)&=\cr
\sum_\pm \prod_l &\delta(\slashed{q}_l \!\pm\! M^{p_1}_{\,l}\!\pm\! \cdots \!\pm\! M^{p_r}_{\,l}
\!\pm\! \delta_{ll_1} \!\pm\!\cdots\!\pm\!\delta_{ll_{t}}) \cr \times &
\delta(\slashed{q}'_l \!\pm\! M^{p'_1}_{\,l}\!\pm\! \cdots \!\pm\! M^{p'_s}_{\,l}
\!\pm\! \delta_{ll_1} \!\pm\!\cdots\!\pm\!\delta_{ll_{t}})
\end{align}
with this caution that the place in lattice and sign of each $\delta_{ll_i}$ is the same in both 
$\delta(\slashed{q}_l \cdots)$ and $\delta(\slashed{q}'_l \cdots)$, as they come from 
the term $\cos(\theta-\theta')$. The total number of terms is then $2^{r+s+t}$. 
At 3rd order in the vacuum block, we have 80 terms, which may be presented in the compact $\Delta$-notation as
\begin{align}\label{130}
\Big[\langle \sqmpb |\vred|\sqmb \rangle_{\bm{0}}&\Big]_{\gamma^3}\!\!\! =
\frac{\gamma^3}{384}\sum_{\mathrm{all}\, p}\left[\Delta\left(\frac{\slashed{q}_l }
{\slashed{q}'_{l},M^{p_1}_{\,l},M^{p_2}_{\,l},M^{p_3}_{\,l}}\right)
+\Delta\left(\frac{\slashed{q}_l ;M^{p_1}_{\,l},M^{p_2}_{\,l},M^{p_3}_{\,l}}
{\slashed{q}'_{l}}\right)\right]
\cr&+\frac{\gamma^3}{128}\sum_{\mathrm{all}\, p}\left[\Delta\left(\frac{\slashed{q}_l ;M^{p_1}_{\,l}}
{\slashed{q}'_{l},M^{p'_1}_{\,l},M^{p'_2}_{\,l}}\right)
+\Delta\left(\frac{\slashed{q}_l ;M^{p_1}_{\,l},M^{p_2}_{\,l}}
{\slashed{q}'_{l};M^{p'_1}_{\,l}}\right)\right]
\cr&+\frac{\gamma^3}{32}\sum_{\mathrm{all}\, p,\,l_1,l_2}\left[\Delta\left(\frac{\slashed{q}_l ;M^{p_1}_{\,l}}
{\slashed{q}'_{l}}; \,\delta_{ll_1},\delta_{ll_{2}}\right)
+\Delta\left(\frac{\slashed{q}_l }{\slashed{q}'_{l};M^{p'_1}_{\,l}}; \,\delta_{ll_1},\delta_{ll_{2}}\right)\right]
\cr&+\frac{\gamma^3}{64}\sum_{\mathrm{all}\, p,\,l_1}\left[\Delta\left(\frac{\slashed{q}_l ;M^{p_1}_{\,l},M^{p_2}_{\,l}}{\slashed{q}'_{l}}; \,\delta_{ll_1}\right)
+\Delta\left(\frac{\slashed{q}_l }{\slashed{q}'_{l};M^{p'_1}_{\,l},M^{p'_2}_{\,l}}; \,\delta_{ll_1}\right)\right]
\cr&+\frac{\gamma^3}{48}\sum_{{l_1},{l_2},{l_3}}\Delta\left(\frac{\slashed{q}_l }{\slashed{q}'_{l}}; \,\delta_{ll_1},\,\delta_{ll_2},\,\delta_{ll_3}\right)
+\frac{\gamma^3}{32}\sum_{\mathrm{all}\, p,\,l_1}\Delta\left(\frac{\slashed{q}_l ;M^{p_1}_{\,l}}{\slashed{q}'_{l};M^{p'_1}_{\,l}}; \,\delta_{ll_1}\right)
\end{align}

At 4th order having $15\times 2^4=240$ terms, the element in the compact notation comes to the form 
\begin{align}\label{131}
\Big[\langle \sqmpb |\vred|\sqmb \rangle_{\bm{0}}&\Big]_{\gamma^4}\!\!\! 
=\frac{\gamma^4}{6144}\sum_{\mathrm{all}\,p}\left[
\Delta\!\left(\!\frac{\slashed{q}_l }{\slashed{q}'_{l};M^{p_1}_{\,l},M^{p_2}_{\,l},M^{p_3}_{\,l},M^{p_4}_{\,l}}\!\right)
\!+\!\Delta\!\left(\!\frac{\slashed{q}_l;M^{p_1}_{\,l},M^{p_2}_{\,l},M^{p_3}_{\,l},M^{p_4}_{\,l} }{\slashed{q}'_{l}}
\!\right)\right]
\cr&+\frac{\gamma^4}{256}\sum_{\mathrm{all}\,p,\,{l_1},{l_2}}\left[
\Delta\left(\frac{\slashed{q}_l;M^{p_1}_{\,l},M^{p_2}_{\,l} }
{\slashed{q}'_{l}};\,\delta_{ll_1},\delta_{ll_{2}}\right)
+\Delta\left(\frac{\slashed{q}_l }{\slashed{q}'_{l};M^{p_1}_{\,l},M^{p_2}_{\,l}};\,\delta_{ll_1},\delta_{ll_{2}}\right)\right]
\cr&+\frac{\gamma^4}{1024}\sum_{\mathrm{all}\,p}\Delta\left(\frac{\slashed{q}_l ;M^{p_1}_{\,l},M^{p_2}_{\,l}}
{\slashed{q}'_{l};M^{p_3}_{\,l},M^{p_4}_{\,l}}\right)
+\frac{\gamma^4}{384}\sum_{{l_1},{l_2},{l_3},{l_4}}\Delta\left(\frac{\slashed{q}_l }
{\slashed{q}'_{l}};\,\delta_{ll_1},\delta_{ll_{2}},\delta_{ll_{3}},\delta_{ll_{4}}\right)
\cr& +\frac{\gamma^4}{1536}
\sum_{\mathrm{all}\,p}\left[\Delta\left(\frac{\slashed{q}_l ;M^{p_1}_{\,l},M^{p_2}_{\,l},M^{p_3}_{\,l}}
{\slashed{q}'_{l};M^{p_4}_{\,l}}\right)
+\Delta\left(\frac{\slashed{q}_l ;M^{p_1}_{\,l}}
{\slashed{q}'_{l};M^{p_2}_{\,l},M^{p_3}_{\,l},M^{p_4}_{\,l}}\right)\right]
\cr&+\frac{\gamma^4}{128}\sum_{\mathrm{all}\,p,\,{l_1},{l_2}}\Delta\left(\frac{\slashed{q}_l;M^{p_1}_{\,l} }
{\slashed{q}'_{l};M^{p_2}_{\,l}};\,\delta_{ll_1},\delta_{ll_{2}}\right)
\cr&+\frac{\gamma^4}{768}\sum_{\mathrm{all}\,p,\,{l_1}}\!\left[\!\Delta\!\left(\frac{\slashed{q}_l;M^{p_1}_{\,l},M^{p_2}_{\,l},M^{p_3}_{\,l}}{\slashed{q}'_{l}};\,\delta_{ll_1}\!\right)\!
+\!\Delta\!\left(\!\frac{\slashed{q}_l}{\slashed{q}'_{l};M^{p_1}_{\,l},M^{p_2}_{\,l},M^{p_3}_{\,l}};\,\delta_{ll_1}\!\right)\!\right]
\cr&+\frac{\gamma^4}{256}\sum_{\mathrm{all}\,p,\,{l_1}}\left[
\Delta\left(\frac{\slashed{q}_l;M^{p_1}_{\,l}}{\slashed{q}'_{l};M^{p_2}_{\,l},M^{p_3}_{\,l}};\,\delta_{ll_1}\right)
+\Delta\left(\frac{\slashed{q}_l;M^{p_1}_{\,l},M^{p_2}_{\,l}}{\slashed{q}'_{l};M^{p_3}_{\,l}};\,\delta_{ll_1}\right)\right]
\cr&+\frac{\gamma^4}{192}\sum_{{p},{l_1},{l_2},{l_3}}\left[\Delta\left(\frac{\slashed{q}_l ;M^{p}_{\,l}}
{\slashed{q}'_{l}};\,\delta_{ll_1},\delta_{ll_{2}},\delta_{ll_{3}}\right)
+\Delta\left(\frac{\slashed{q}_l }
{\slashed{q}'_{l};M^{p}_{\,l}};\,\delta_{ll_1},\delta_{ll_{2}},\delta_{ll_{3}}\right)\right]
\end{align}

\section{Numerical Factor }
Here the numerical factor appearing in (\ref{69}) is derived. By definitions
\begin{align}\label{132}
A&=\sum _p \cos (M^p_{~l} \theta^l)\\
\label{133}
B&=\sum _p \cos (M^p_{~l} \theta'^l)\\
\label{134}
C&=\sum _l \cos(\theta^l-\theta'^l)
\end{align}
we can expand the exponent as follows
\begin{align}\label{135}
e^{\gamma (\frac{1}{2}A+\frac{1}{2}B+C)}&= 
\sum_{n=0}^\infty \frac{\gamma^n}{n!}\sum_{\ell=0}^n \frac{n!}{\ell! (n-\ell)!} 
\left(\frac{A}{2}+\frac{B}{2}\right)^{n-\ell} C^\ell
\cr&= \sum_{n=0}^\infty \frac{\gamma^n}{n!}\sum_{\ell=0}^n \frac{n!}{\ell! (n-\ell)!} 
\sum_{m=0}^{n-\ell} \frac{(n-\ell)!}{m! (n-\ell-m)!} 
\left(\frac{A}{2}\right)^{m} \left(\frac{B}{2}\right)^{n-\ell-m} C^\ell 
\cr&=\sum_{n=0}^\infty \sum_{\ell=0}^n \sum_{m=0}^{n-\ell}
\frac{\gamma^n}{m! (n-\ell-m)! \ell !} 
\left(\frac{A}{2}\right)^{m} \left(\frac{B}{2}\right)^{n-\ell-m} C^\ell
\end{align}
Setting $n-\ell-m=m'$, and changing the phase in Fourier integrals
\begin{align}\label{136}
\cos \alpha=\frac{1}{2} (e^{i\alpha} + e^{-i\alpha})
\end{align}
we find the numerical factor in (\ref{69}) as 
\begin{align}\label{137}
\frac{1}{2^{2m+2m'+\ell}}\frac{1}{m!\, m'!\, \ell!}
\end{align}

\section{More by Rules}
As applications of the rules for $\gamma$-expansion of the matrix-elements,
here more examples are presented. 

In vacuum block, we consider the state with two units of currents in the first plaquette,
as $|\bm{q_{2}}\cdot\bm{M} \rangle=|\bm{2}\rangle$ with $\bm{q_2}=(2,0,...,0)$, represented by 
\begin{align}\label{138}
	\rec_2
\end{align}
Then by the rules we find for the matrix-element $\langle \bm{0}|\vred|\bm{2}\rangle_{\bm{0}}$:
\begin{align}\label{139}
	\langle \bm{0}|\vred|\bm{2}\rangle_{\bm{0}}:
	\begin{cases}
		\gamma^2: ~~~ \bm{0} ~\displaystyle{\mathop{\pmb{\longrightarrow}}^0}~ \rectwocirc_2  ~~~~~~~~ \frac{1}{2^4}\frac{1}{2!} \cr
		\gamma^4: \begin{cases}
			\bm{0}~\displaystyle{\mathop{\pmb{\longrightarrow}}^0}~ \rectwocirc_2+ \twocirc &~ \frac{1}{2^8}\frac{1}{4!} C^1_4 (3\np-2) \cr
			\bm{0}+ \twocirc~\displaystyle{\mathop{\pmb{\longrightarrow}}^0}~ \rectwocirc_2  &~ \frac{1}{2^8}\frac{1}{2! 2!} 2\np  \cr
			\bm{0} + \backforth~\displaystyle{\mathop{\pmb{\longrightarrow}}^0}~ \rectwocirc_2+ \backforth &~ \frac{1}{2^6}\frac{1}{2! 2!} 2\nl 
		\end{cases} \cr
		\gamma^6: \begin{cases}
			\bm{0}~\displaystyle{\mathop{\pmb{\longrightarrow}}^0}~ \rectwocirc_2 + \twocirc~\twocirc &~ \frac{1}{2^{12}}\frac{1}{6!} C^2_6 (12\np^2-22\np+11) \cr
			\bm{0}+ \twocirc~\twocirc~\displaystyle{\mathop{\pmb{\longrightarrow}}^0}~ \rectwocirc_2 &~ \frac{1}{2^{12}}\frac{1}{2! 4!} C^2_4 (2\np^2-\np) \cr
			\bm{0}+\twocirc~\displaystyle{\mathop{\pmb{\longrightarrow}}^0}~ \rectwocirc_2 +\twocirc &~ \frac{1}{2^{12}}\frac{1}{2!4!} C^1_2 C^1_4 \np (3\np-2) \cr
			\bm{0}+ \backforth~\displaystyle{\mathop{\pmb{\longrightarrow}}^0}~ \rectwocirc_2 +\twocirc + \backforth &~ \frac{1}{2^{10}}\frac{1}{2!4!} C^1_2 C^1_4 \nl (3\np-2) \cr
			\bm{0}+\twocirc + \backforth~\displaystyle{\mathop{\pmb{\longrightarrow}}^0}~ \rectwocirc_2  + \backforth &~ \frac{1}{2^{10}}\frac{1}{2!2!2!} C^1_2 C^1_2 \nl \np \cr
			\bm{0}~\displaystyle{\mathop{\pmb{\longrightarrow}}^0}~ \rectwocirc_2 +\backforth + \backforth &~ \frac{1}{2^8}\frac{1}{2!4!} C^2_4 (2\nl ^2-\nl)\cr
			\bm{0}+\dashreccirc\mathop{\pmb{\longrightarrow}}^0\recdashreccirc_2  &~ \frac{1}{2^8}\frac{1}{4!} C^2_4  \,4 
		\end{cases}
	\end{cases}
\end{align} 
leading to
\begin{align}\label{140}
	\langle \bm{0}|\vred|\bm{2}\rangle_{\bm{0}}=& \frac{\gamma^2}{32}
	+\left(-\frac{1}{768}+\frac{\nl}{128}+\frac{\np}{256}\right)\gamma^4
	\cr& +\left(\frac{779}{196608}-\frac{5\nl}{6144}+\frac{\nl^2}{1024}-\frac{11\np}{49152}+\frac{\nl\np}{1024}+\frac{\np^2}{4096}\right)\gamma^6+\cdots 
\end{align}
Also for the element below we have 
\begin{align}\label{141}
	\langle \bm{1}|\vred|\bm{2}\rangle_{\bm{0}}:
	\begin{cases}
		\gamma^3:~~~ \recdashcirc  ~\displaystyle{\mathop{\pmb{\longrightarrow}}^0}~ \rectwocirc_2  ~~~~~ \frac{1}{2^6}\frac{1}{2!} \cr
		\gamma^5: 
		\begin{cases}
			\recdashcirc  ~\displaystyle{\mathop{\pmb{\longrightarrow}}^0}~ \rectwocirc_2 + \twocirc &~ \frac{1}{2^{10}}\frac{1}{4!} C^1_4 (3\np-2) \cr
			\recdashcirc + \twocirc ~\displaystyle{\mathop{\pmb{\longrightarrow}}^0}~ \rectwocirc_2  &~ \frac{1}{2^{10}}\frac{1}{2!3!}C^1_3 (2\np-1) \cr
			\recdashcirc+\backforth ~\displaystyle{\mathop{\pmb{\longrightarrow}}^0}~ \rectwocirc_2 + \backforth &~ \frac{1}{2^8}\frac{1}{2!2! } C^1_2 \nl \cr
			\recdashrec ~\displaystyle{\mathop{\pmb{\longrightarrow}}^0}~ \recdashreccirc_2  &~ \frac{1}{2^6}\frac{1}{4! } C^2_4 \,4
	\end{cases}\end{cases}
\end{align}
leading to
\begin{align}\label{142}
	\langle \bm{1}|\vred|\bm{2}\rangle_{\bm{0}}=& \frac{\gamma^3}{128}
	+\left(\frac{185}{12288}+\frac{\nl}{512}+\frac{\np}{1024}\right)\gamma^5 + \cdots 
\end{align}

As the last example by $|\bm{2}\rangle$ we consider
\begin{align}\label{143}
	\langle \bm{2}|\vred|\bm{2}\rangle_{\bm{0}}:
	\begin{cases}
		\gamma^4:~~~\rectwocirc_2  ~\displaystyle{\mathop{\pmb{\longrightarrow}}^0}~ \rectwocirc_2  ~~~~ \frac{1}{2^8}\frac{1}{2!2!} \cr
		\gamma^6:
		\begin{cases}
			\rectwocirc_2  ~\displaystyle{\mathop{\pmb{\longrightarrow}}^0}~ \rectwocirc_2+ \twocirc &~ \frac{1}{2^{12}}\frac{1}{2!3!} C^1_4 (3\np-2) \cr
			\rectwocirc_2 + \twocirc ~\displaystyle{\mathop{\pmb{\longrightarrow}}^0}~ \rectwocirc_2  &~ \frac{1}{2^{12}}\frac{1}{2!3!} C^1_4 (3\np-2) \cr
			\rectwocirc_2  + \backforth ~\displaystyle{\mathop{\pmb{\longrightarrow}}^0}~ \rectwocirc_2+ \backforth &~ \frac{1}{2^{10}}\frac{1}{2! 2! 2!} 2\nl \cr
			\recdashreccirc_2 ~\displaystyle{\mathop{\pmb{\longrightarrow}}^0}~ \recdashreccirc_2  &~ \frac{1}{2^{8}}\frac{1}{4!} C^2_4  \,4 
	\end{cases}\end{cases}
\end{align}
leading to
\begin{align}\label{144}
	\langle \bm{2}|\vred|\bm{2}\rangle_{\bm{0}}=& \frac{\gamma^4}{1024}
	+\left(\frac{47}{12288}+\frac{\nl}{4096}+\frac{\np}{8192}\right)\gamma^6 +\cdots 
\end{align}
As the next example, we consider the state with three units of currents in the first plaquette  
$|\bm{q_{3}}\cdot \bm{M}\rangle=|\bm{3}\rangle$ with $\bm{q_3}=(3,0,...,0)$, represented by  
\begin{align}\label{145}
	\rec_3
\end{align}
The rules for the matrix-element $\langle \bm{0}|\vred|\bm{3}\rangle_{\bm{0}}$ give:
\begin{align}\label{146}
	\langle \bm{0}|\vred|\bm{3}\rangle_{\bm{0}}:
	\begin{cases}
		\gamma^3: ~~~\bm{0} ~\displaystyle{\mathop{\pmb{\longrightarrow}}^0}~\recthreecirc_3 
		~~~~ \frac{1}{2^6}\frac{1}{3!} \cr
		\gamma^5:
		\begin{cases} 
			\bm{0}~\displaystyle{\mathop{\pmb{\longrightarrow}}^0}~ \recthreecirc_3+ \twocirc &~ \frac{1}{2^{10}}\frac{1}{5!} C^1_5 (4\np-3) \cr
			\bm{0}+ \twocirc~\displaystyle{\mathop{\pmb{\longrightarrow}}^0}~ \recthreecirc_3 &~ \frac{1}{2^{10}}\frac{1}{2! 3!} 2\np \cr
			\bm{0} + \backforth~\displaystyle{\mathop{\pmb{\longrightarrow}}^0}~ \recthreecirc_3+ \backforth &~ \frac{1}{2^8}\frac{1}{2! 3!} 2\nl 
		\end{cases}
	\end{cases}
\end{align}
leading to
\begin{align}\label{147}
	\langle \bm{0}|\vred|\bm{3}\rangle_{\bm{0}}= \frac{\gamma^3}{384}
	+\left(-\frac{1}{8192}+\frac{\nl}{1536}+\frac{\np}{3072}\right)\gamma^5 + \cdots 
\end{align}
As another example, we consider the state with one unit of current in two neighbor plaquettes,
for which we take the first plaquette and its right-side neighbor. By the numbering 
in Fig.~3 and made as $|\bm{q_{1,1}}\cdot \bm{M}\rangle=|\bm{1,1}\rangle$ with
$\bm{q_{1,1}}=(\underbrace{1,\cdots,0}_{\np},1,0,...,0)$, represented by
\begin{align}\label{148}
	\rec\!\rec
\end{align} 
It is obvious that the net current in the shared link is zero. 
In this case, we have the following:
\begin{align}\label{149}
	\langle \bm{0}|\vred|\bm{1,1}\rangle_{\bm{0}}:
	\begin{cases}
		\gamma^2:~~~\bm{0} ~\displaystyle{\mathop{\pmb{\longrightarrow}}^0}~\recdashcirc\!\recdashcirc  ~~~~ \frac{1}{2^4}\frac{1}{2!} C^1_2 \cr
		\gamma^4: 
		\begin{cases}
			\bm{0}~\displaystyle{\mathop{\pmb{\longrightarrow}}^0}~ \recdashcirc\!\recdashcirc + \twocirc &~ \frac{1}{2^8}\frac{1}{4!} C^1_4 6(\np-1) \cr
			\bm{0}+ \twocirc~\displaystyle{\mathop{\pmb{\longrightarrow}}^0}~ \recdashcirc\!\recdashcirc   &~ \frac{1}{2^8}\frac{1}{2! 2!}C^1_2 2\np \cr
			\bm{0} + \backforth~\displaystyle{\mathop{\pmb{\longrightarrow}}^0}~ \recdashcirc\!\recdashcirc + \backforth &~ \frac{1}{2^6}\frac{1}{2! 2!} C^1_2 2\nl 
		\end{cases}
	\end{cases}
\end{align}
giving
\begin{align}\label{150}
	\langle \bm{0}|\vred|\bm{1,1}\rangle_{\bm{0}}= \frac{\gamma^2}{16}
	+\left(-\frac{1}{256}+\frac{\nl}{64}+\frac{\np}{128}\right)\gamma^4 + \cdots 
\end{align}
Another example is by the state with one unit of current in the opposite direction of $|\bm{1}\rangle$,
that is $|\bm{q_{-1}\cdot M}\rangle=|\bm{-1}\rangle$ with $\bm{q_{-1}}=(-1,0,...,0)$, for which,
with special care about the direction of arrows, we draw  
\begin{align}\label{151}
	\langle \bm{1}|\vred|\bm{-1}\rangle_{\bm{0}}:
	\begin{cases}
		\gamma^2:~~~		\recdashcirc  ~\displaystyle{\mathop{\pmb{\longrightarrow}}^0}~\recdashcircinv  
~~~~~~ \frac{1}{2^4} \cr
\gamma^4: \begin{cases}
		\recdashcirc  ~\displaystyle{\mathop{\pmb{\longrightarrow}}^0}~ \recdashcircinv + \twocirc &~ \frac{1}{2^8}\frac{1}{3!} C^1_3 (2\np-1) \cr
		\recdashcirc + \twocirc ~\displaystyle{\mathop{\pmb{\longrightarrow}}^0}~ \recdashcircinv    &~ \frac{1}{2^8}\frac{1}{3!}C^1_3 (2\np-1) \cr
		\recdashcirc+\backforth ~\displaystyle{\mathop{\pmb{\longrightarrow}}^0}~ \recdashcircinv + \backforth &~ \frac{1}{2^6}\frac{1}{2! } C^1_2 \nl 
		\end{cases}
	\end{cases}
\end{align}
leading to
\begin{align}\label{152}
	\langle \bm{1}|\vred|\bm{-1}\rangle_{\bm{0}}= \frac{\gamma^2}{16}
	+\left(-\frac{1}{256}+\frac{\nl}{64}+\frac{\np}{128}\right)\gamma^4 + \cdots 
\end{align}

As the other example transition between $|\bm{1}\rangle$ and one unit of current in another
plaquette than the first one, showing as $|\bm{1'}\rangle$, is represented as 
\begin{align}\label{153}
	\rec'
\end{align}
for which we find for the matrix-element
\begin{align}\label{154}
	\langle \bm{1}|\vred|\bm{1'}\rangle_{\bm{0}}:
	\begin{cases}
		\gamma^2:~~~\recdashcirc  ~\displaystyle{\mathop{\pmb{\longrightarrow}}^0}~\recdashcirc'   ~~~~ \frac{1}{2^4} \cr
		\gamma^4: \begin{cases}
			\recdashcirc  ~\displaystyle{\mathop{\pmb{\longrightarrow}}^0}~ \recdashcirc'  + \twocirc &~ \frac{1}{2^8}\frac{1}{3!} C^1_3 (2\np-1) \cr
			\recdashcirc + \twocirc ~\displaystyle{\mathop{\pmb{\longrightarrow}}^0}~ \recdashcirc'  &~ \frac{1}{2^8}\frac{1}{3!}C^1_3 (2\np-1) \cr
			\recdashcirc+\backforth ~\displaystyle{\mathop{\pmb{\longrightarrow}}^0}~ \recdashcirc' + \backforth &~ \frac{1}{2^6}\frac{1}{2! } C^1_2 \nl 
		\end{cases}
	\end{cases}
\end{align}
leading to
\begin{align}\label{155}
	\langle \bm{1}|\vred|\bm{1'}\rangle_{\bm{0}}= \frac{\gamma^2}{16}
	+\left(-\frac{1}{256}+\frac{\nl}{64}+\frac{\np}{128}\right)\gamma^4 + \cdots 
\end{align}
As a final example in the vacuum block, we consider $|\bm{1}\rangle$ and the state
with two units of currents in opposite directions as $|\bm{-2}\rangle$, 
represented by 
\begin{align}\label{156}
	\recinv_{-2}
\end{align}
Then the rules, with special care about the direction of arrows, give the following:  
\begin{align}\label{157}
	\langle \bm{1}|\vred|\bm{-2}\rangle_{\bm{0}}:
	\begin{cases}
		\gamma^3:~~\recdashcirc  ~\displaystyle{\mathop{\pmb{\longrightarrow}}^0}~ \rectwocircinv_{-2}  ~~~~~ \frac{1}{2^6}\frac{1}{2!} \cr
		\gamma^5:
		\begin{cases}
			\recdashcirc  ~\displaystyle{\mathop{\pmb{\longrightarrow}}^0}~ \rectwocircinv_{-2} + \twocirc &~ \frac{1}{2^{10}}\frac{1}{4!} C^1_4 (3\np-2) \cr
			\recdashcirc + \twocirc ~\displaystyle{\mathop{\pmb{\longrightarrow}}^0}~ \rectwocircinv_{-2}  &~ \frac{1}{2^{10}}\frac{1}{2!3!}C^1_3 (2\np-1) \cr
			\recdashcirc+\backforth ~\displaystyle{\mathop{\pmb{\longrightarrow}}^0}~ \rectwocircinv_{-2} + \backforth &~ \frac{1}{2^8}\frac{1}{2!2! } C^1_2 \nl 
	\end{cases}\end{cases}
\end{align}
leading to
\begin{align}\label{158}
	\langle \bm{1}|\vred|\bm{-2}\rangle_{\bm{0}}= \frac{\gamma^3}{128}
	+\left(-\frac{7}{12288}+\frac{\nl}{512}+\frac{\np}{1024}\right)\gamma^5 + \cdots 
\end{align}

Samples of the non-vacuum block of $\bm{k_1}$ are already given. Here as extra examples 
in the non-vacuum block, first we consider the co-blocks of vector $\bm{k_2}=(2,0,...,0)$ 
with two units of currents on the first link. 
The lowest order of this block occurs when $\bm{q'}=\bm{q}=\bm{0}$, given by
\begin{align}\label{159}
	\langle \bm{k_2}|\vred|\bm{k_2}\rangle_{\bm{2}}:
	\begin{cases}
		\gamma^2:~
		\backforthhor_2 ~\displaystyle{\mathop{\pmb{\longrightarrow}}^0}~\backforthhor_2 ~~~~
		~~~~~~~~~~ \frac{1}{8} \cr
		\gamma^4:
		\begin{cases}
			\backforthhor_2  ~\displaystyle{\mathop{\pmb{\longrightarrow}}^0}~ \backforthhor_2 +\twocirc &~ \frac{1}{2^6 2!2!}C^1_2 \np \cr
			\backforthhor_2 +\twocirc ~~\displaystyle{\mathop{\pmb{\longrightarrow}}^0}~\backforthhor_2 &~ \frac{1}{2^6 2!2!}C^1_2 \np \cr
			\backforthhor_2 +\backforth ~\displaystyle{\mathop{\pmb{\longrightarrow}}^0}~ \backforthhor_2+\backforth &~ \frac{1}{2^4 4!}C^1_4 (3\nl-2) 
		\end{cases}
	\end{cases}
\end{align}
leading to 
\begin{align}\label{160}
	\langle \bm{k_2}|\vred|\bm{k_2}\rangle_{\bm{2}}=  \frac{\gamma^2}{8}
	+\left(-\frac{1}{48}+\frac{\nl}{32}+\frac{\np}{64}\right)\gamma^4+\cdots
\end{align}
As a co-block of $\bm{k_2}$ we consider $\bm{k_{2;1}}=\bm{k_2}+\slashed{\bm{q}}_{\bm{1}}$,
which has three units of currents on the first link, and two units of currents on other links of the first plaquette.
Similar to (\ref{62}) we may represent it as 
\begin{align}\label{161}
	\mathop{\rec}_{\linkcur\,2}
\end{align}
The graphical representation for $\bm{q'}=\bm{0}$ and $\bm{q}=\bm{q_1}$ will be given as follows:
\begin{align}\label{162}
	\langle \bm{k_2}|\vred|\bm{k_{2;1}}\rangle_{\bm{2}}:
	\begin{cases}
		\gamma^3: ~~\backforthhor_2~\displaystyle{\mathop{\pmb{\longrightarrow}}^0}~\mathop{\recdashcirc}_{\backforthhor\,2} 
		~~~~~~~ \frac{1}{32} \cr
		\gamma^5:
		\begin{cases}
			\backforthhor_2  ~\displaystyle{\mathop{\pmb{\longrightarrow}}^0}~\mathop{\recdashcirc}_{\backforthhor\,2} +\twocirc &~ 
			\frac{1}{2^8 2! 3!}C^1_3 (2\np-1) \cr
			\backforthhor_2 +\twocirc~\displaystyle{\mathop{\pmb{\longrightarrow}}^0}~ \mathop{\recdashcirc}_{\backforthhor\,2}  &~ 
			\frac{1}{2^8 2!2!}C^1_2 \np \cr
			\backforthhor_2 +\backforth
			~\displaystyle{\mathop{\pmb{\longrightarrow}}^0}~ \mathop{\recdashcirc}_{\backforthhor\,2} +\backforth &~ \frac{1}{2^6 4!}C^1_4 (3\nl-2) 
		\end{cases}
	\end{cases}
\end{align}
all together leading to
\begin{align}\label{163}
	\langle \bm{k_2}|\vred|\bm{k_{2;1}}\rangle_{\bm{2}}=  \frac{\gamma^3}{32}
	+\left(-\frac{19}{3072}+\frac{\nl}{128}+\frac{\np}{256}\right)\gamma^5+\cdots
\end{align}
The graphical representation for the last example of this non-vacuum block is
$\bm{q'}=\bm{q}=\bm{q_1}$:
\begin{align}\label{164}
	\langle \bm{k_{2;1}}|\vred|\bm{k_{2;1}}\rangle_{\bm{2}}:\!
	\begin{cases}
		\gamma^4\!:~ \displaystyle{\mathop{\recdashcirc}_{\backforthhor\,2}}  ~\displaystyle{\mathop{\pmb{\longrightarrow}}^0}~\mathop{\recdashcirc}_{\backforthhor\,2}   ~~~~~~~~~~~\frac{1}{2^7} 
		\vspace{2mm}\cr
		\gamma^6\!:\!
		\begin{cases}
			\displaystyle{\mathop{\recdashcirc}_{\backforthhor\,2}} ~\displaystyle{\mathop{\pmb{\longrightarrow}}^0}~ \displaystyle{\mathop{\recdashcirc}_{\backforthhor\,2}}
			\!+\!\twocirc & \frac{1}{2^{10}}\frac{1}{3!2!}C^1_3  (2\np-1) \vspace{2mm}\cr
			\displaystyle{\mathop{\recdashcirc}_{\backforthhor\,2}} +\twocirc 
			~\displaystyle{\mathop{\pmb{\longrightarrow}}^0}~\displaystyle{\mathop{\recdashcirc}_{\backforthhor\,2}}
			&  \frac{1}{2^{10}}\frac{1}{3!2!}C^1_3  (2\np-1) \vspace{2mm}\cr
			\displaystyle{\mathop{\recdashcirc}_{\backforthhor\,2}} +\backforth ~\displaystyle{\mathop{\pmb{\longrightarrow}}^0}~ \displaystyle{\mathop{\recdashcirc}_{\backforthhor\,2}}\!+\!\backforth & \frac{1}{2^8} \frac{1}{4!}C^1_4 (3\nl-2) 
			\vspace{2mm}\cr
			\displaystyle{\mathop{\recdashcirc}_{\backforthhor\,2}}
			~\displaystyle{\mathop{\pmb{\longrightarrow}}^0}~ \displaystyle{\mathop{\recdashcirc}_{\backforthhor\,2}}
			& \frac{1}{2^6} \frac{1}{6!} C^2_6 \,8
		\end{cases}
	\end{cases}
\end{align}
leading to
\begin{align}\label{165}
	\langle \bm{k_{2;1}}|\vred|\bm{k_{2;1}}\rangle_{\bm{2}}=&  \frac{\gamma^4}{128}
	+\left(\frac{5}{6144}+\frac{\nl}{512}+\frac{\np}{1024}\right)\gamma^6+\cdots
\end{align}

As the last example, we consider the block by the representative state
$\bm{k_{1,1'}}$ in which two separated links have one unit of currents, showing them as
\begin{align}\label{166}
	\displaystyle{\mathop{\raisebox{-1mm}{\linkcur}}^{\raisebox{1mm}{$~~~\linkcur'$}}}
\end{align}
Then by the rules we have
\begin{align}\label{167}
	\langle \bm{k_{1,1'}}|\vred|\bm{k_{1,1'}}\rangle_{\bm{1,1'}}:
	\begin{cases}
		\gamma^2:~~
		\displaystyle{\mathop{\raisebox{-2mm}{\backforthhor}}^{\raisebox{0.5mm}{$~~~\backforthhor'$}}}
		~\displaystyle{\mathop{\pmb{\longrightarrow}}^0}~ 
		\displaystyle{\mathop{\raisebox{-2mm}{\backforthhor}}^{\raisebox{0.5mm}{$~~~\backforthhor'$}}}
		~~~~~~~~~ C^1_2 \frac{1}{8} \vspace{1mm}\cr
		\gamma^4:
		\begin{cases}
			\displaystyle{\mathop{\raisebox{-2mm}{\backforthhor}}^{\raisebox{0.5mm}{$~~~\backforthhor'$}}} ~\displaystyle{\mathop{\pmb{\longrightarrow}}^0}~ \displaystyle{\mathop{\raisebox{-2mm}{\backforthhor}}^{\raisebox{0.5mm}{$~~~\backforthhor'$}}} +\twocirc &~ \frac{1}{2^6 2!2!}C^1_2~C^1_2 \np \vspace{2mm}\cr
			\displaystyle{\mathop{\raisebox{-2mm}{\backforthhor}}^{\raisebox{0.5mm}{$~~~\backforthhor'$}}} +\twocirc ~\displaystyle{\mathop{\pmb{\longrightarrow}}^0}~ \displaystyle{\mathop{\raisebox{-2mm}{\backforthhor}}^{\raisebox{0.5mm}{$~~~\backforthhor'$}}} &~ \frac{1}{2^6 2!2!}C^1_2 ~C^1_2\np \vspace{2mm}\cr
			\displaystyle{\mathop{\raisebox{-2mm}{\backforthhor}}^{\raisebox{0.5mm}{$~~~\backforthhor'$}}} +\backforth ~\displaystyle{\mathop{\pmb{\longrightarrow}}^0}~ \displaystyle{\mathop{\raisebox{-2mm}{\backforthhor}}^{\raisebox{0.5mm}{$~~~\backforthhor'$}}}+\backforth &~ \frac{1}{2^4 4!}C^1_4 (6\nl-6) 
		\end{cases}
	\end{cases}
\end{align}
leading to
\begin{align}\label{168}
	\langle \bm{k_{1,1'}}|\vred|\bm{k_{1,1'}}\rangle_{\bm{1,1'}}=&  \frac{\gamma^2}{4}
	+\left(-\frac{1}{16}+\frac{\nl}{16}+\frac{\np}{32}\right)\gamma^4+\cdots
\end{align}

\end{appendices}

\vskip .5cm
\textbf{Acknowledgment:}
The helpful comments by M. Khorrami on the manuscript are gratefully acknowledged. 
The authors also would like to thank the anonymous referee for providing detailed and useful 
comments leading to directions to improve this work. 
This work is supported by the Research Council of Alzahra University.


\begin{thebibliography}{99}

\bibitem{fey} R.P. Feynman, ``Space-Time Approach to Quantum Electrodynamics",
Phys. Rev. \textbf{76} (1949) 769.

\bibitem{kinoshita} T. Aoyama, M. Hayakawa, T. Kinoshita, and M. Nio, ``Quantum Electrodynamics Calculation of Lepton
Anomalous Magnetic Moments: Numerical Approach to the Perturbation Theory of QED", 
Prog. Theor. Exp. Phys. \textbf{2012} (2012) 01A107. 

\bibitem{wilson} K.G. Wilson, 
``Confinement of Quarks",
Phys. Rev. D \textbf{10} (1974) 2445.

\bibitem{kogut} J.B. Kogut, 
``An Introduction to Lattice Gauge Theory and Spin Systems",
Rev. Mod. Phys. \textbf{51} (1979) 659.

\bibitem{savit} R. Savit, 
``Topological Excitations in U(1)-Invariant Theories",
Phys. Rev. Lett. \textbf{39} (1977) 55.

\bibitem{banks} T. Banks, R. Myerson, and J.B. Kogut,
``Phase Transitions in Abelian Lattice Gauge Theories",
Nucl. Phys. B \textbf{129} (1977) 493.

\bibitem{villain} J. Villain, 
“Theory of One- and Two-Dimensional Magnets with an 
Easy Magnetization Plane II. The Planar, Classical, Two-Dimensional Magnet”, 
J. Phys. (France) \textbf{36} (1975) 581.

\bibitem{gatt} C. Gattringer,
``New Developments for Dual Methods in Lattice Field Theory at Non-Zero Density",
PoS LATTICE 2013 \textbf{002} (2014), 1401.7788 [hep-lat].

\bibitem{zach1} M. Zach, M. Faber, and P. Skala,
``Investigating Confinement in Dually Transformed U(1) Lattice Gauge Theory",
Phys. Rev. D \textbf{57} (1998) 123-131, hep-lat/9705019

\bibitem{zach2} M. Zach, M. Faber, and P. Skala,
``Flux Tubes and Their Interaction in U(1) Lattice Gauge Theory",
Nucl. Phys. B \textbf{529} (1998) 505-522, hep-lat/9709017

\bibitem{panero} M. Panero,
``A Numerical Study of Confinement in Compact QED",
JHEP \textbf{0505} (2005) 066, hep-lat/0503024

\bibitem{boris}
O. Borisenko, V. Chelnokov, M. Gravina, and A. Papa,
``Deconfinement and Universality in the 3D U(1) Lattice Gauge Theory at Finite 
Temperature: Study in the Dual Formulation",
JHEP \textbf{09} (2015) 062, 1507.00833 [hep-lat].

\bibitem{casel} M. Caselle, M. Panero, and D. Vadacchino,
``Width of the Flux Tube in Compact U(1) Gauge Theory in Three Dimensions",
JHEP \textbf{02} (2016) 180, 1601.07455 [hep-lat].

\bibitem{vadfat} N. Vadood and A.H. Fatollahi, 
``On U(1) Gauge Theory Transfer-Matrix in Fourier Basis",
Commun. Theor. Phys. \textbf{71} (2019) 921-926, 1807.03150 [hep-lat].

\bibitem{luscher} M. Luscher, 
``Construction of a Selfadjoint, Strictly Positive Transfer Matrix for Euclidean Lattice Gauge Theories",
Commun. Math. Phys. \textbf{54} (1977) 283-292.  

\bibitem{seiler} K. Osterwalder and E. Seiler, 
``Gauge Field Theories on a Lattice",
Ann. Phys. \textbf{110} (1978) 440-471.

\bibitem{normfixing} N.~Vadood and A.H. Fatollahi,
``Lost in Normalization", 
Europhys. Lett. \textbf{131} (2020) 41003,
1803.05497 [hep-lat].

\bibitem{rohrlich} J.M. Jauch and F. Rohrlich, ``The Theory of Photons and Electrons", 
Springer, 2nd edition (1980), ch.~9.

\bibitem{peskin} M.E. Peskin and D.V. Schroeder, ``An Introduction to Quantum Field Theory",
CRC Press (2018), ch.~4.

\bibitem{ryder} L.H. Ryder, ``Quantum Field Theory", Cambridge University Press, 2nd edition (1996), ch.~6.

\bibitem{huang} K. Huang, ``Statistical Mechanics", Wiley 2nd edition (1987).

\bibitem{krawan} H.A. Kramers and G.H. Wannier, 
``Statistics of the Two-Dimensional Ferromagnet. Part I", 
Phys. Rev. \textbf{60} (1941) 252.

\bibitem{onsager} L. Onsager,
``Crystal Statistics. I. A Two-Dimensional Model with an Order-Disorder Transition", 
Phys. Rev. \textbf{65} (1944) 117.

\end{thebibliography}
\end{document}